\documentclass{aa}  
\usepackage{graphicx}
\usepackage{txfonts}
\usepackage{multirow}
\usepackage{hhline}
\usepackage{amssymb}
\usepackage{subcaption}
\usepackage{xcolor}

\newcommand\kms{km s$^{-1}$}
\newcommand\HttCO{H$^{13}$CO$^+$}
\newcommand\HCO{HCO$^+$}

\usepackage{natbib}
\bibpunct{(}{)}{;}{a}{}{,} 
\usepackage{tablefootnote}
\usepackage{ulem}

\usepackage{hyperref}
\hypersetup{
    colorlinks = true,
    citecolor = {blue},
    linkcolor = {blue},
    urlcolor = {blue},
    linkbordercolor = {white}}

\begin{document} 

   \title{CASCADE: Filamentary accretion flows in Cygnus X DR20}


   \author{M.~Sawczuck
          \inst{1},
          H.~Beuther\inst{1},
          S.~Suri\inst{1,2},
          F.~Wyrowski\inst{3},
          K.M.~Menten\inst{3}\thanks{In memory of Karl Menten, who suddenly passed away before completing this work. His invaluable advice and contributions will be deeply missed.}, 
          J.M.~Winters\inst{4},
          L.~Bouscasse\inst{4},
          N.~Schneider\inst{5},
          T.~Csengeri\inst{6},
          C.~Gieser\inst{1,7},
          S.~Li\inst{1},
          D.~Semenov\inst{1},
          I.~Skretas\inst{3},
          M.R.A.~Wells\inst{1},
          }

   \institute{\inst{1} Max Planck Institute for Astronomy, K\"onigstuhl 17, 69117 Heidelberg, Germany, email: name@mpia.de\\
   \inst{2} Department of Astrophysics, University of Vienna, T\"urkenschanzstrasse 17,1180 Vienna, Austria\\
   \inst{3} Max-Planck-Institut f\"ur Radioastronomie, Auf dem H\"ugel 69, 53121 Bonn, Germany\\
   \inst{4} IRAM, 300 rue de la Piscine, Domaine Universitaire de Grenoble, 38406 St.-Martin-d’Hères, France\\
   \inst{5}  I. Physik. Institut, University of Cologne, Cologne, Germany\\
   \inst{6} Laboratoire d’astrophysique de Bordeaux, Univ. Bordeaux, CNRS, B18N, allée Geoffroy Saint-Hilaire, 33615 Pessac,France\\
   \inst{7} Max Planck Institute for Extraterrestrial Physics, Gießenbachstraße 1, 85749 Garching bei München, Germany
             }

   \date{Received September 15, 1996; accepted March 16, 1997}

\abstract
  {Filamentary gas flows are an important process for funneling gas from cloud scales onto star-forming cores.}
  {We investigate the role of filaments in high-mass star formation, whether gas flows from large to small scales along them, and what their properties might reveal about the region they are found in.}
  {The Max Planck IRAM Observatory Program (MIOP), the Cygnus Allscale Survey of Chemistry and Dynamical Environments (CASCADE), includes high spatial resolution ($\sim$3$''$) data of \HCO$(1-0)$ and \HttCO$(1-0)$ emission in the star-forming DR20 region in the Cygnus X complex. In this data we identify filaments with the structure identification algorithm DisPerSE. We further analyze these filaments using Gaussian fits to the spectra to determine the line peak velocity and full width half maximum along them. The Python package FilChaP was used to determine filament widths.}
  {We find projected velocity gradients inside several filaments between 0.4 to 2.4\,\kms\ over projected length-scales of 0.1\,pc toward star-forming cores. This can be interpreted as a sign of gas flowing along the filaments toward the cores. The filament width distributions exhibit median values between 0.06 and 0.14\,pc depending on the core, the tracer, and the method. Standard deviations are approximately 0.02 to 0.06\,pc. These values are roughly in agreement with the filament width of 0.1\,pc typically found in nearby low-mass star-forming regions.}
  {This first analysis of filamentary properties within the Cygnus X CASCADE program reveals potential signatures of gas flows along filaments onto star-forming cores. Furthermore, the characteristics of the filaments in this high-mass star-forming region can be compared to those of filaments in low-mass star-forming regions typically studied before. Extending such studies to the entire CASCADE survey will enhance our knowledge of high-mass filament properties on solid statistical grounds.}


\authorrunning{M. Sawczuck et al.}
   \maketitle
%

\section{Introduction}

The filamentary nature of molecular clouds has been observed for decades \citep[e.g.,][]{arzoumanian2011,andre2014,pineda2022,hacar_initial_2022}, and the ubiquity of filaments has been revealed by \textit{Herschel} observations (e.g., \citealt{andre_filamentary_2010,motte2010,schisano2020}) of nearby molecular clouds. Filaments are found over a large range of scales, from small substructures of clouds \citep[$\sim$ 0.1 pc, e.g.,][]{hacar2018,li2022} up to giant structures associated with galactic spiral arms \citep[several kiloparsecs, e.g.,][]{ragan2012,goodman2014,zucker_skeleton_2015,wang2015,abreu2016,mattern2018}. They are also identified in various environments, from dense high-mass star-forming regions \citep[e.g.,][]{hill2011} to the atomic interstellar medium (ISM) \citep[e.g.,][]{clark_magnetically_2014,syed_atomic_2020} \citep[for a recent review see][]{hacar_initial_2022}. This suggests that filaments may precede and lead to star formation (e.g., \citealt{andre_filamentary_2014,pineda2022}). The general picture is that large-scale supersonic flows compress gas into such filamentary structures, which further fragment into pre-stellar cores and ultimately protostars when gravity takes over \citep[e.g.,][]{inoue2009,heitsch2009,hennebelle2013,klessen2010,tielens_molecular_2021}. Since high densities are observed at filament intersections, referred to as “hubs,” high-mass star formation is thought to take place in these locations, with gas flowing along the filaments \citep[e.g.,][]{schneider2012,peretto2014,trevino-morales_dynamics_2019,kumar_unifying_2020}. The DR21 ridge in Cygnus X is a well-studied example of such a process (e.g., \citealt{hennemann2012}). It was shown that parsec-scale sub-filaments, connected to the ridge, provide a significant mass reservoir to further build up the DR21 ridge, probably guided by the magnetic field \citep{schneider2010}. On much smaller scales ($<$0.1\,pc), \citet{csengeri2011} found velocity-coherent flows toward some of the massive star-forming cores in Cygnus X. In this paper, we aim to investigate such gas flows at scales of star-forming regions (parsec to sub-parsec) and characterize their physical properties.

\begin{figure*}[h]
\sidecaption
	\includegraphics[width=12cm ,keepaspectratio]
	{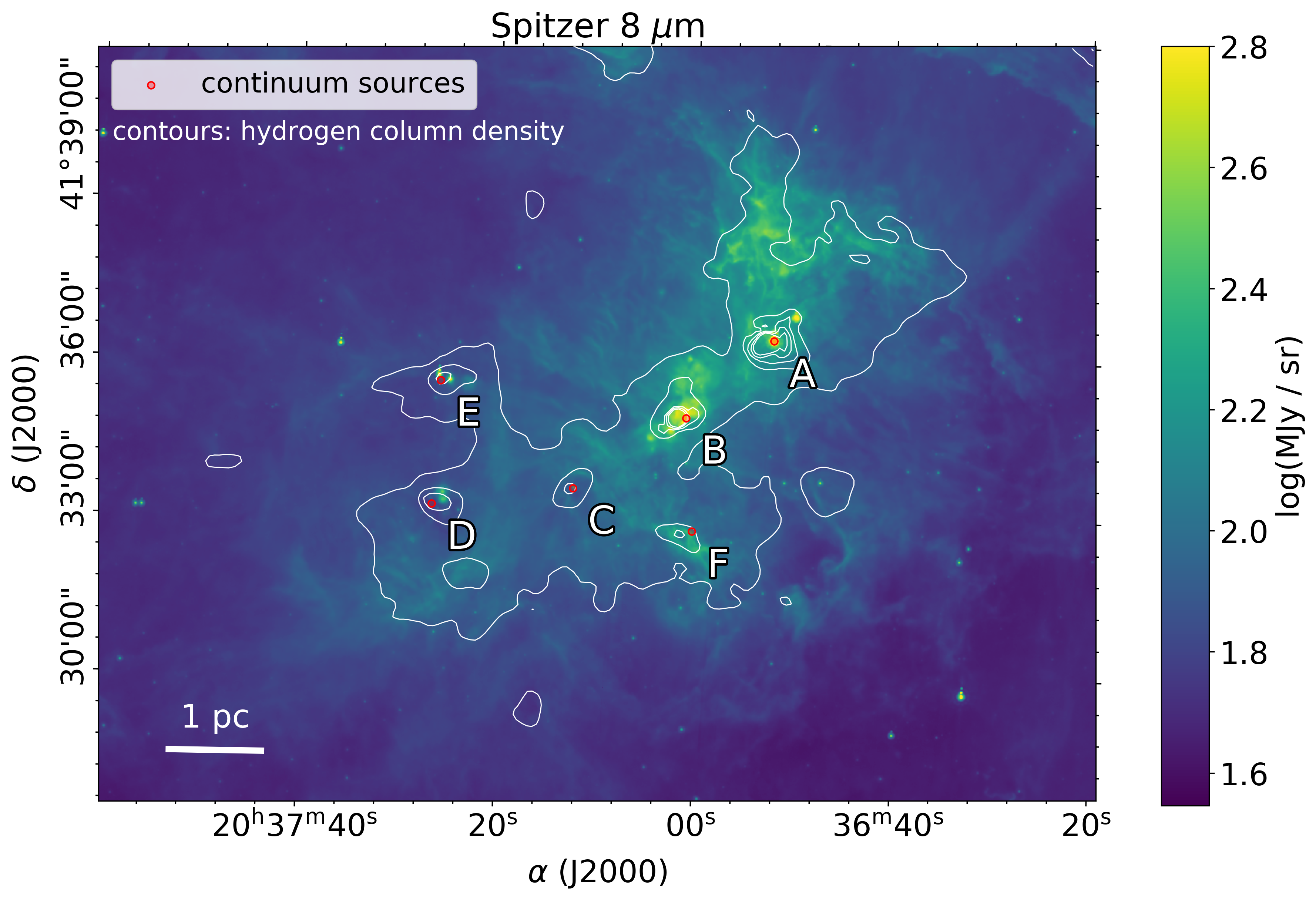}
	\caption[Overview of DR20.]{Overview of the DR20 region in Spitzer 8 $\mu$m emission in color-scale with white contours showing the hydrogen column density in five steps from 1 to 5 $\times$ 10$^{22}$\,H$_2$\,cm$^{-2}$, derived with Herschel data by \cite{marsh_multitemperature_2017}. The red dots indicate the NOEMA 3.6 mm continuum intensity peaks, which are labeled from A to F \citep{beuther_cygnus_2022}.}
	\label{plot_dr20_overview}
\end{figure*}

Furthermore, several studies so far have found filament widths around 0.1 pc, independent of the resolution and the fitting methods of the radial profiles, therefore suggesting a typical filament width (e.g., \citealt{arzoumanian_characterizing_2011,andre_filamentary_2014,suri_carma-nro_2019}). \cite{andre_filamentary_2014} and \citet{andre2022} summarized possible physical processes that could lead to this; for example, filaments may be accreting additional material from their surroundings while contracting, which would constrain their central diameter on the order of the effective Jeans length, i.e., $\sim$0.1\,pc. However, in a recent review, \citet{hacar_initial_2022} presented a meta-analysis of filament widths with a broader spread in measured filament width. Hence, it remains an open question whether a universal filament width exists or not. While the majority of prior filament studies have concentrated on low-mass star-forming clouds, there exists a paucity of investigations examining high-mass star-forming clouds. Single-dish continuum studies typically found broader filament widths in high-mass regions (e.g., \citealt{hennemann2012,schisano2014}), which may also be related to the limited angular resolution, whereas some studies within Orion also found narrower filament widths (e.g., \citealt{hacar2018,monsch2018}).
We now investigate the filament properties in a high-mass star-forming region in the Cygnus X complex and look for possible differences in the evolutionary star formation stages they are found in. Given the typically larger distances of high-mass star-forming regions, high angular resolution is crucial to disentangle the structures.

For our investigation, we chose the Cygnus X region. It is a nearby \citep[1.46\,kpc,][]{rygl_parallaxes_2012}, very luminous star-forming region, which includes various stages of long-lasting (at least for the past few million years) and on-going massive star formation events \citep[e.g.,][]{schneider2006,reipurth_star_2008,wright2015}. 
We use high spatial resolution data obtained by combining the NOrthern Extended Millimeter Array (NOEMA) data with the Institut de Radioastronomie Millim\'etrique IRAM\,30\,m single-dish data taken for the Max Planck IRAM Observatory Program (MIOP), the Cygnus Allscale Survey of Chemistry and Dynamical Environments (CASCADE; \cite{beuther_cygnus_2022} present an overview and first results). This combination allows us to look at filaments at (sub-)parsec scales (down to approximately 10$^{-2}$ pc or 3700 AU), where we aim to study the connection of the larger, diffuse scales to the smaller, denser regions and embedded cores.
The data cover a large area (640 arcmin$^2$), i.e., all important subregions of Cygnus X, with a very broad band pass of $\sim$ 15.51 GHz, covering a plethora of spectral lines; see \citet{beuther_cygnus_2022}, who outline the wealth of data.

Only noting this wealth of spectral lines available in CASCADE, we focus here on the dense gas-tracing molecule \HCO$(1-0)$ and its isotopologue \HttCO$(1-0)$. While the former also traces more extended gas emission around the star-forming regions, the latter more closely follows the dust continuum emission and hence traces the densest core regions.

This work concentrates on the pilot region of CASCADE, the DR20 subregion at a distance of 1.46 kpc \citep{rygl_parallaxes_2012} with a total mass estimate of $\sim$ 2500 $M_\odot$ \citep{beuther_cygnus_2022}. It shows strong centimeter and millimeter continuum emission, with 6 mm intensity peaks labeled A to F (see Fig. \ref{plot_dr20_overview}). These peaks show different evolutionary stages of star formation, with a gradient in evolutionary stage from A (most evolved) to F (very young) \citep[e.g.,][]{beuther_cygnus_2022}. A and B are associated with evolved (ultracompact) H{\sc ii} regions with radio and strong mid-infrared emission and elevated dust temperatures around 25 to 30 K, with B having slightly lower temperatures than A \citep{marsh_multitemperature_2017,beuther_cygnus_2022}. Additionally, B exhibits strong compact molecular emission and a 6.7\,GHz Class II CH$_3$OH maser \citep{ortiz-leon_global_2021}. Meanwhile, clumps C to F are more quiescent at radio and mid-infrared; they are also colder at about 20 K and are therefore thought to be younger, less-evolved star-forming regions \citep{beuther_cygnus_2022}. These different regions, which are found to be in various evolutionary stages of star formation, make the DR20 subregion a very interesting region to investigate potential filamentary gas flows connecting to the star-forming cores.

\begin{figure*}[h]
	\centering
        \begin{subfigure}{0.47\textwidth}
  		\centering
		\includegraphics[width=1\linewidth ,keepaspectratio]
	       {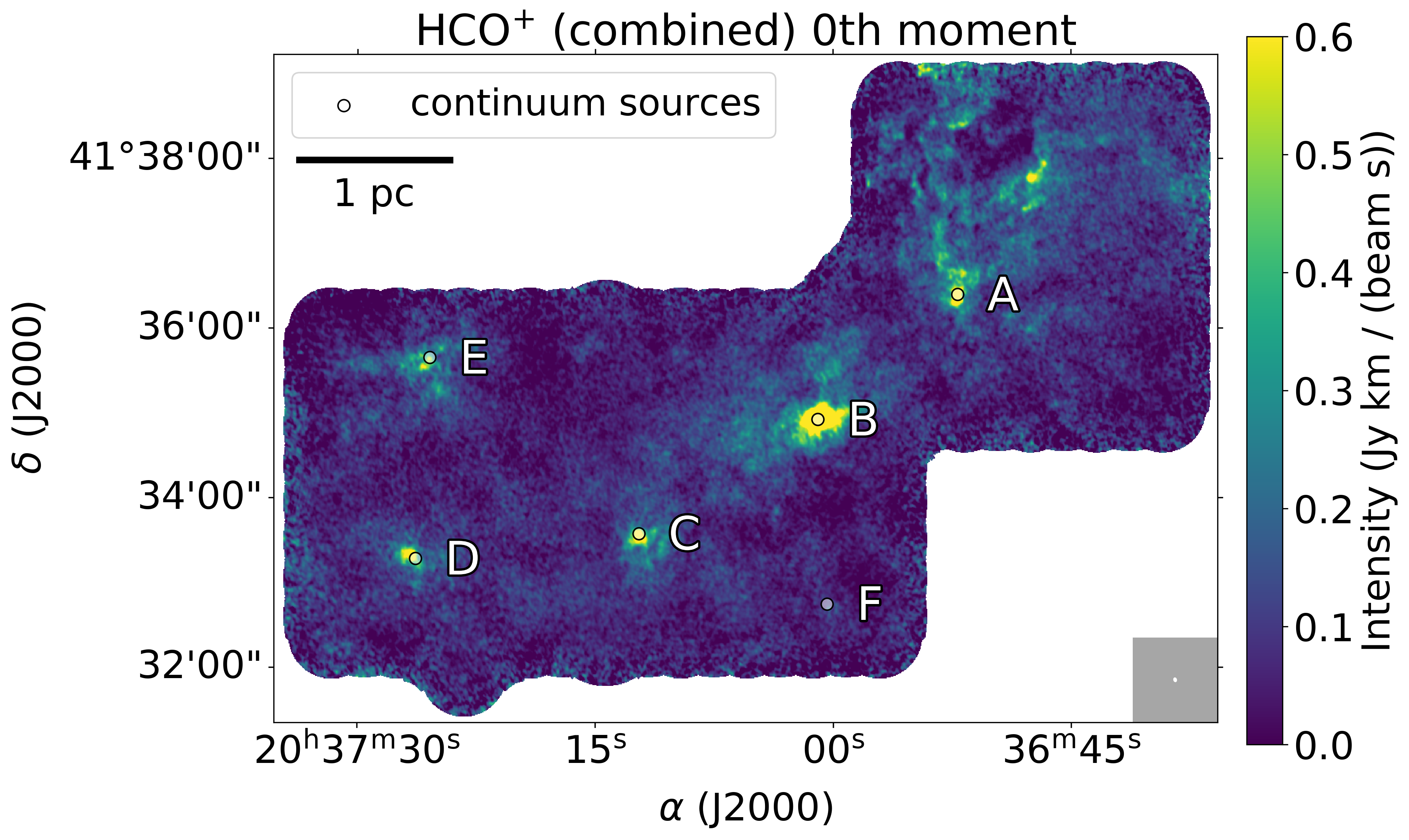}
	\end{subfigure}%
        \begin{subfigure}{0.47\textwidth}
  		\centering
		\includegraphics[width=1\linewidth ,keepaspectratio]
		{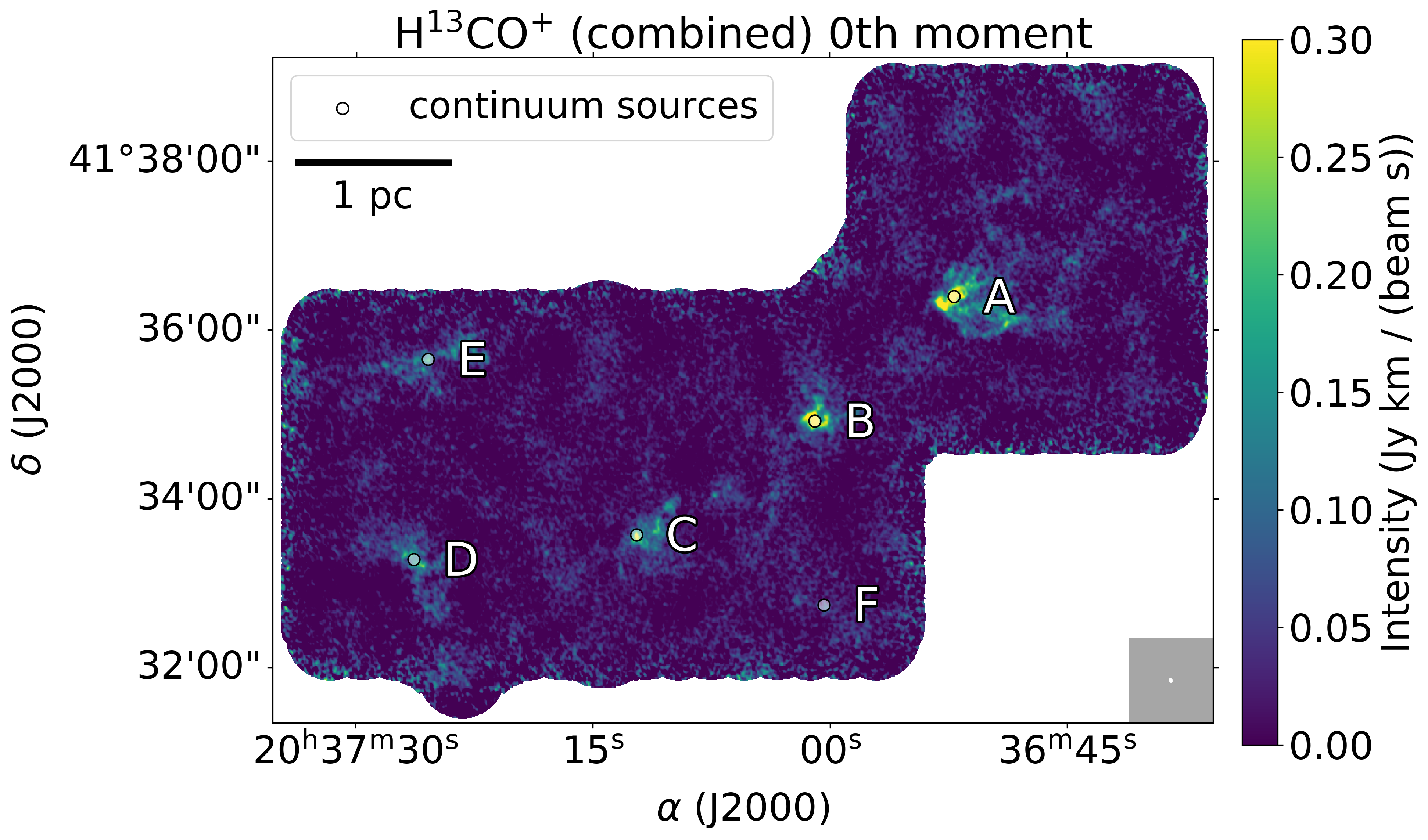}
	\end{subfigure}%
	\caption[Zeroth moment map of combined \HCO\ and \HttCO\ data.]{Integrated intensities (from $-$8.4 km s$^{-1}$ to 6 km s$^{-1}$) for \HCO$(1-0)$ (left) and \HttCO$(1-0)$ (right). The intensity peaks of the NOEMA 3.6\,mm continuum emission (continuum sources) are labeled A to F. The NOEMA synthesized beam is shown in the bottom-right.}
	\label{plot_HCO+_H13CO+_maps}
\end{figure*}

The combined NOEMA and IRAM\,30\,m data have a velocity resolution of 0.8 km s$^{-1}$, making it possible to investigate the gas kinematics along the filaments. By investigating these properties of the filament structures in the different star formation stages of DR20, we try to answer whether filaments connect the large and small scales; whether gas flows along the filaments into the cores; whether filament properties depend on the evolutionary stage of star formation; and whether a typical filament width may also be found in this high-mass region.

First, we summarize the properties of the data used (Sect. \ref{sect_obs}), followed by the methods used to identify filaments; then, we determine their kinematics and width distributions (Sect. \ref{sect_meth}). Next, we show the identified filaments and their resulting properties (Sect. \ref{sect_results}). After a discussion of the results (Sect. \ref{sect_disc}), we end with a summary and conclusion (Sect. \ref{sect_concl}).
   
\section{Observations}
\label{sect_obs}

The data we use for this work are taken from "The Cygnus Allscale Survey of Chemistry and Dynamical Environments" (CASCADE\footnote{\url{https://www2.mpia.de/CASCADE}}). 
The data and the reduction procedures are described in detail in
\cite{beuther_cygnus_2022}, which we summarize here briefly.
CASCADE combines NOEMA 3.6\,mm interferometric data with IRAM 30 m single-dish radio data of the same target region Cygnus X. While the NOEMA data give high spatial resolution for small-scale structures ($\sim$3$''$ correspond to $\sim$4380\,AU at the distance of DR20), the IRAM 30 m complements the missing short spacings to map extended structures of several parsec size. The combination of these two datasets therefore allows us to connect large to small spatial scales, which is one of the goals of CASCADE. The observations cover $\sim$16\,GHz bandwidth in the 3.6\,mm window, containing many different spectral lines (combined NOEMA+30\,m data) and also good continuum measurements (NOEMA-only data). The DR20 subregion, which is the focus of this work, is composed of three mosaics. Each mosaic consists of 78 NOEMA pointings. All regions are observed in the NOEMA C and D configurations, with typically ten antennas in the array, with baselines between $\sim$ 15 m and $\sim$ 356 m. The 3.6 mm continuum data were obtained only with NOEMA with a broad band pass of $\sim$ 15.51 GHz and a synthesized beam of 3.45$''$ $\times$ 2.77$''$, corresponding to a linear spatial resolution of approximately 4400\,AU. The spectral lines were observed over a band width of $\sim$ 8 GHz in the upper and lower sideband at a central nominal frequency of $\sim$82.028 GHz. Important ground state lines as well as unique deuterated molecule lines are covered. The available spectral lines and their parameters are reported in Table 1 and A.1 in \cite{beuther_cygnus_2022}. The NOEMA data with the higher spectral resolution ($\sim$0.2\,km\,s$^{-1}$) were resampled to match the spectral resolution ($\sim$0.8\,km\,s$^{-1}$) of the IRAM\,30\,m data before merging the datasets. The velocity at the local standard of rest ($v_{\text{LSR}}$) of DR20 is $-$2.0\,km\,s$^{-1}$ (e.g., \citealt{beuther_cygnus_2022}). The parameters of the HCO$^+$ and H$^{13}$CO$^+$ spectral lines used in this work are presented in Table \ref{tab_beuther_line_parameters}.

\begin{table}
\caption{Parameters of the spectral lines used.}             
\label{tab_beuther_line_parameters}      
\centering                          
\begin{tabular}{c c c c c c}        
\hline\hline                 
Line & \HCO(1-0) & \HttCO(1-0) \\  \hline
Frequency & 89.189 GHz & 86.754 GHz \\   
Beam (comb.) & 3.2$''\times 2.4''$ & 3.3$''\times 2.5''$ \\
Beam (30 m) & 28.2$''$ & 29$''$ \\
Res. (comb.) & 4100 AU, 0.02 pc & 4200 AU, 0.02 pc \\
Res. (30 m) & 41000 AU, 0.21 pc & 42000 AU, 0.21 pc \\
\hline                                   
\end{tabular}
\tablefoot{The synthesized beam values for the combined data are taken from \cite{beuther_cygnus_2022}. “Res.” refers to the approximate linear spatial resolution, calculated using the distance to the DR20 subregion \citep[1.46 kpc,][]{rygl_parallaxes_2012}.}
\end{table}
We identify filaments in the merged data of NOEMA and the IRAM 30 m (referred to as “combined data”), which are used for further analysis. Although a higher spectral resolution would be desirable, the available higher-resolution NOEMA-only data suffer severely from missing flux, which results in unreliable spectral structures (e.g., \citealt{plunkett2023}). Hence, using the combined data is mandatory for our analysis. For comparison we also identify filaments in the lower spatial resolution IRAM 30 m single-dish spectral line data on its own (Table \ref{tab_beuther_line_parameters}), as well as in the Herschel hydrogen column density data with $12''$ resolution derived by \cite{marsh_multitemperature_2017}.


\section{Methods}
\label{sect_meth}

With the structure identification algorithm \textit{Discrete Persistent Structures Extractor} \citep[DisPerSE,][]{sousbie_persistent_2011}, we identified filamentary structures in the 3D (position-position-velocity, PPV) data cubes. Apart from the combined NOEMA and IRAM 30 m data, we analyze the lower spatial resolution IRAM\,30\,m single-dish data separately to compare the filaments to those identified in the combined data. We did the same with the 2D Herschel hydrogen column density data taken from \cite{marsh_multitemperature_2017}. The Herschel data cover a larger area of DR20 than the NOEMA and 30 m observations, showing us how the smaller filament structures connect to larger ones. Using the combined NOEMA and IRAM 30 m data, we studied the properties of the identified filaments with the Python \textit{Filament Characterization Package} \citep[FilChaP,][]{suri_astrosurifilchap_2018}. FilChaP provides filament widths along the filaments, as well as statistical results of filament width distributions in different star formation evolutionary stages. Additionally, it gives us the possibility to study how well different functional forms fit the filaments' radial profiles (see section \ref{subsect_meth_Filchap}).

\subsection{DisPerSE}

DisPerSE \citep{sousbie_persistent_2011} extracts persistent topological features, such as peaks, voids, walls, and filamentary structures, using discrete Morse theory\footnote{The manual can be found at \url{http://www2.iap.fr/users/sousbie/web/html/index4f3e.html?category/Manual}.}. Morse theory is based on the concept of ascending and/or descending k-manifolds, which partition space into k-dimensional domains defined by the gradient of a function. These k-manifolds can be associated with astrophysical objects; for example, ascending 1-manifolds can be interpreted as filaments. DisPerSE was originally developed for research in cosmology but is now widely used in other fields, especially to identify filamentary structures in star formation.

DisPerSE uses the concept of persistence, which is a measure of robustness of topological features, i.e., their contrast with respect to their surroundings. If we set the persistence threshold higher, fewer intensity maxima, saddle points, and arcs connecting features, which are filaments, are identified. We chose a threshold so that only the most significant topological features remain, which is between 3 to 6$\sigma$, depending on the dataset (HCO$^+$, H$^{13}$CO$^+$ and Herschel column density data, Table \ref{tab_noise}).
For the CASCADE data, we determined the noise as the standard deviation of the masked last velocity channel, which is noise-dominated and has no source emission. The mask excludes the higher noise at the edges of the maps. Since for radio interferometers, i.e., the combined data, the noise is not Gaussian, the standard deviation is not entirely accurate. Additionally, we measured the noise in only one velocity channel and assumed it is similar for all. To make up for these inaccuracies, we took a persistence threshold at a conservative high level of 6$\sigma$. For the IRAM\,30\,m single-dish data, the noise is approximately Gaussian, and we chose a 4$\sigma$ persistence threshold. In the Herschel data, the noise is determined as the standard deviation of the signal in a 50 pix$^2$ rectangle at a position with no significant source emission within the image. Since there is no region completely without emission, the standard deviation is too high. Therefore, we took a comparatively lower persistence threshold of 3$\sigma$. All of these thresholds have been additionally inspected in the persistence plots given by DisPerSE. These show the persistence of all persistence pairs over their background densities, so that the most significant features stand out of the general distribution on the y-axis and can be separated from it with a horizontal line, i.e., a persistence threshold. Additionally, we visually checked the filaments resulting from the different thresholds.

\begin{table}
\caption{Noise of the different datasets.}             
\label{tab_noise}      
\centering                          
\begin{tabular}{c c c c}        
\hline\hline                 
 \multicolumn{4}{c}{$\sigma$ (persistence threshold)} \\
 & \HCO\ & \HttCO\ & Herschel \\
 & [mJy beam$^{-1}$] & [mJy beam$^{-1}$] & [10$^{20}$ cm$^{-2}$] \\    
\hline                        
   Combined & 6.78 ($\times$6) & 6.86 ($\times$6) & \multirow{2}{*}{5.22 ($\times$3)} \\ \hhline{---~}     
   30 m & 285 ($\times$4) & 300 ($\times$4) & \\
\hline                                   
\end{tabular}
\tablefoot{The noise of the different CASCADE datasets is calculated as the standard deviation of the signal in the respective edge-masked last velocity channel. For the Herschel hydrogen column density data, the noise is calculated as the standard deviation of the signal in a 50 pix$^2$ rectangle at a position with no significant source emission. In brackets, the multiplicator to calculate the persistence threshold used as DisPerSE input is shown.}
\end{table}

We applied these chosen thresholds (Tab.\,\ref{tab_noise}) and the same mask used for the noise calculation to exclude the high noise at the edges of the maps.
Namely, we used the main program “mse” to compute Morse-smale complexes, with the option “upSkl” to dump the “up” skeleton, i.e., arcs linking maxima to saddle-points, to extract the filamentary structures. The order of the parameters is “mse,” followed by the mask, the cutting of the threshold, and “upSkl.”

We tested several additional setting options, the most important being “periodicity," which can be used to change the default fully periodic to nonperiodic boundary conditions; “breakdown," which breaks filaments into smaller parts and keeps them from overlapping; and “assemble," which combines filaments connected to each other at an angle lower than a specifiable number. Our tests show that due to these options, some filaments disappear and/or new ones appear, which visually do not represent the data as well. The differences are minimal, but to avoid them while gaining the merits of breakdown and assemble, we chose to combine short filaments ourselves.

While the filaments in the Herschel continuum data were extracted from the 2D spatial map, those in the 3D spectral line cubes (PPV) were extracted from the 3D data including the velocity information. Therefore, we can also measure velocity changes along the filaments (e.g., Fig.~\ref{plot_MIOP_filaments_0mom}). However, since DISPERSE typically identifies only one velocity component along a given spine due to the large selected persistence thresholds, and because several velocity components may be present (e.g., Fig.~\ref{plot_along_fit_exmpl}), here below we analyze the velocity structure in more depth by fitting Gaussians -- often several Gaussian components -- along the identified skeleton spines. We note that in this approach multiple Gaussian components measured toward one filament are assigned to a single spatial structure, whereas small spatial offsets between different velocity components are possible (e.g., \citealt{hacar2013}). For the current analysis, we stick to the conservative high thresholds; however, future analysis may investigate smaller spatial substructures potentially identified with other filament identification algorithms.

\subsection{Velocity and linewidth}
\label{subsect_meth_vel}

\begin{figure*}[h] 
	\centering 
	\begin{subfigure}{0.33\textwidth}
  		\centering
		\includegraphics[width=1\linewidth ,keepaspectratio]
		{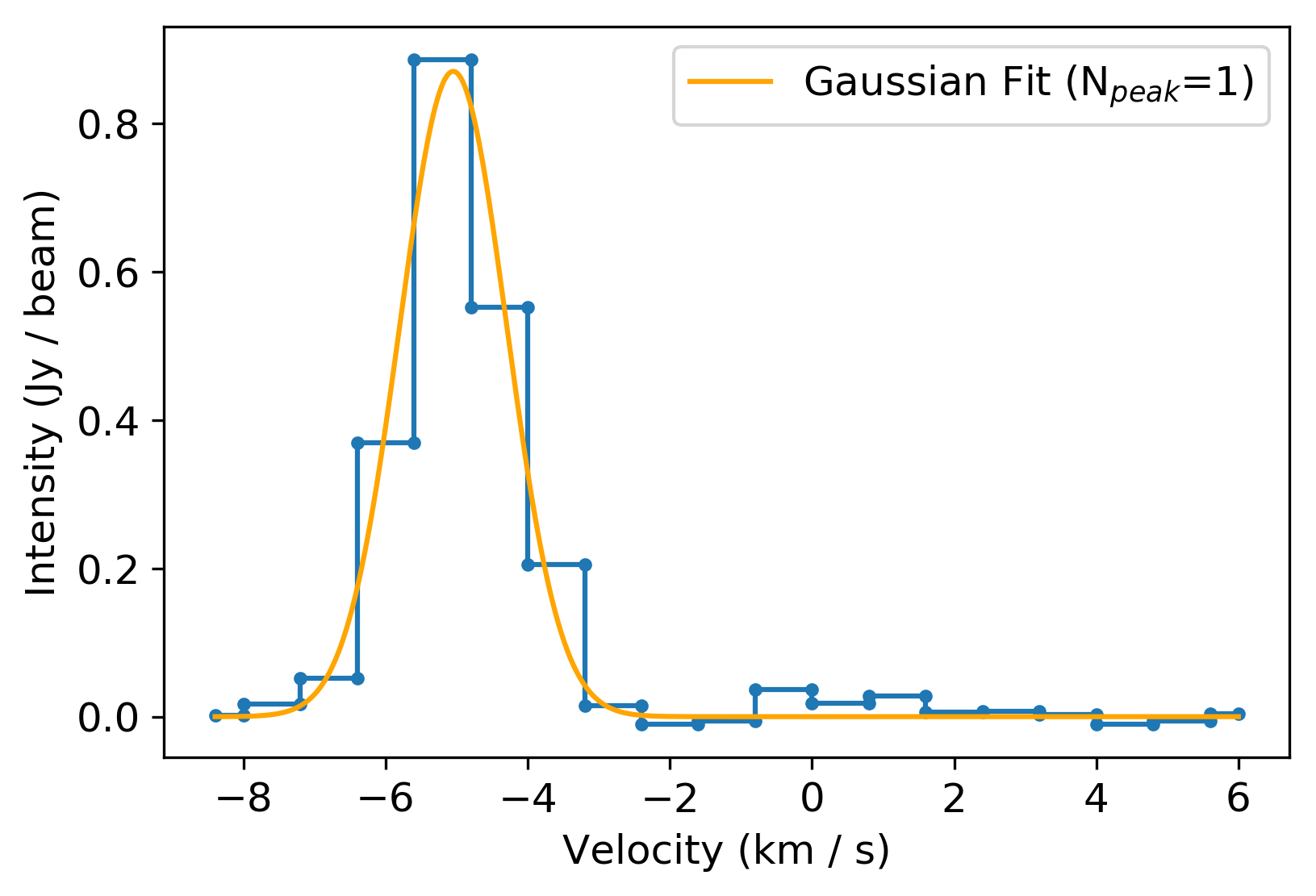}
	\end{subfigure}%
	\begin{subfigure}{0.33\textwidth}
		\centering
		\includegraphics[width=1\linewidth ,keepaspectratio]
		{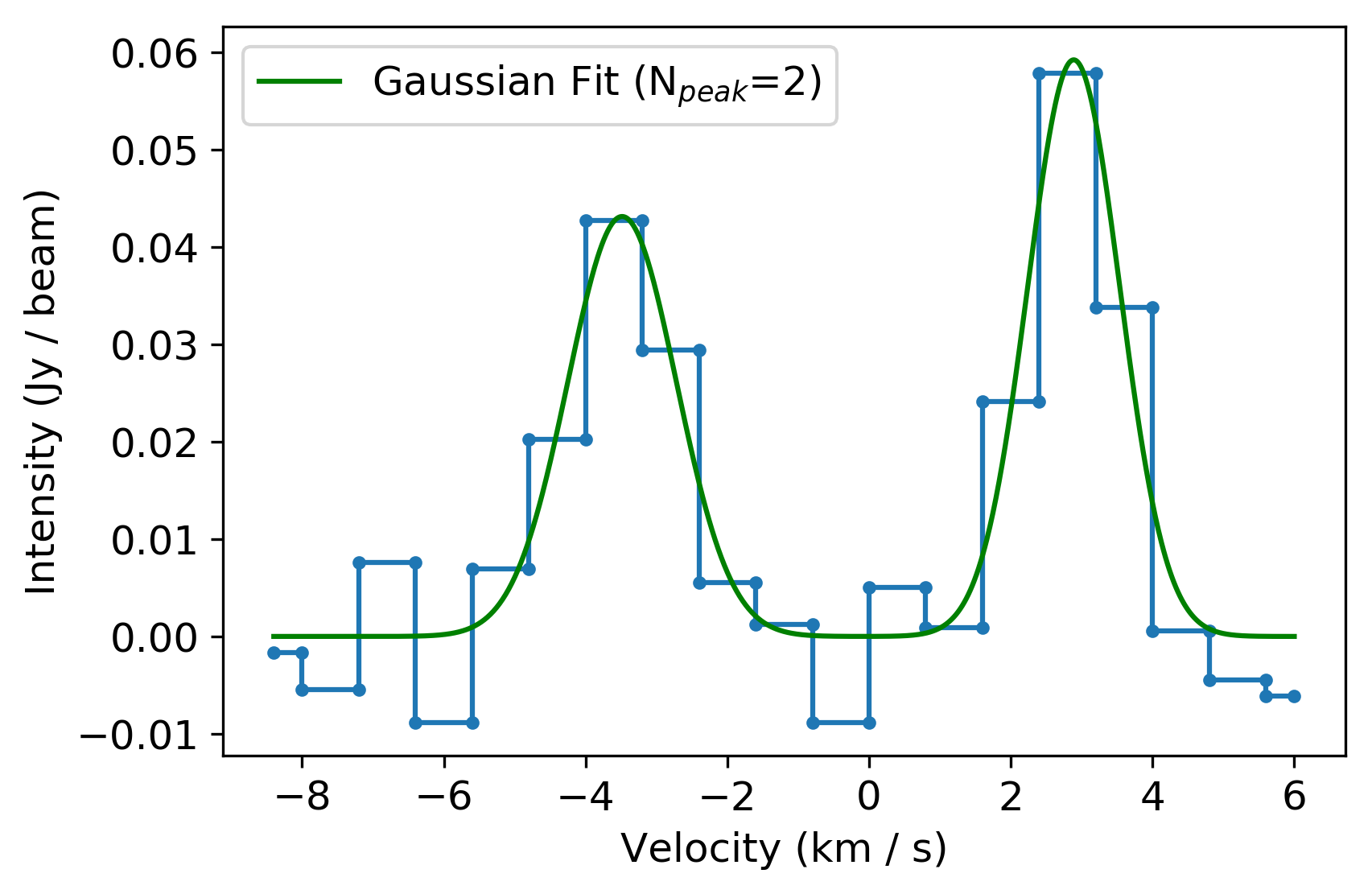}
	\end{subfigure}
	\begin{subfigure}{0.33\textwidth}
		\centering
		\includegraphics[width=1\linewidth ,keepaspectratio]
		{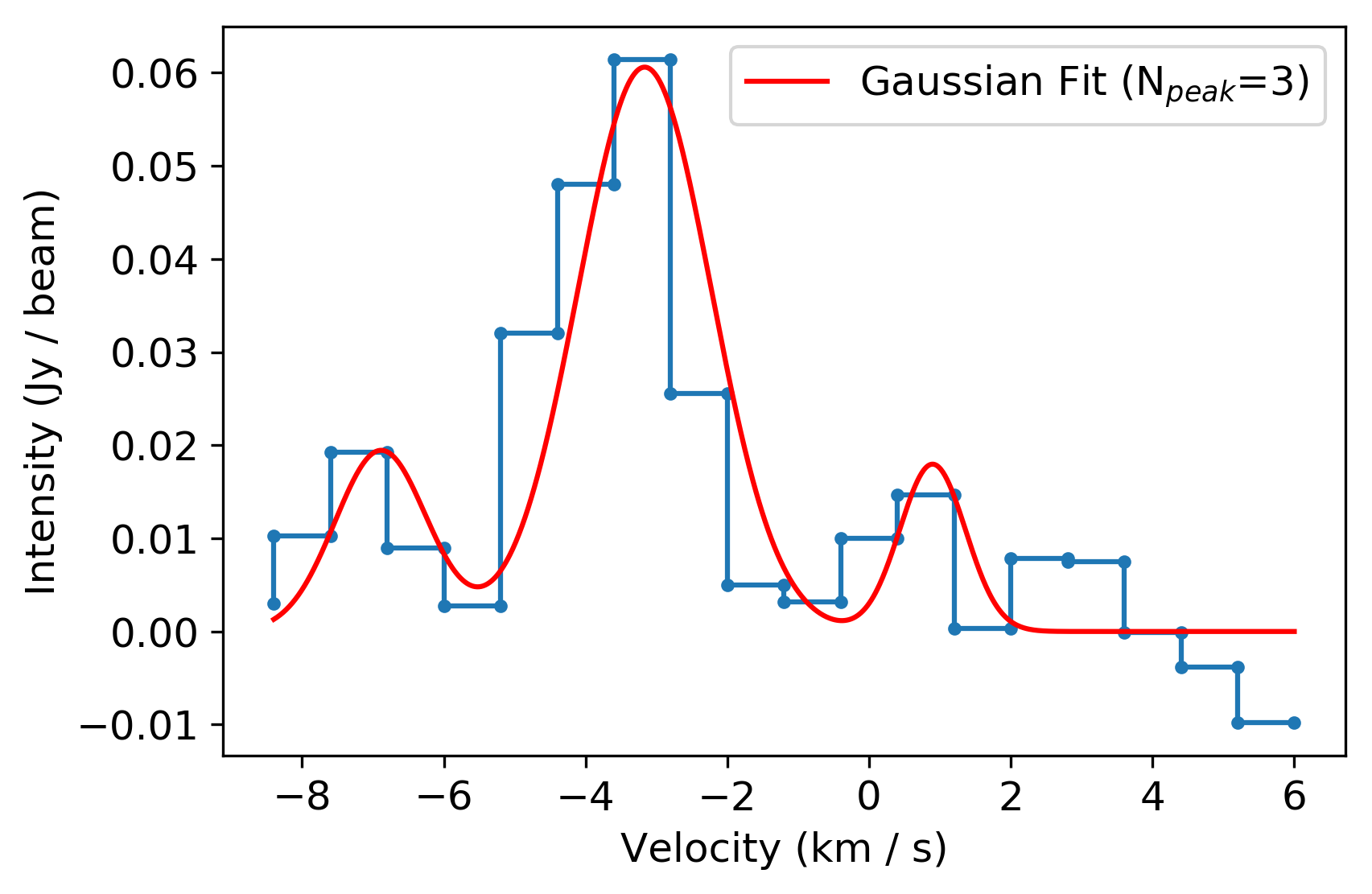}
	\end{subfigure}
\caption[Example Gaussian fits of spectra within filaments.]{Examples of the Gaussian fits to averaged spectra within filaments connected to core A in the combined NOEMA+30\,m \HCO$(1-0)$ data. The peak positions and FWHM seen in, for example, Figs.~\ref{plot_along_chosen_HCO} result from such fits.}
\label{plot_along_fit_exmpl}
\end{figure*}

To visualize the gas velocity and linewidth along a filament, we investigated all spectra along that filament. To improve the signal-to-noise ratio, we consistently averaged three adjacent spectra following the crest of the spine of the filament (corresponding to $\sim$2/3 of the beam width). We determined the velocity of the spectral peaks in the average spectrum and tried to simultaneously fit a number of Gaussian components equal to the number of identified spectral peaks (to a maximum of four). For the fitting, the curve\_fit function of the \textsf{scipy.optimize} Python package is used. We tried a Gaussian fit with one component less until the error of the full width half maximum (FWHM) of the fitted peaks reached below 2\,km\,s$^{-1}$. An example of a Gaussian fit with one to three peaks each can be seen in Fig. \ref{plot_along_fit_exmpl}.
Furthermore, a fit was discarded if the amplitude of the fitted main peak was lower than 3$\sigma$ (with $\sigma$ = 6.78 mJy beam$^{-1}$ for HCO$^+$ and $\sigma$ = 6.86 mJy beam$^{-1}$ for H$^{13}$CO$^+$, Table \ref{tab_noise}). After visual inspection we further excluded fits where the FWHM > 4 km s$^{-1}$, since in these cases an unresolved second peak was likely included in the first. Depending on the molecule (HCO$^+$ or its H$^{13}$CO$^+$ isotopologue) and the core, the number of discarded fits varies. For the rarer H$^{13}$CO$^+$, especially on the more extended structures, additional fits were discarded with a mean value around 54\%. For the main isotopologue HCO$^+$, only about 10\% of the fits were discarded. In Figs.~\ref{plot_along_chosen_HCO}, \ref{plot_along_chosen_H13CO} and Appendix \ref{add_figs} (linewidth and velocity within all analyzed filaments), the discarded fits correspond to missing points in the plots.

\subsection{FilChaP}
\label{subsect_meth_Filchap}

FilChaP is a Python package developed by \cite{suri_astrosurifilchap_2018} in order to determine filament properties, in particular their width. It picks up coordinates of a line perpendicular to a given filament, and extracts the intensity profile along this line. This is done several times along each filament, where the distance of the lines, or slices, as well as the length of the line can be given as input settings. Additionally, the distance along the filament over which these intensity profiles should be averaged can be set. The resulting average intensity profile can then be fit to determine the filament width. This is often done with Plummer functions, which define a cylinder with a dense and flat inner portion and a power-law decline at larger radii \citep{plummer1911,arzoumanian2011,suri_carma-nro_2019}:

\begin{align}
	\Sigma_{p}(r) = A_{p}\frac{\rho_c R_{\text{flat}}}{\left[1 + \left(\frac{r}{R_{\text{flat}}}\right)^2\right]^{\left(\frac{p-1}{2}\right)}},
\label{eq_plummer}
\end{align} 

\noindent with $\Sigma_{p}(r)$ being the column density, $\rho_c$ the central density, $R_{\text{flat}}$ the filament's inner radius within which the density is uniform, $p$ the power-law index, and $A_{p}$ a finite constant for $p>1$ that depends on the filament's inclination angle $i$ to the plane of the sky: 
$ A_{p}$~=~$ \frac{1}{\cos\,i} \, \int_{-\infty}^{\infty}\frac{du}{(1+u^{2})^{{p}/2}} $
\citep{arzoumanian_characterizing_2011}.
The inclination angle $i$ is assumed to be zero for simplicity in FilChaP. The FWHM, i.e., the filament width, can be calculated as (e.g., \citealt{hacar_initial_2022}):

\begin{align}
\mathrm{FWHM} = 2 R_{\text{flat}} \sqrt{2^{2/(p-1)}-1},
\end{align}

\noindent which assumes an isolated filament with no background column density.
For an isothermal hydrostatic cylinder, the power-law index is $p$ = 4 \citep{ostriker1964}, but observations show a better agreement with $p$ $\sim$ 1.5-2.5 \citep{arzoumanian_characterizing_2011}. Profiles with $p <$ 4 are expected for cylinders not in equilibrium, possibly magnetized, externally pressurized, non-isothermal, polytropic or rotating (see \citealt{hacar_initial_2022} for a full discussion).

FilChaP fits the determined average profile of neighboring intensities with Plummer functions with power-law indices of $p$ = 2 and $p$ = 4, as well as a Gaussian function \citep{suri_carma-nro_2019}:

\begin{align}
	\Sigma_{\mathrm{p}}(r) = A\mathrm{e}^{-(r-r_0)^2/2\sigma^2}, 
\label{eq_Gauss}
\end{align}

\noindent where $\sigma$ is the standard deviation of the Gaussian function. The filament width then corresponds to

\begin{align}
\mathrm{FWHM} = 2 \sqrt{2 ln(2)} \cdot \sigma.
\end{align}

Furthermore, FilChaP computes the filament width via the second moment of the distribution $I$, where the $n$th moment is given by

\begin{align}
  m_n = \frac{1}{N}\frac{\sum_{i}^{N} I_i \left(x_i-\bar{x}\right)^n}{\sigma^n},
\label{eq_2ndmoment}
\end{align}

\noindent where $\bar{x}$ is the intensity weighted mean position of the profile, ${I_i}$ is the intensity at position $x_i$, and $\sigma$ is the intensity weighted standard deviation of the profile. The second moment is the variance of the distribution, and therefore the width of the profile can be calculated with $\sqrt{m_2} \times 2\sqrt{2ln2}$.\\

The de-convolved width is

\begin{align}
    \mathrm{FWHM_{deconv}} = \sqrt{\mathrm{FWHM}^2-B^2},
\end{align}

\noindent with $B$ the observed beam-width (a mean of the two beam axes for the NOEMA+30\,m data, Table \ref{tab_beuther_line_parameters}). \cite{zucker_radfil_2018} used this deconvolution procedure on FWHM values of Gaussian as well as Plummer functions. They found that this deconvolution can lead to an overestimation of the filament width that increases significantly as the spatial resolution of the data decreases. However, the resolution of our data is a factor of 10 higher than that of their best test data. They also note that this overestimation is a little less significant for the Plummer function compared to the Gaussian function.


We used FilChaP on the combined HCO$^+$ and H$^{13}$CO$^+$ data and chose settings similar to \cite{suri_carma-nro_2019}. For the distance from one slice to the next, they chose 1.5 beam sizes to ensure statistical independence of each slice, i.e., no pixels were used in several slices. They then averaged the intensity profiles over a length of three beam sizes; therefore, they averaged two intensity profiles each. We tested different combinations of these two settings, i.e., slightly smaller and larger separations between the slices as well as slightly smaller and larger average distances. However, the overall distribution of the filament width did not change significantly. Hence, we also used a separation of 1.5 beam sizes, corresponding to $\sim$ 0.031 pc, and we also averaged over two slices (three beam sizes). For the length of the perpendicular line, we chose 100 pixels corresponding to $\sim$ 0.73 pc, sufficient to estimate the widths of the filaments. 
The length should not be too short to ensure the closest significant minima that determine the fitting range are covered, but it should also not be too long, as additional surrounding emission can influence the baseline subtraction \citep[for a description of the baseline subtraction, see Appendix A of][]{suri_carma-nro_2019}.

To exclude slices where the fitting does not perform well, 
we decided not to use reduced $\chi^2_{R}$ criteria (e.g.,  \citealt{suri_carma-nro_2019}) because the substructure as well as the low signal-to-noise ratios along the slices affect these too strongly in our data. Instead, we chose to exclude slices based on the calculated parameters. Since the determined peak position of the fit should be very close to zero, corresponding to the center of the filament, we excluded fits where the peak position is further away from zero than $\pm$ 0.05 pc for HCO$^+$ and $\pm$ 0.038 pc for H$^{13}$CO$^+$. Visual inspection shows that, in such cases, the wrong peak or several peaks were fit. To determine an appropriate value for the threshold, we made a histogram of all determined peak positions with signal-to-noise-ratio S/N $>$ 3 and chose a value where the number of positions indicates a sudden drop, while also trying to maximize the amount of non-excluded data that passes the visual inspection. We decided to use the same value on the negative and positive side. Furthermore, we chose the error of the parameter giving the profile width -- which is the standard deviation ($\sigma$) for the Gaussian fit and $R_{\text{flat}}$ for the Plummer fits -- as an exclusion criterion. For that, we used the same method as for the peak positions, i.e., finding a drop in their distributions and visual inspection. The determined threshold values are $\sigma$ = 0.0065 pc, $R_{\text{flat}}$ = 0.0065 pc ($p$ = 2), and $R_{\text{flat}}$ = 0.012 pc ($p$ = 4) for \HCO, and $\sigma$ = 0.0065 pc, $R_{\text{flat}}$ = 0.0063 pc ($p$ = 2), and $R_{\text{flat}}$ = 0.011 pc ($p$ = 4) for \HttCO. Determined filament widths with errors larger than these thresholds are excluded from the results. Sometimes one of the fits gives a large error while the others do not, leading to slightly different sample sizes for the different methods (Tab.~\ref{tab_Filchap_all_AF}). Since the second moment does not have a formal error, we include its result only when all other methods are included as well. 

Further results achieved with FilChap are goodness-of-fit values for the different fits, making it possible to investigate the equations that could describe our identified filaments best. Shoulders, i.e., local minima in the second derivative of the averaged intensity profile with a significance of 5$\sigma$, are also identified. \cite{suri_carma-nro_2019} find the number of shoulders to correlate with the determined filament width, suggesting that this might show the complex substructure of filaments influencing the widths. 
Features of FilChaP separating it from other methods are presented in the Appendix (Sect. \ref{sect_app_filchap}).


\section{Results}
\label{sect_results}

\begin{figure*}[h!]
	\centering
	\begin{subfigure}{.5\textwidth}
  		\centering
		\includegraphics[width=0.97\linewidth ,keepaspectratio]
		{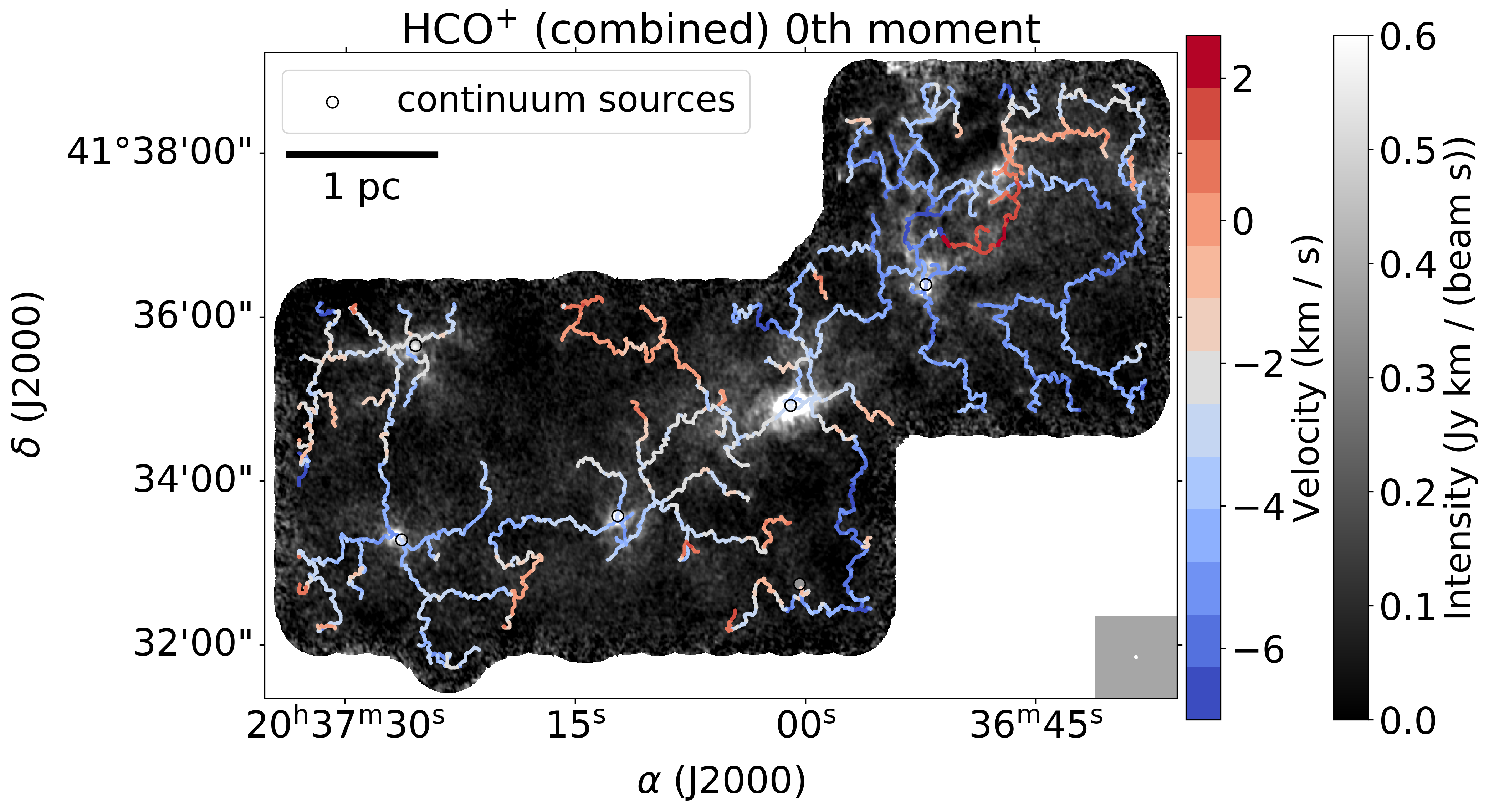}
	\end{subfigure}%
	\begin{subfigure}{.5\textwidth}
		\centering
		\includegraphics[width=0.97\linewidth ,keepaspectratio]
		{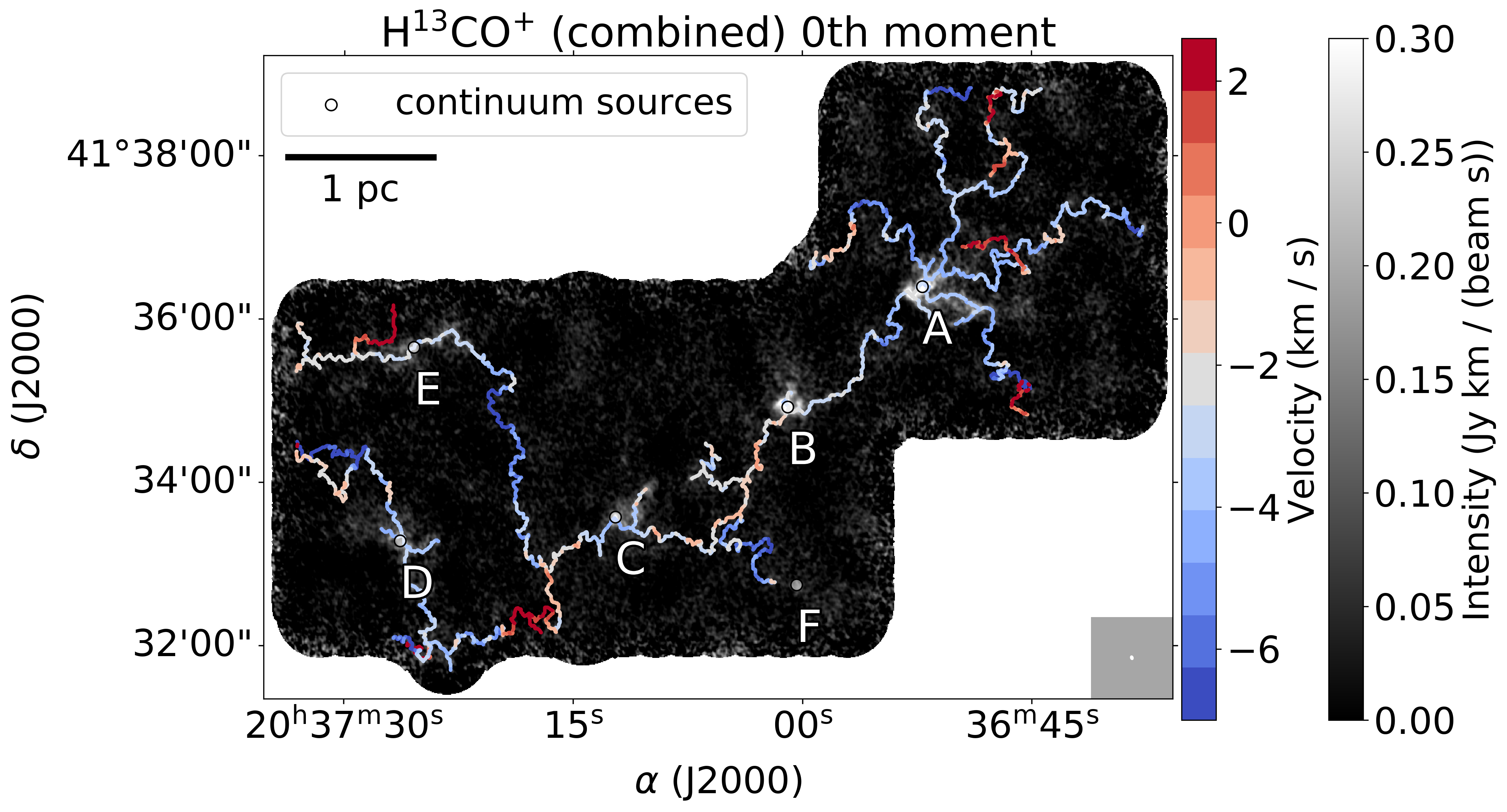}
	\end{subfigure}
	\vskip\baselineskip
	\begin{subfigure}{.5\textwidth}
  		\centering
		\includegraphics[width=0.97\linewidth ,keepaspectratio]
		{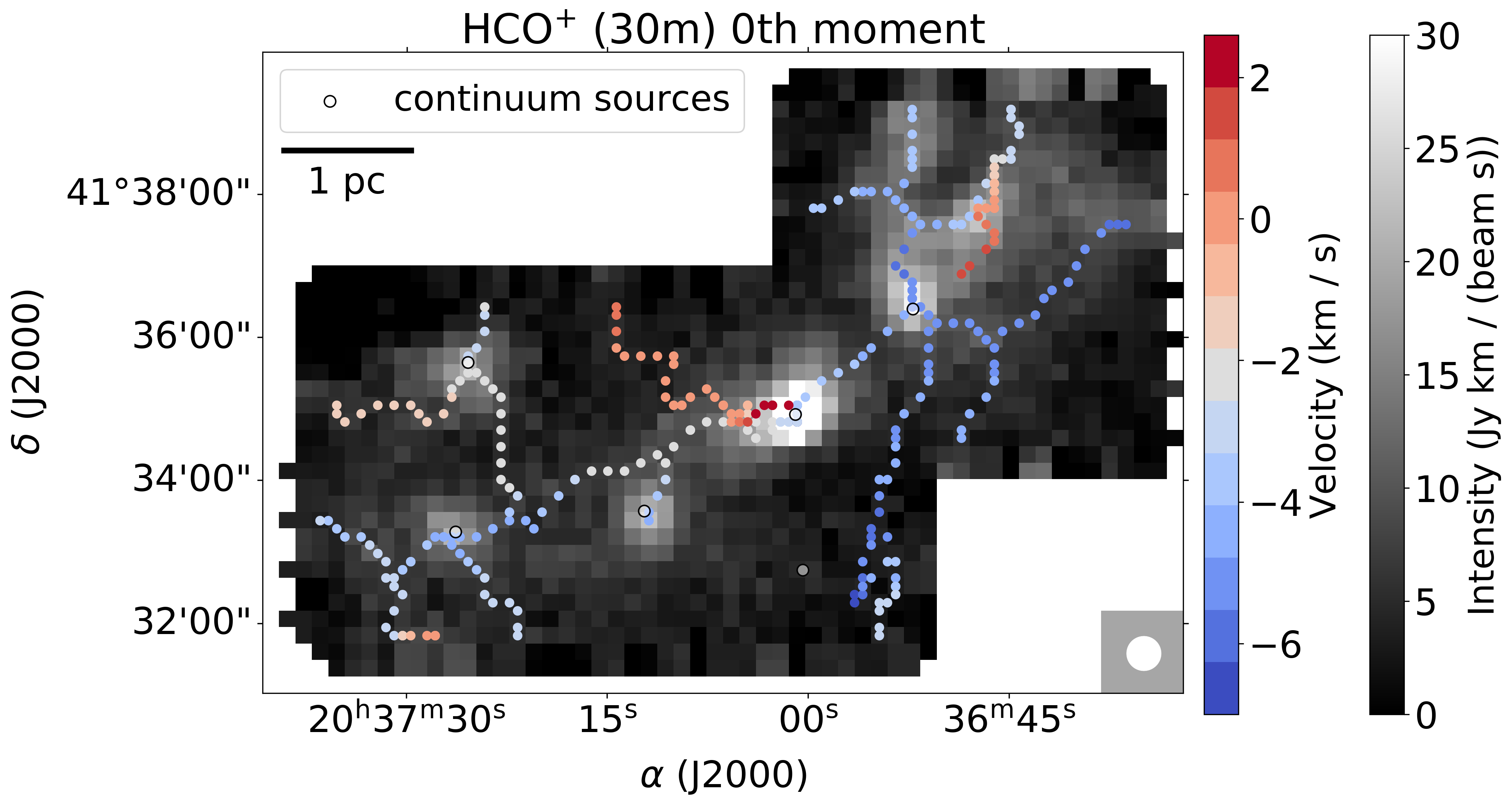}
	\end{subfigure}%
	\begin{subfigure}{.5\textwidth}
		\centering
		\includegraphics[width=0.97\linewidth ,keepaspectratio]
		{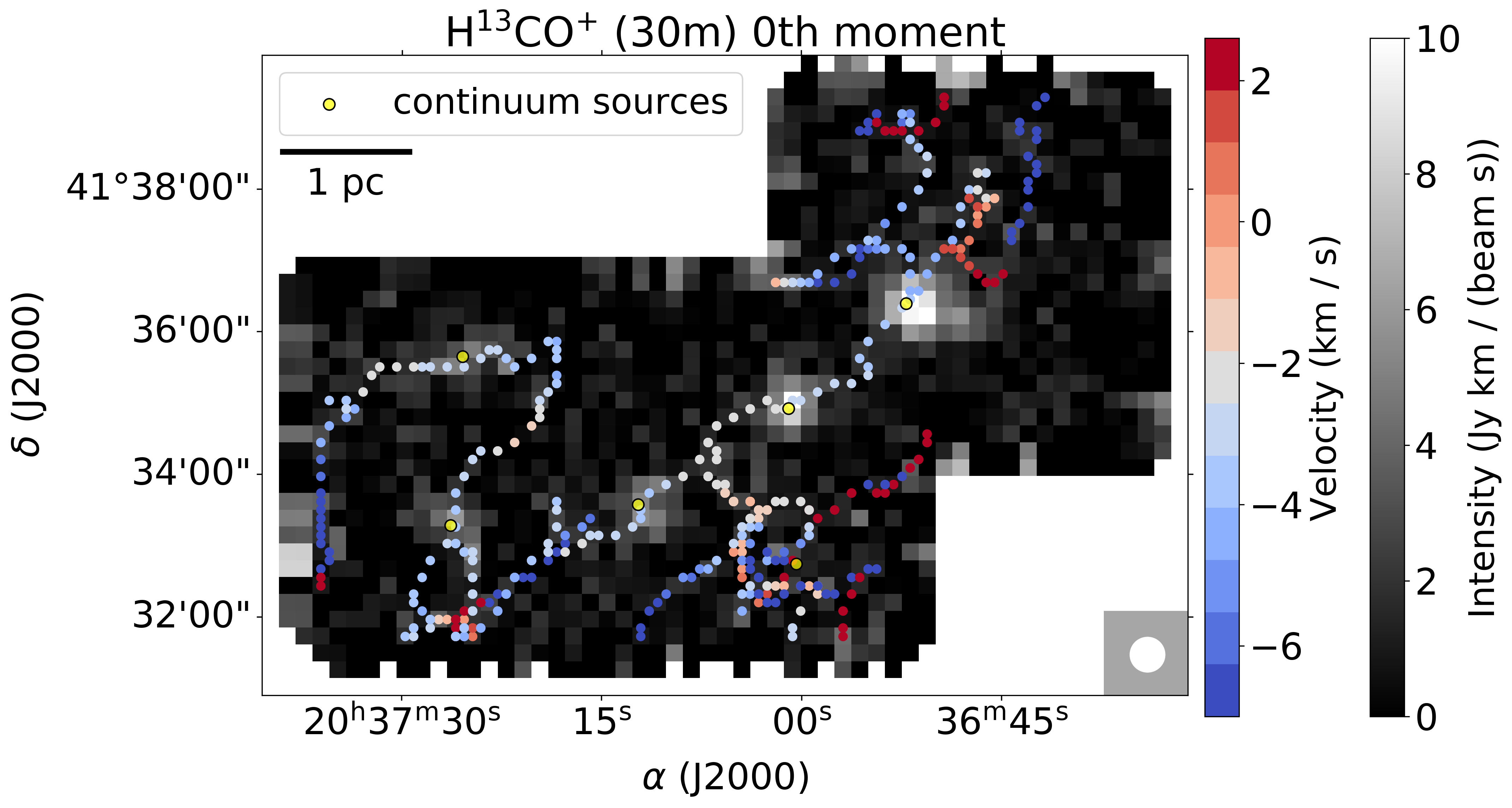}
	\end{subfigure}
\caption[Identified filaments in combined and 30 m \HCO\  and \HttCO.]{Filaments identified with DisPerSE in all velocity channels (from -8.4 km s$^{-1}$ to 6 km s$^{-1}$) overlaid on the respective zeroth moment maps, i.e., the intensity integrated over all velocity channels, of all CASCADE data types. \textit{Top}: Combined \HCO$(1-0)$ (left) and \HttCO$(1-0)$ (right). \textit{Bottom}: Single-dish \HCO$(1-0)$ (left) and \HttCO$(1-0)$ (right). The intensity maxima of the 3.6\,mm continuum emission are labeled A-F in the upper right panel. The beam is shown in the bottom-right of each panel.}
\label{plot_MIOP_filaments_0mom}
\end{figure*}

\begin{figure}[h]
	\centering
	\includegraphics[width=1\linewidth ,keepaspectratio]
	{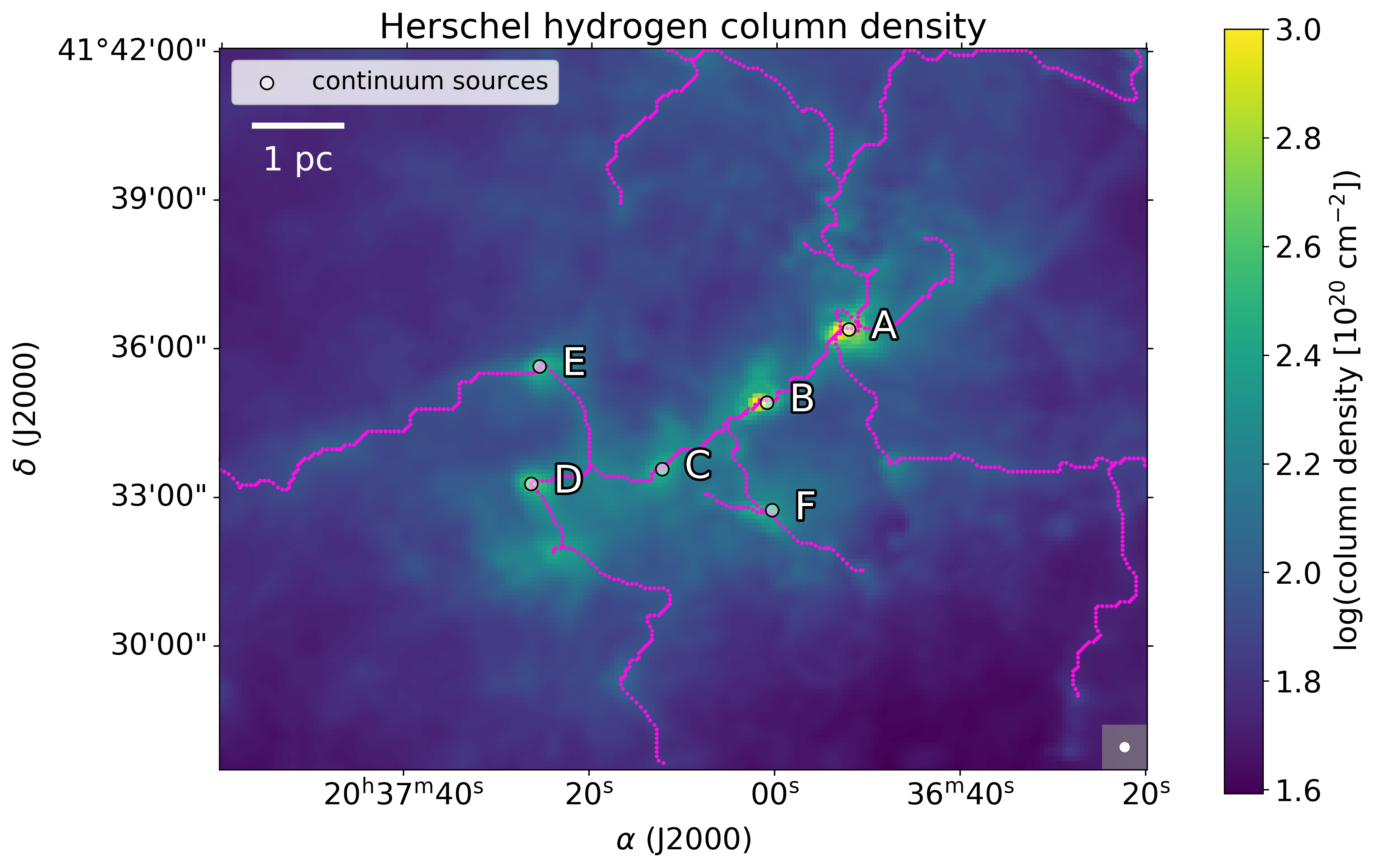}
	\caption[Identified filaments in Herschel hydrogen column density.]{Filaments identified with DisPerSE in the Herschel hydrogen column density data \citep{marsh_multitemperature_2017}. To also show the lower hydrogen column densities, the color scale bar is truncated at $3\times 10^{20}$\,H$_2$\,cm$^{-2}$, but the peak positions reach much higher column densities $>10^{22}$\,cm$^{-2}$ (Fig.~\ref{plot_dr20_overview}). The intensity maxima of the 3.6\,mm continuum emission are labeled A-F. The beam is shown in the bottom-right.}
	\label{plot_cdens_fil}
\end{figure} 

\begin{figure*}[h]
	\centering
	\begin{subfigure}{.44\textwidth}
  		\centering
		\includegraphics[width=0.97\linewidth ,keepaspectratio]
		{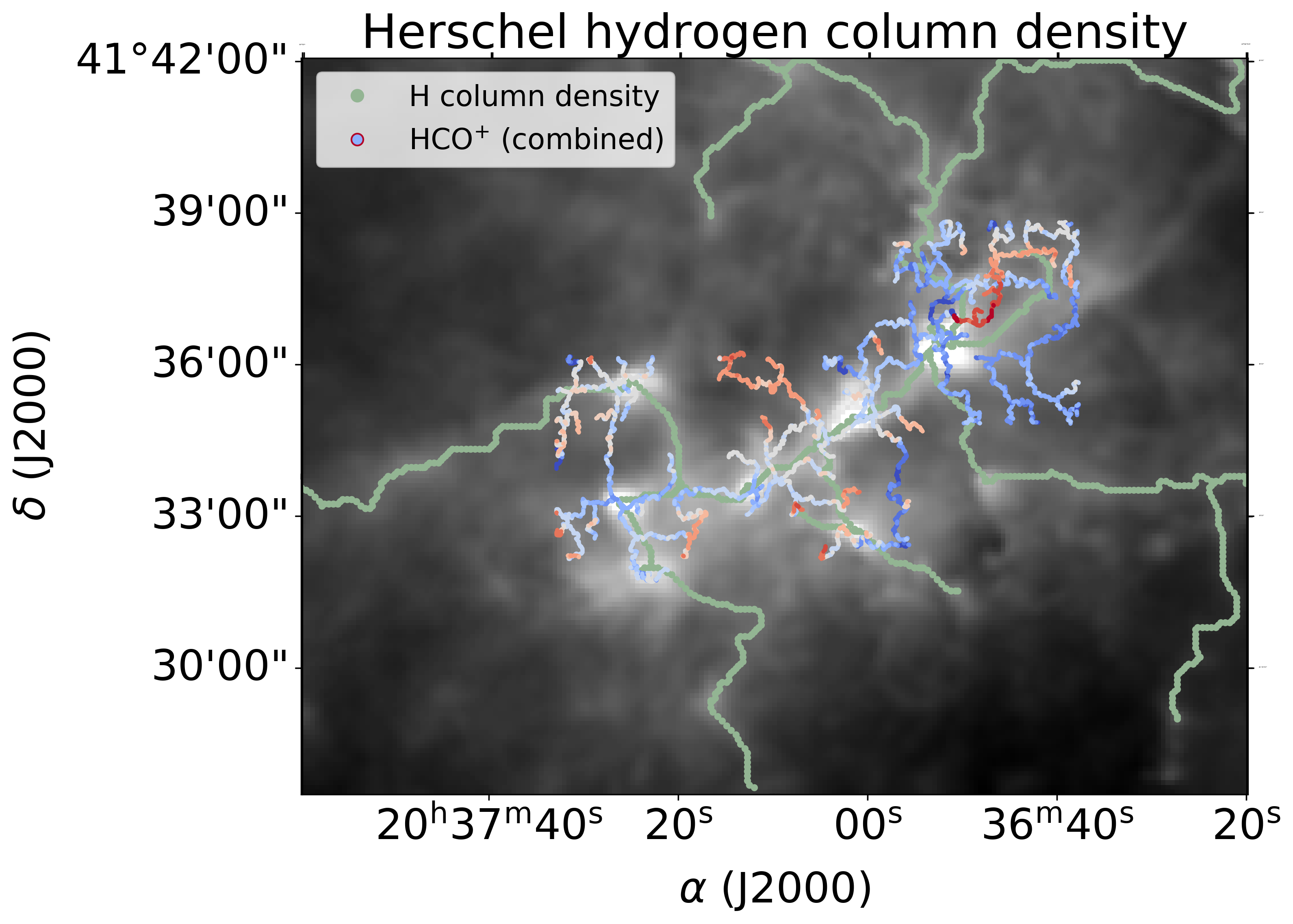}
	\end{subfigure}%
	\begin{subfigure}{.56\textwidth}
		\centering
		\includegraphics[width=0.97\linewidth ,keepaspectratio]
		{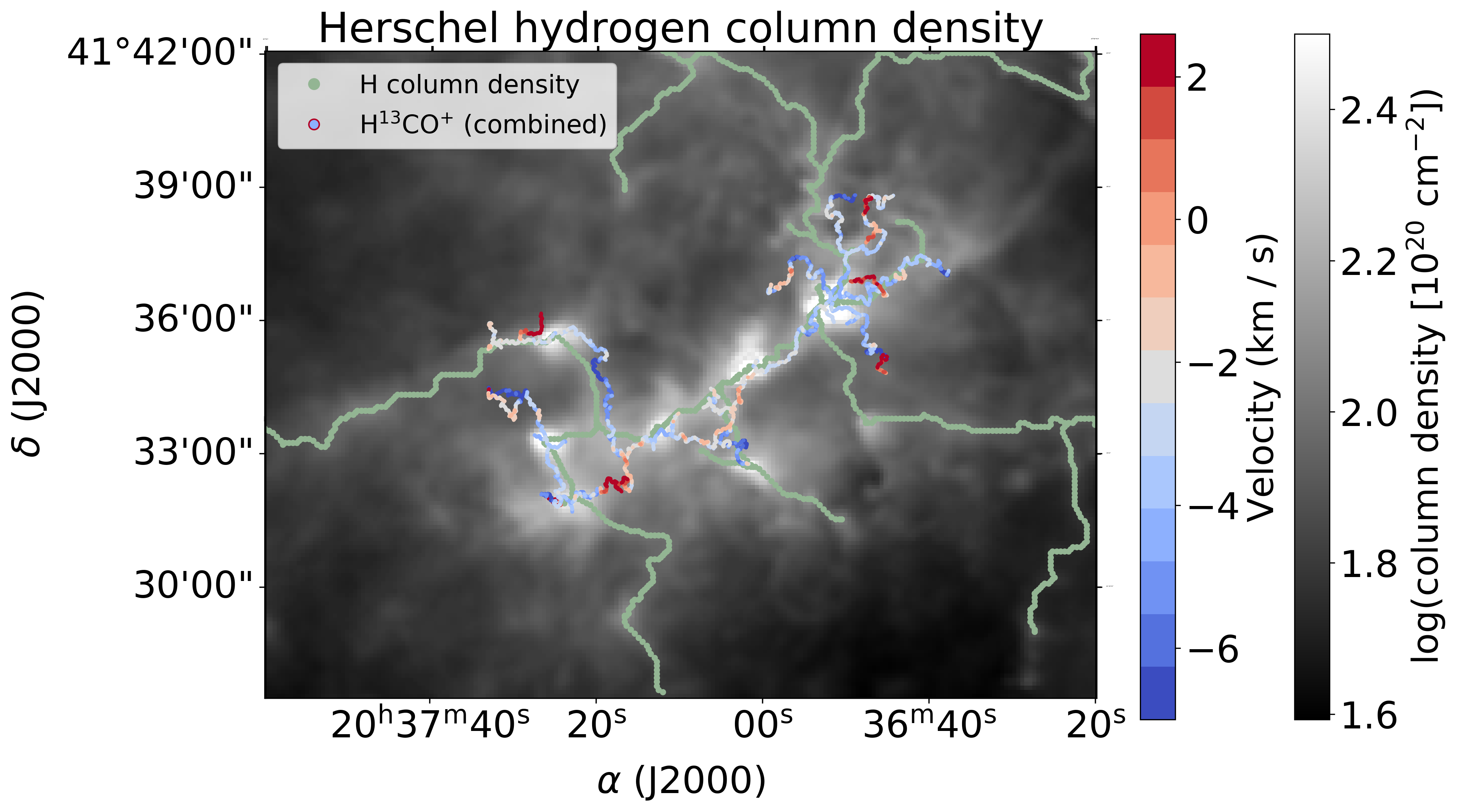}
	\end{subfigure}
\caption[Overlay of MIOP data (combined \HCO\  and \HttCO) filaments and Herschel hydrogen column density filaments and map.]{Filaments identified in the combined CASCADE data overlaid on the Herschel hydrogen column density image and its filaments (see Figs.~\ref{plot_MIOP_filaments_0mom} and \ref{plot_cdens_fil}). To also show the lower hydrogen column densities, the color scale bar is truncated at $\sim$2.5$\times 10^{20}$\,H$_2$\,cm$^{-2}$, but the peak positions reach much higher column densities $>10^{22}$\,cm$^{-2}$ (Fig.~\ref{plot_dr20_overview}). \textit{Left}: \HCO$(1-0)$. \textit{Right}: \HttCO$(1-0)$.}
\label{plot_overlay_cdens}
\end{figure*}

We present the integrated emission maps and the identified filaments of the CASCADE data in Fig. \ref{plot_MIOP_filaments_0mom}. Both the combined NOEMA+30\,m and the 30m-only data are shown to investigate the filament properties at different spatial scales. The filaments are identified in each velocity channel, which is represented by their color. While the overall filamentary structure is similar between the single-dish and interferometer data, one difference between them is the long filament that connects cores A to D in the low-resolution 30 m data, whereas the high-resolution data rather exhibit individual filament networks around each core. The identified filaments of the Herschel hydrogen column density data are displayed in Fig. \ref{plot_cdens_fil}. An overlay of the filaments identified in the CASCADE data with those in the Herschel data (Fig.~\ref{plot_overlay_cdens}) shows that some of the filaments coincide. Especially most of the large-scale filaments in the outer regions of the Herschel data connect to the small-scale filaments in the CASCADE data closer to the cores, for both \HCO\ and \HttCO\ (Fig. \ref{plot_overlay_cdens}). The correspondence of the Herschel filaments seems to be slightly better with the H$^{13}$CO$^+$ structures than the HCO$^+$ filaments. This is plausible because, while the HCO$^+$ traces more of the extended diffuse emission, the lower optical depth of H$^{13}$CO$^+$ should trace better the denser structures that are also visible in the Herschel dust continuum data.

The number of identified filaments for each dataset is listed in Table \ref{tab_fil_numb}. Note that different DisPerSE settings lead to different filament lengths and therefore numbers. However, the numbers can be relatively interpreted, confirming that the higher spatial resolution of the combined data compared to the single-dish data, as well as the higher sensitivity to low-density material of \HCO\ compared to \HttCO, lead to a higher number of (identified) filaments.

\begin{table}
\caption{Number of filaments identified with DisPerSE.}             
\label{tab_fil_numb}      
\centering                          
\begin{tabular}{c c c c}        
\hline\hline                 
 \multicolumn{4}{c}{Number of identified filaments} \\
 & \HCO\ & \HttCO\ & Herschel \\ 
\hline                        
   Combined & 366 & 95 & \multirow{2}{*}{56} \\ \hhline{---~}     
   30 m & 36 & 25 & \\
\hline                                   
\end{tabular}
\end{table}

The identification of partially different structures in the HCO$^+(1-0)$ and H$^{13}$CO$^+(1-0)$ lines depends on several aspects. As mentioned before, HCO$^+$ traces additional extended emission whereas H$^{13}$CO$^+$ follows better the higher column densities also traced by the Herschel dust continuum data. The high gas column densities, especially toward the positions of the densest cores, cause high optical depth, in particular for the main isotopologue HCO$^+$, which can result in self-absorption dips at the respective velocities in the spectra. Complex spectral profiles may also be caused by infall motions in optically thick lines (e.g., \citealt{myers1996}). We compared the HCO$^+(1-0)$ and H$^{13}$CO$^+(1-0)$ spectra along the filamentary structures and identified different features. Toward the highest column density cores, some HCO$^+$ spectra indeed exhibit velocity shifts compared to where the more optically thin H$^{13}$CO$^+$ peaks (Fig.~\ref{comparison} left spectrum). However, such velocity shifts between H$^{13}$CO$^+$ and HCO$^+$,  likely caused by high optical depth, are typically spatially constrained only to regions close to the main cores. With increasing distance to the cores, and by that decreasing column density, one typically finds well agreeing velocities between HCO$^+(1-0)$ and H$^{13}$CO$^+(1-0)$. One finds single component spectra for both lines but one sometimes also finds multiple components in the main isotopologue HCO$^+$ but only a single component in H$^{13}$CO$^+$ (Fig.~\ref{comparison},  middle and right spectra). If the velocity of the single component in H$^{13}$CO$^+$ agrees then with one of the HCO$^+$ components, this is indicative for HCO$^+$ tracing multiple gas components along those lines of sight. And only the higher column density component is also seen in the rarer H$^{13}$CO$^+$ isotopologue. 

\begin{figure}[h]
	\centering
	\includegraphics[width=0.97\linewidth ,keepaspectratio]
		{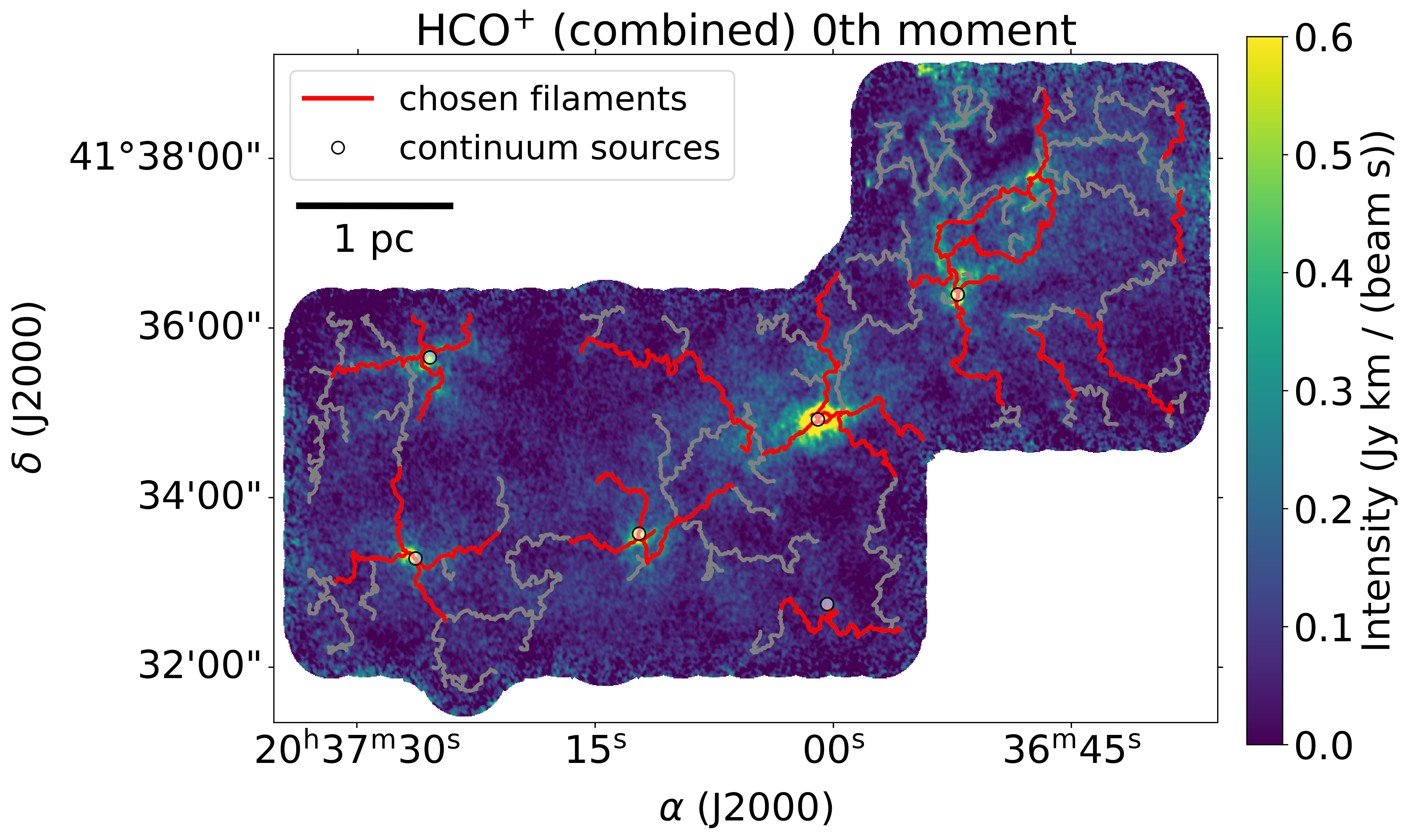}
	\includegraphics[width=0.97\linewidth ,keepaspectratio]
		{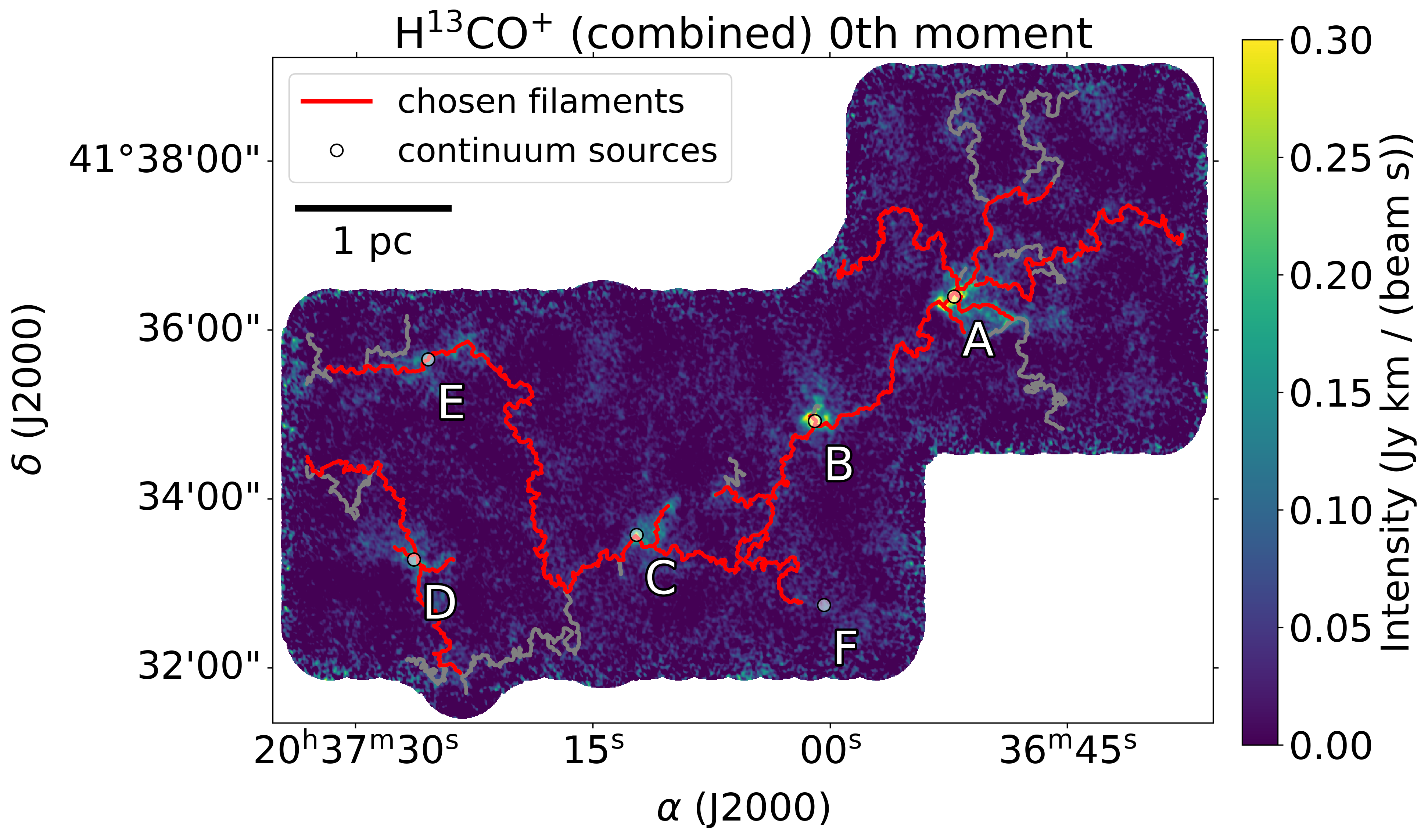}
\caption[Filaments in combined \HCO\ and \HttCO\ data selected for further analysis.]{Selected filaments for further analysis shown in red in the combined \HCO$(1-0)$ (top) and \HttCO$(1-0)$ (bottom) data. The six main cores labeled A to F are shown in the bottom panel.}
\label{plot_chosen_filaments}
\end{figure}

While analyzing all identified filaments is beyond the scope of the paper, for further analysis, we selected several representative filaments from the combined \HCO\ and \HttCO\ data. We combined adjacent and coherent short filaments from the DisPerSE analysis to obtain longer coherent structures. We define the “origin” of these new filaments as their coordinates closest to the core peak position derived from the millimeter continuum data. These origin coordinates are identical for all filaments connected to each core. In addition, in \HCO\ we also chose some filaments in more remote areas, which are not directly connected to a core, to see if their properties differ from those connected to cores. In this case their “origin” was chosen arbitrarily as the last coordinates on one of their ends.
The selected filaments are presented in Fig. \ref{plot_chosen_filaments}.

\subsection{Spectral line parameters of filaments}
\label{subsect_res_vel}

\begin{figure*}[h!]
	\centering
		\begin{subfigure}{1\linewidth}
		\centering
		\includegraphics[width=0.86\linewidth ,keepaspectratio]
		{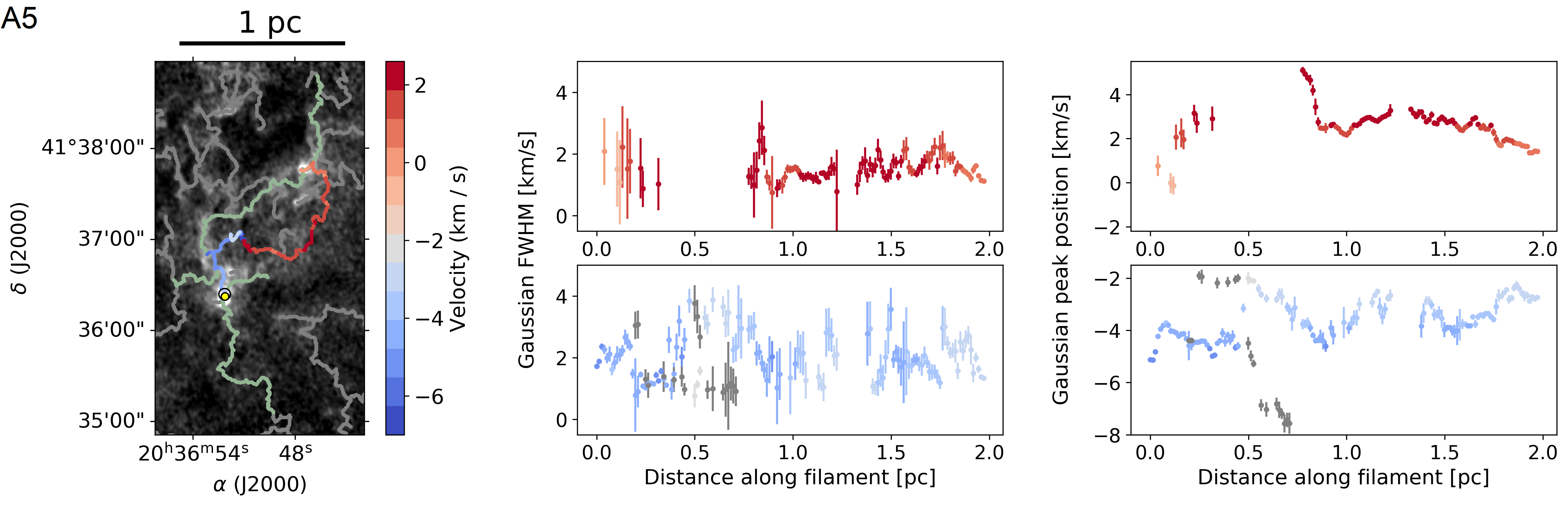}
	\end{subfigure}
	\begin{subfigure}{1\linewidth}
		\centering
		\includegraphics[width=0.86\linewidth ,keepaspectratio]
		{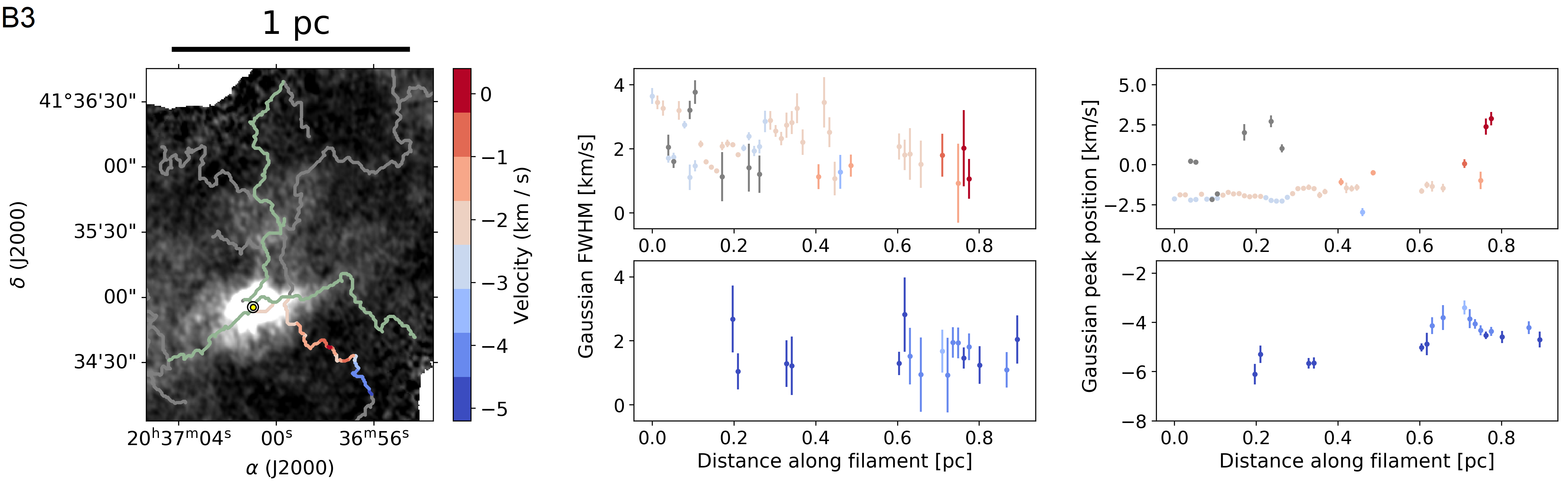}
	\end{subfigure}
	\begin{subfigure}{1\linewidth}
		\centering
		\includegraphics[width=0.86\linewidth ,keepaspectratio]
		{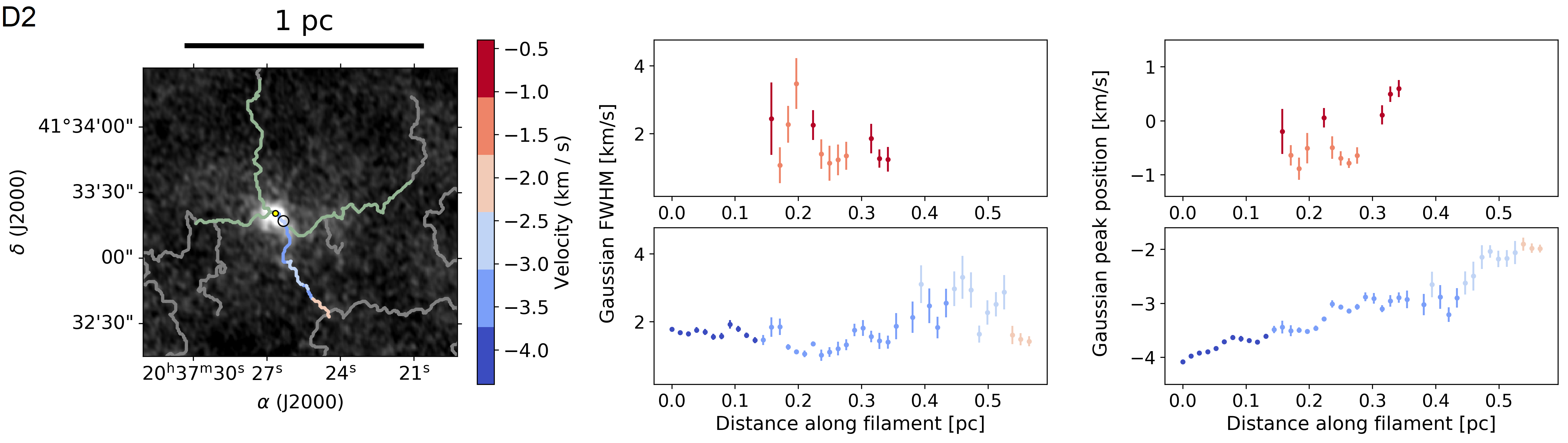}
	\end{subfigure}
	\begin{subfigure}{1\linewidth}
		\centering
		\includegraphics[width=0.86\linewidth ,keepaspectratio]
		{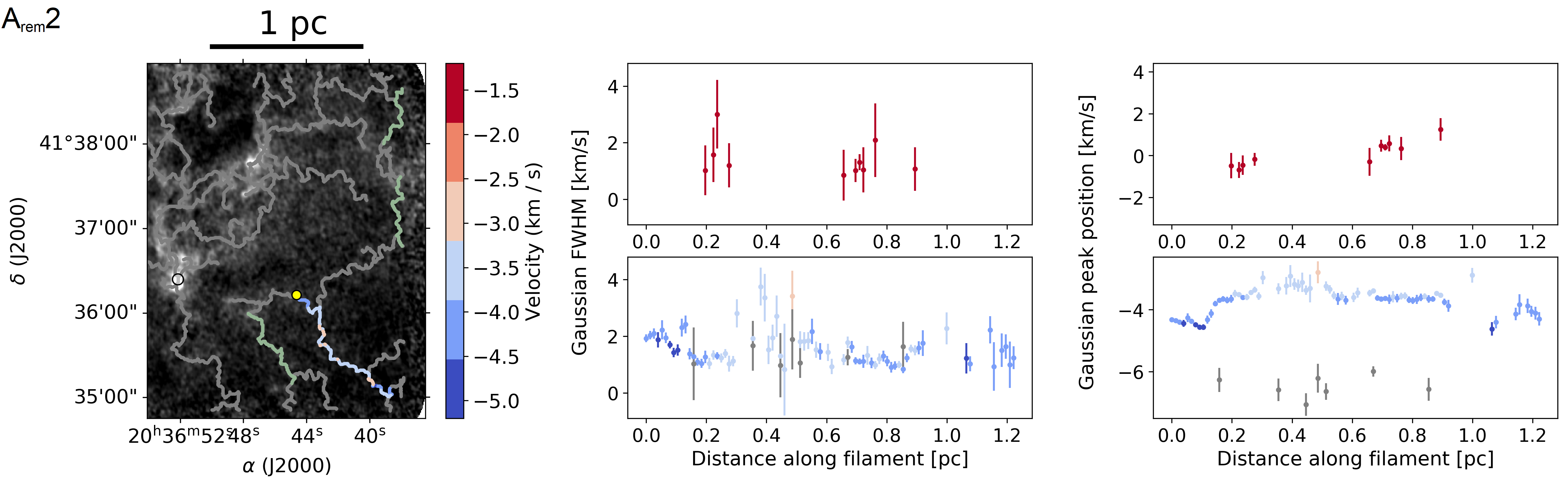}
	\end{subfigure}
\caption[FWHM and peak positions of the Gaussian fits to the spectra within filaments connected to cores A, B, D, and a remote area in \HCO.]{\textit{From top to bottom}: Cores A, B, D, and a remote area close to core A, respectively (A5, B3, D2, A$_{\text{rem}}$2). \textit{Images}: \HCO$(1-0)$ zeroth moment map (see top left of Fig. \ref{plot_MIOP_filaments_0mom}). The white dot within the core indicates the coordinates of the continuum source, while the yellow dot indicates the filament “origin,” as described in the main text. The filament corresponding to the plots is shown in its velocity colors. Other analyzed filaments are shown in green, their plots can be found in supplementary material at Zenodo (Figs.~B.1 and B.16). \textit{Plots}: Gaussian fitted HCO$^+$ FWHM (left) and peak positions (right) of the spectra within the corresponding filament, plotted over the distance from the core. The yellow dot in the image corresponds to a distance of zero. If there is a significant amount of data points ($\geq$ 10) for a second spectral Gaussian peak, the positions of the higher velocity peaks (top) and the lower velocity peaks (bottom) are shown in separate plots. The peaks that do not clearly belong to either of the two components are shown in gray. The other data points are colored by their fitted peak positions.}
\label{plot_along_chosen_HCO}
\end{figure*}

\begin{figure*}
	\centering
		\begin{subfigure}{1\linewidth}
		\centering
		\includegraphics[width=0.9\linewidth ,keepaspectratio]
		{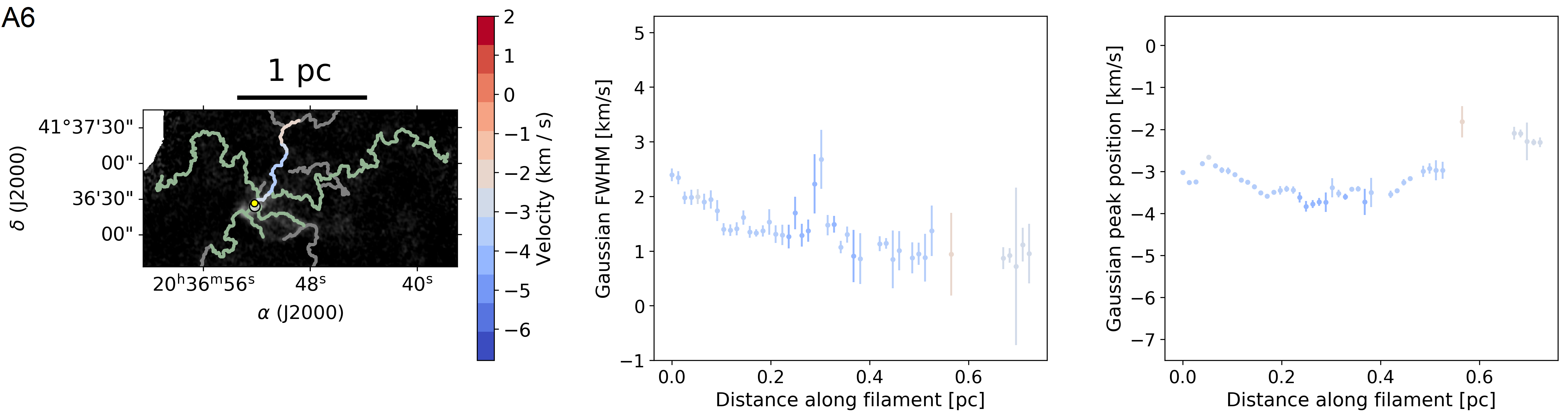}
	\end{subfigure}
	\begin{subfigure}{1\linewidth}
		\centering
		\includegraphics[width=0.9\linewidth ,keepaspectratio]
		{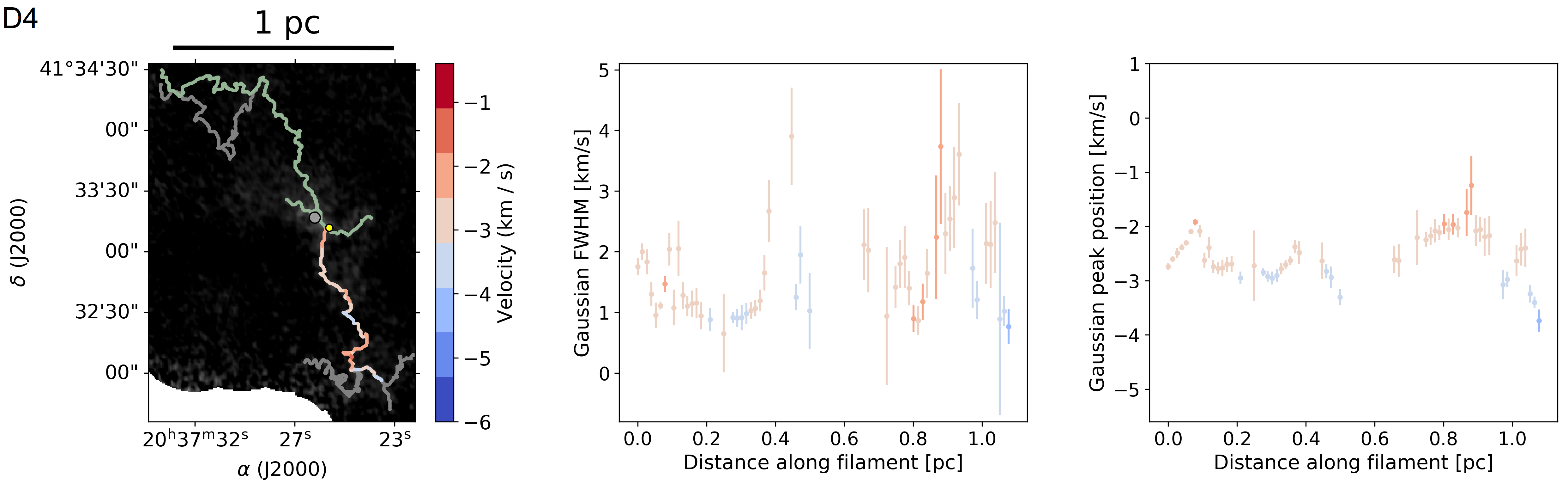}
	\end{subfigure}
\caption[FWHM and peak positions of the Gaussian fits to the spectra within filaments connected to cores A and D in \HttCO$(1-0)$.]{\textit{Rows}: Cores A and D (filaments A6 and D4) measured in H$^{13}$CO$^+$. Left panels: Images of the \HttCO\ zeroth moment maps (see top right of Fig. \ref{plot_MIOP_filaments_0mom}). Middle and right panels: Same as Fig.\,\ref{plot_along_chosen_HCO}, but for one velocity component.}
\label{plot_along_chosen_H13CO}
\end{figure*}

For the selected filaments, we determined peak velocity, linewidth and their corresponding uncertainties along the filament lengths, as described in Section \ref{subsect_meth_vel}. We show some of the results in Figs.~\ref{plot_along_chosen_HCO} and \ref{plot_along_chosen_H13CO}). The filaments corresponding to the plots in their row are colored by the velocity of the channel they were identified in. The red color indicates the analyzed filaments in this region (Fig. \ref{plot_chosen_filaments}), and all the additional plots can be found in the supplementary material available at Zenodo. The filament origin together with the intensity peak of the line emission overlaps with the continuum source for half the cores, and is slightly offset in cores D, E, and F in \HCO, and in cores B, D, and F in \HttCO\ by a distance on the order of 10$^{-2}$ pc. While such offsets for HCO$^+$ could be caused by high optical depth, this is less likely for H$^{13}$CO$^+$. Offsets between the latter and the continuum emission may relate to chemical effects; however, this is beyond the scope of this paper. We number the filaments of each core and write this number and the core label together to identify a filament; for example, the third filament connected to core A is referred to as “A3." In the plots in the Appendix showing all the analyzed filaments, this number corresponds to the row the filament is shown in. In \HCO\ there are some filaments in remote areas, which are labeled according to the core they are close to, i.e., A$_{\text{rem}}$ and B$_{\text{rem}}$.

\begin{figure*}
	\centering
	\begin{subfigure}{1\linewidth}
		\centering
		\includegraphics[width=0.97\linewidth ,keepaspectratio]
		{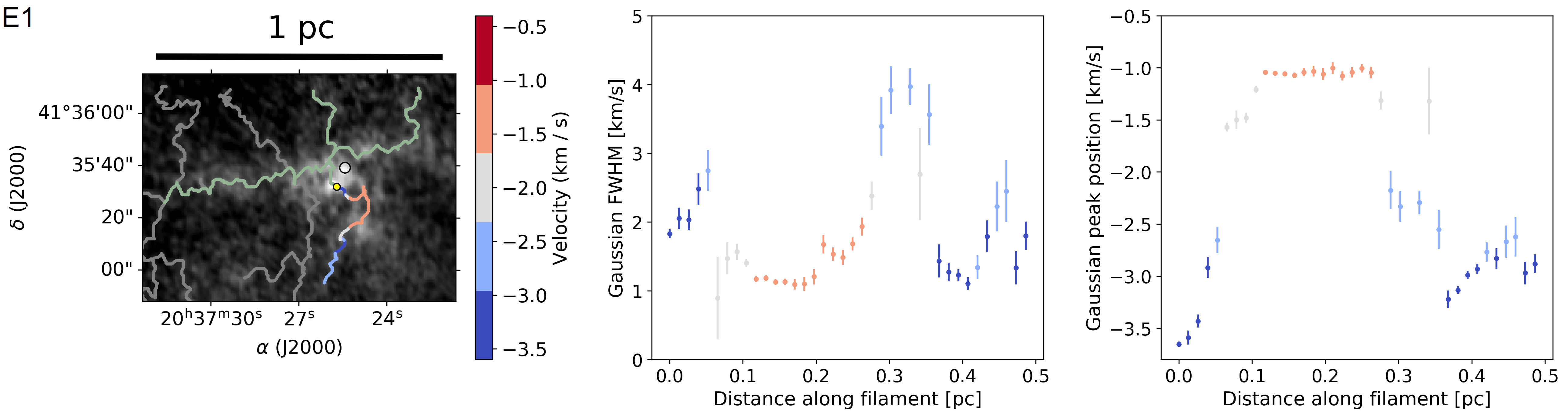}
	\end{subfigure}
	\begin{subfigure}{0.19\linewidth}
		\centering
		\includegraphics[width=1\linewidth ,keepaspectratio]
		{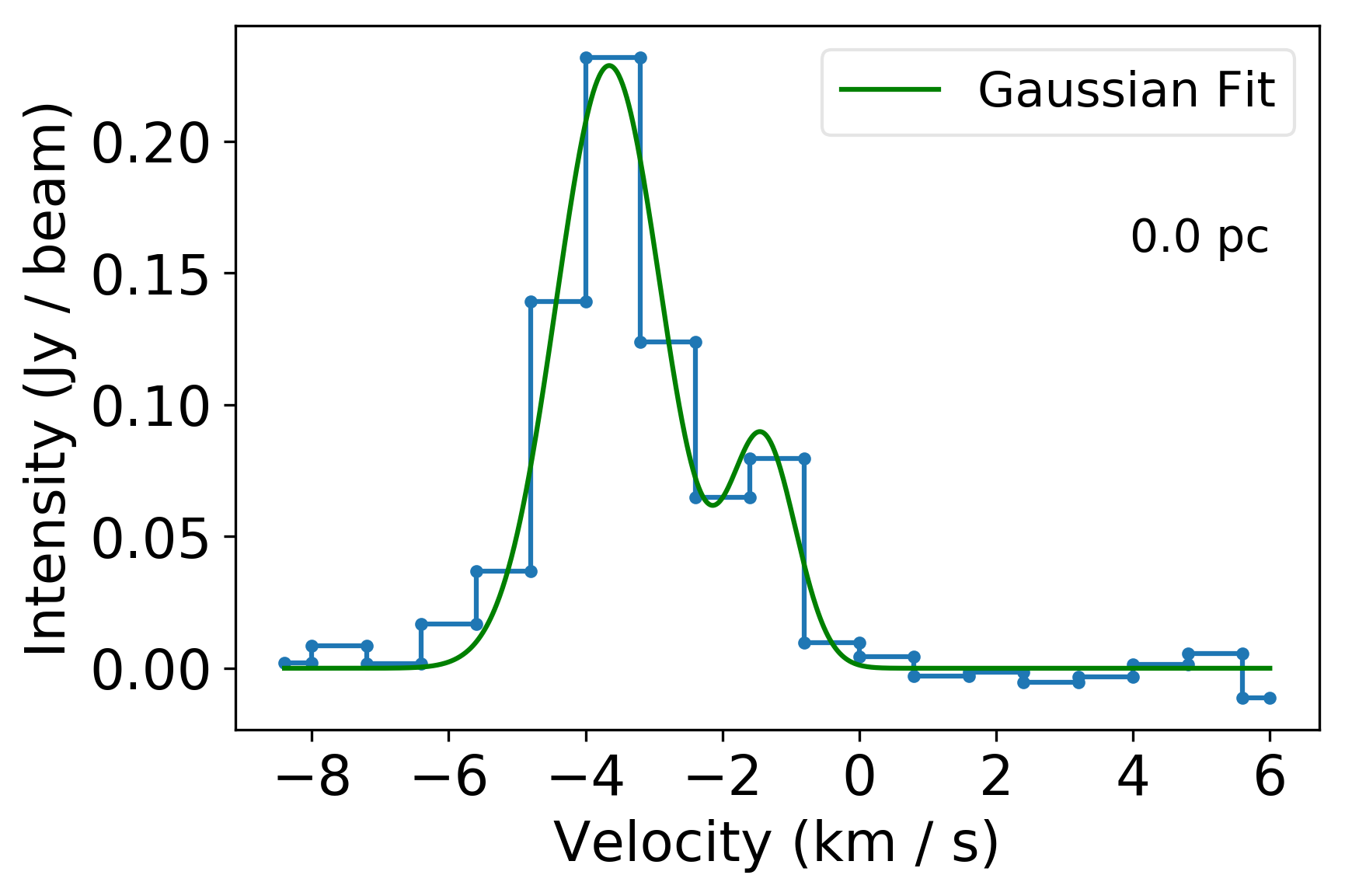}
	\end{subfigure}
	\begin{subfigure}{0.19\linewidth}
		\centering
		\includegraphics[width=1\linewidth ,keepaspectratio]
		{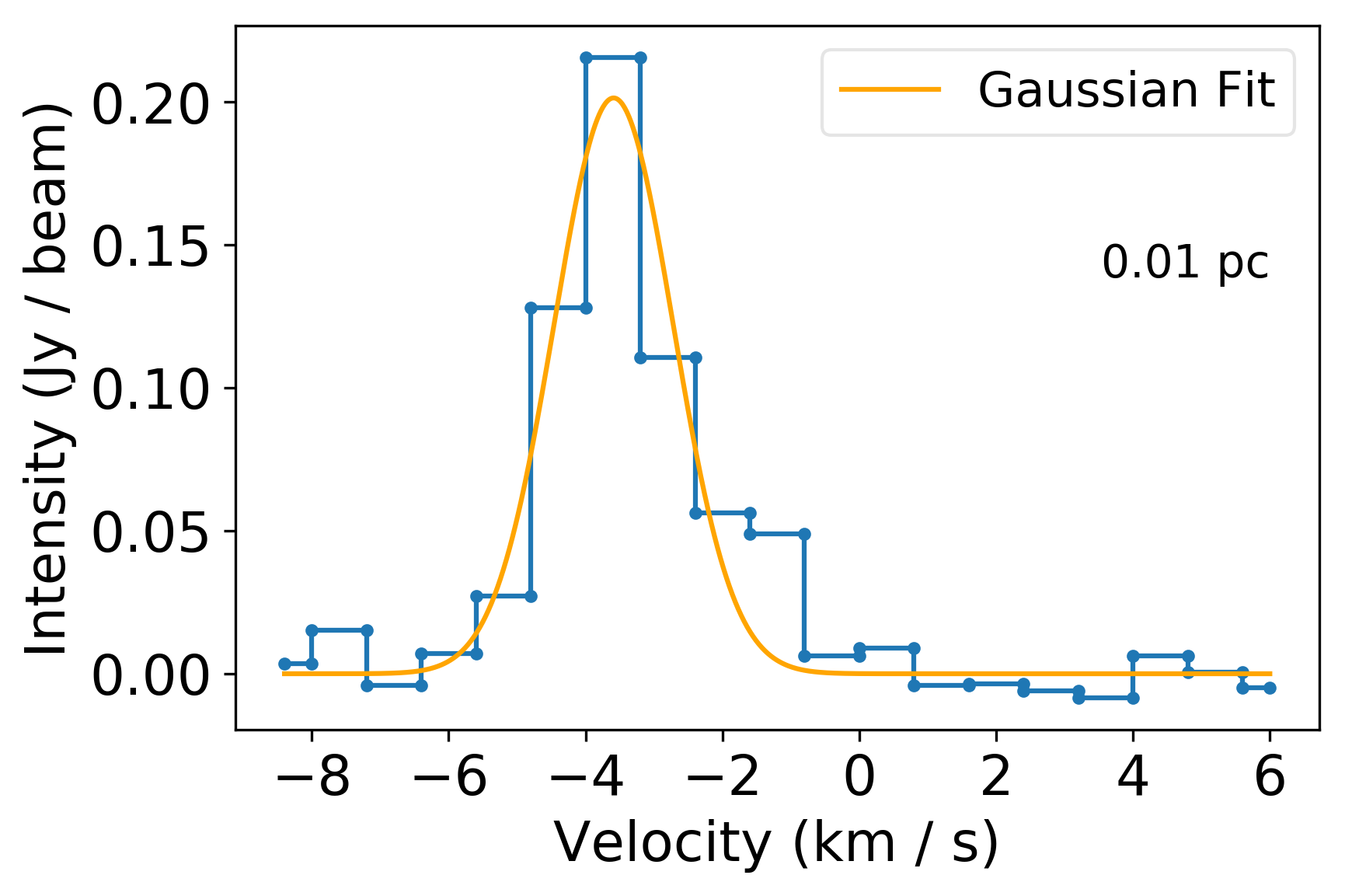}
	\end{subfigure}
	\begin{subfigure}{0.19\linewidth}
		\centering
		\includegraphics[width=1\linewidth ,keepaspectratio]
		{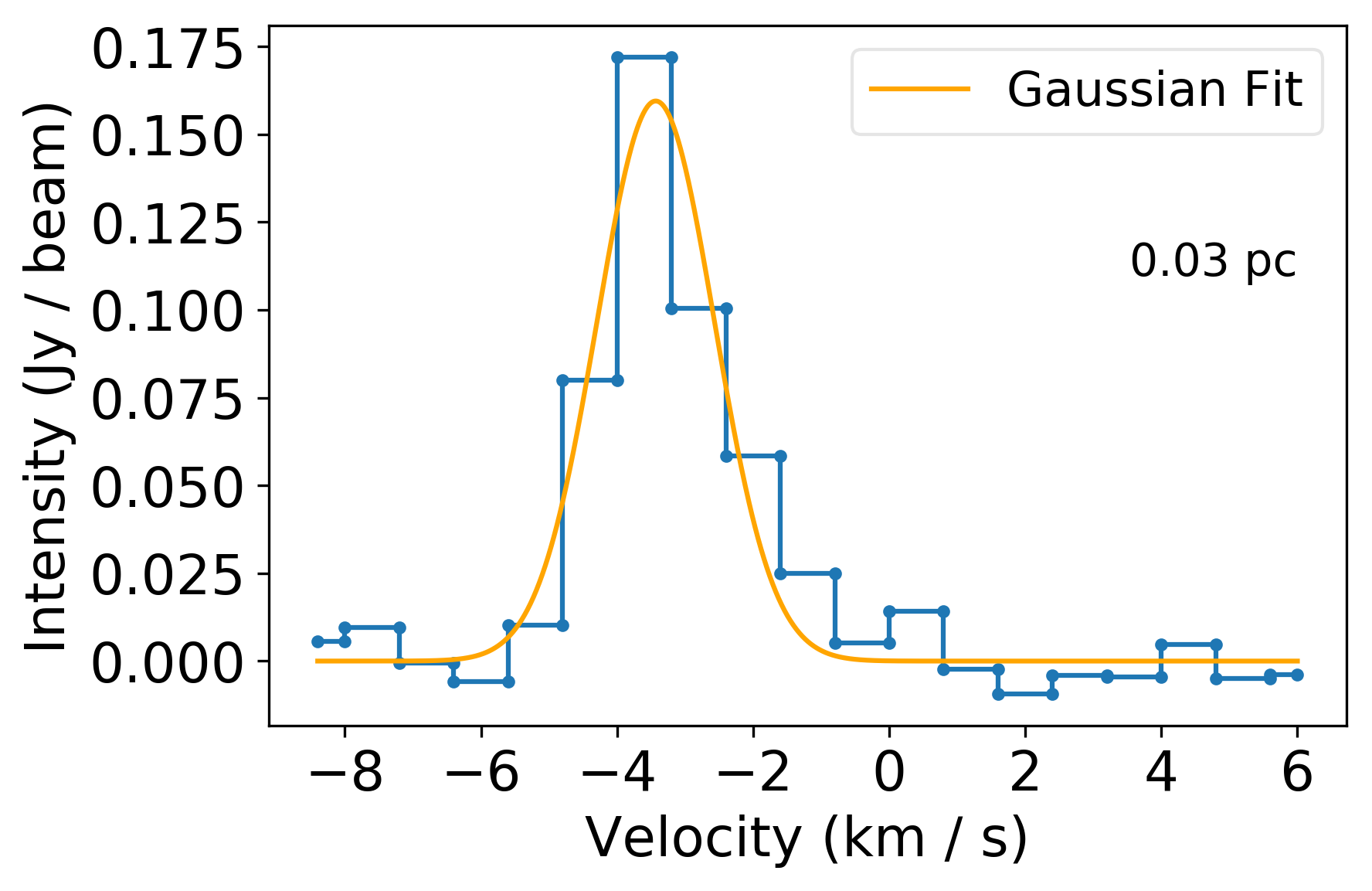}
	\end{subfigure}
	\begin{subfigure}{0.19\linewidth}
		\centering
		\includegraphics[width=1\linewidth ,keepaspectratio]
		{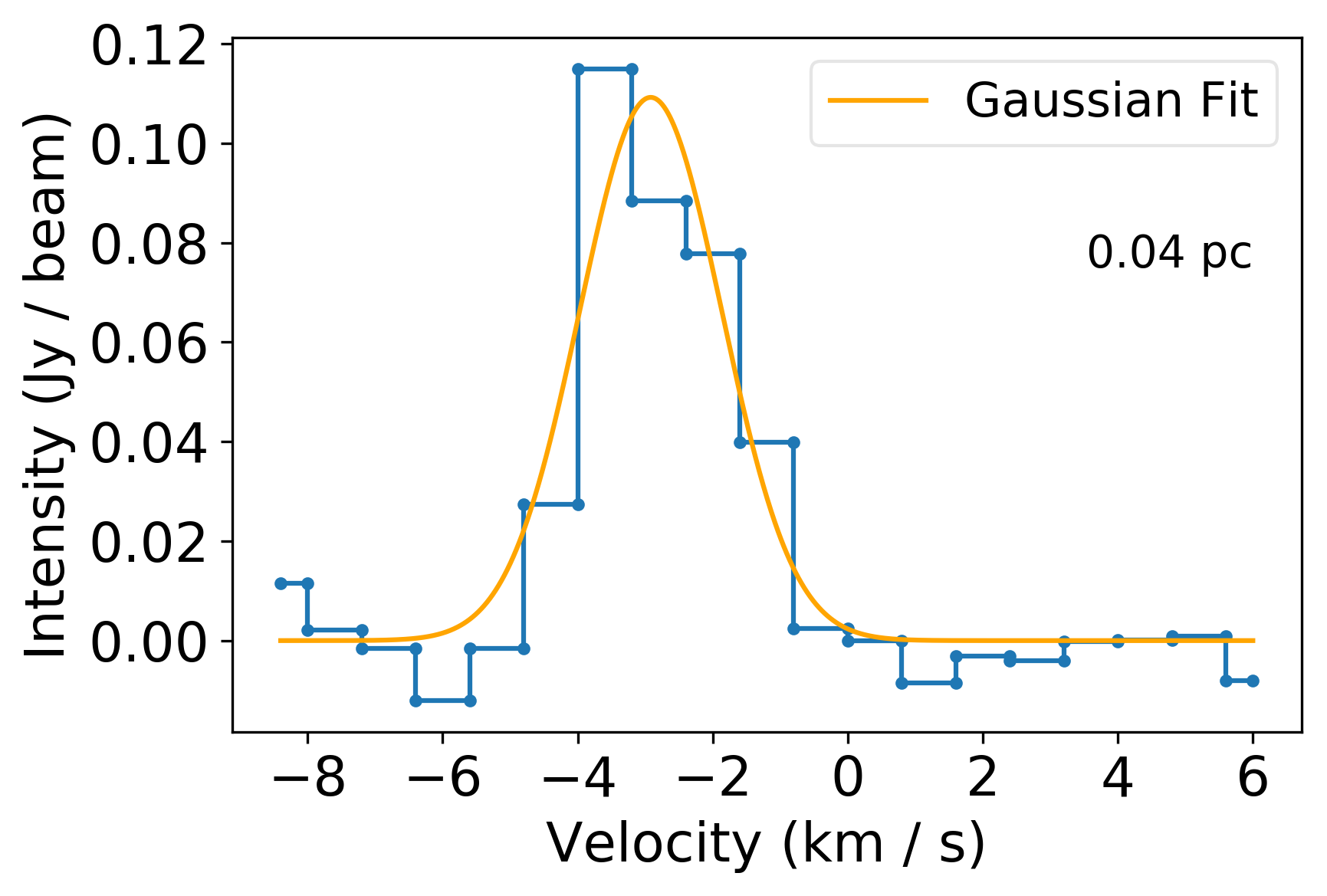}
	\end{subfigure}
	\begin{subfigure}{0.19\linewidth}
		\centering
		\includegraphics[width=1\linewidth ,keepaspectratio]
		{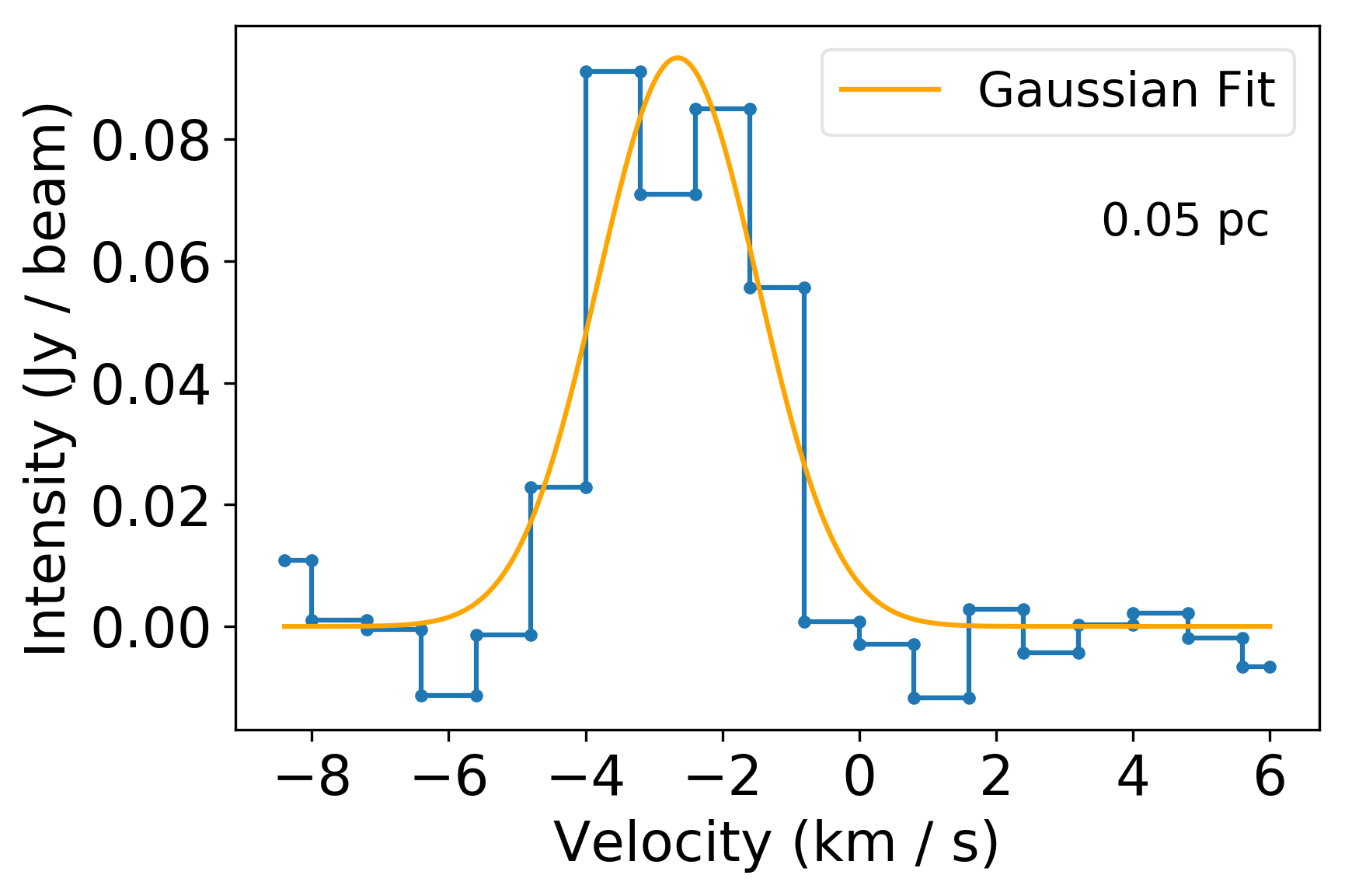}
	\end{subfigure}
	\begin{subfigure}{0.19\linewidth}
		\centering
		\includegraphics[width=1\linewidth ,keepaspectratio]
		{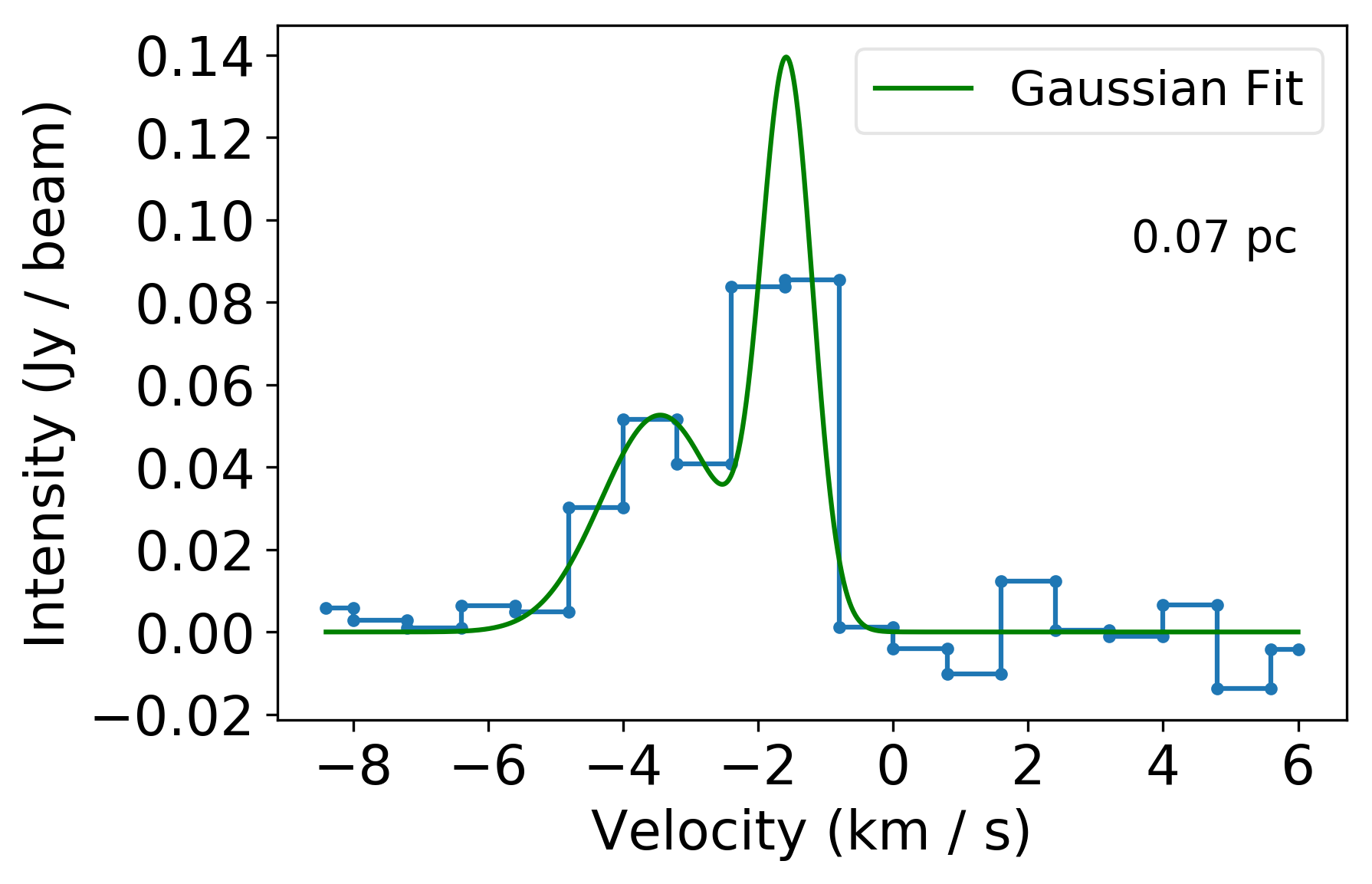}
	\end{subfigure}
	\begin{subfigure}{0.19\linewidth}
		\centering
		\includegraphics[width=1\linewidth ,keepaspectratio]
		{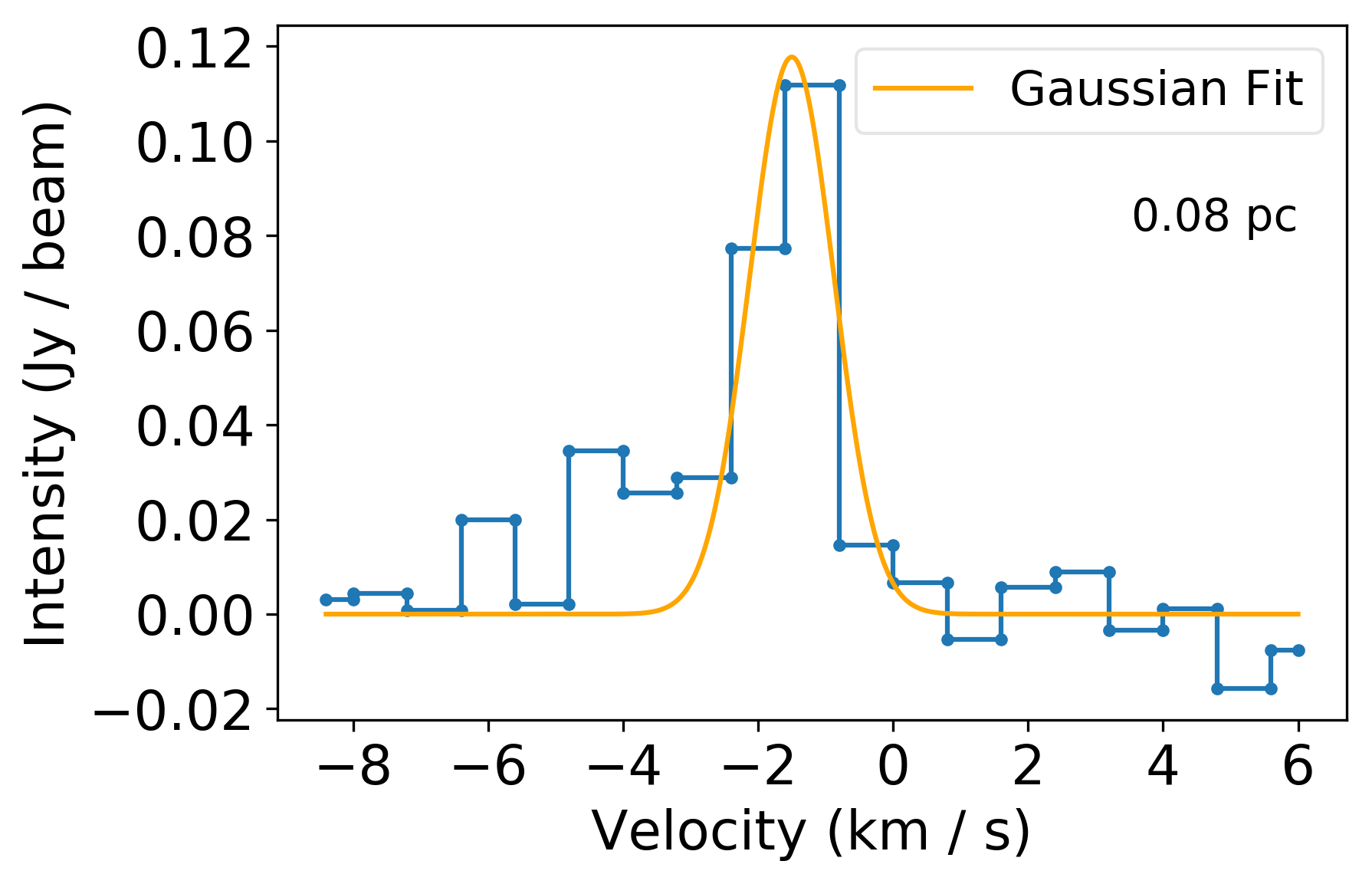}
	\end{subfigure}
	\begin{subfigure}{0.19\linewidth}
		\centering
		\includegraphics[width=1\linewidth ,keepaspectratio]
		{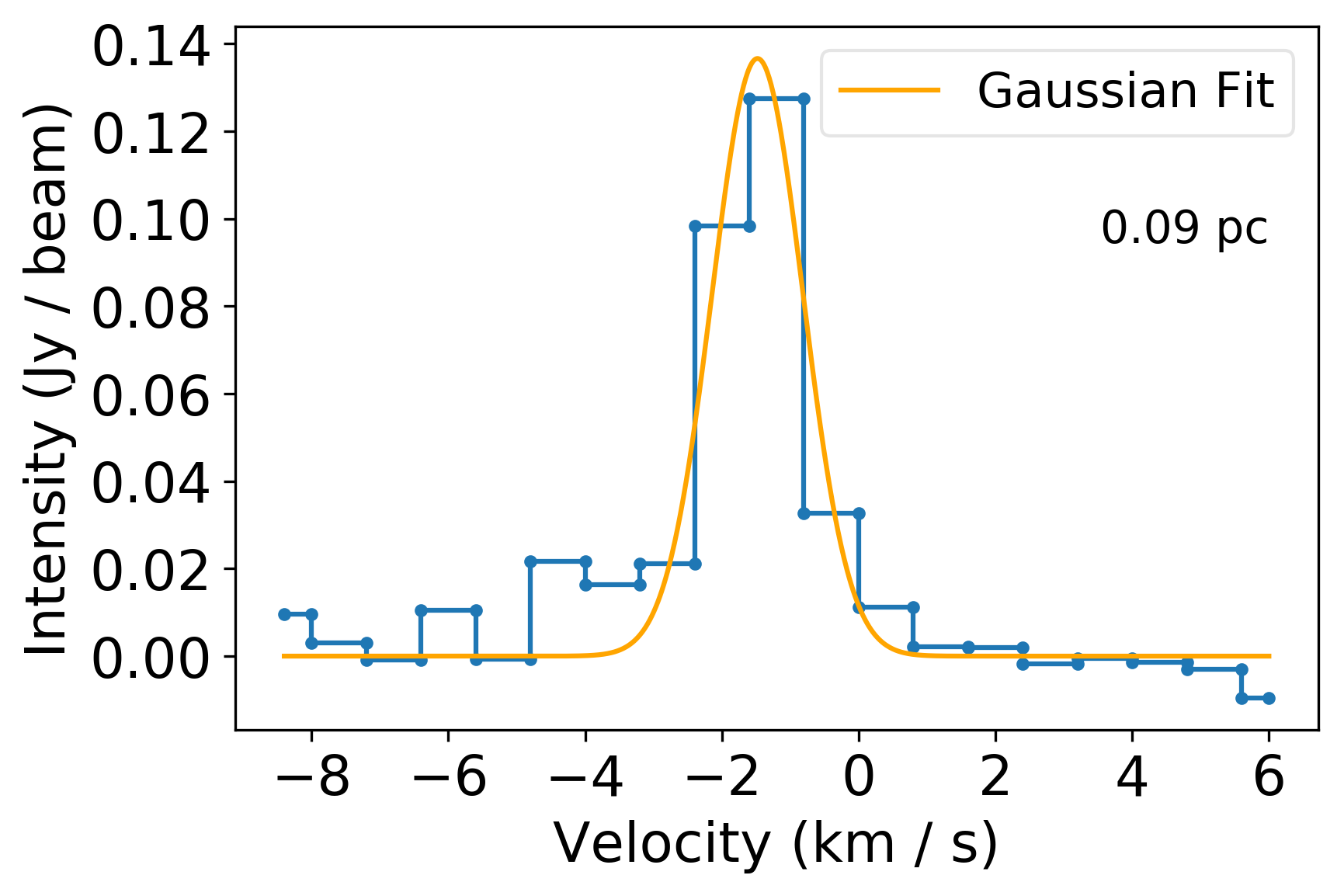}
	\end{subfigure}
	\begin{subfigure}{0.19\linewidth}
		\centering
		\includegraphics[width=1\linewidth ,keepaspectratio]
		{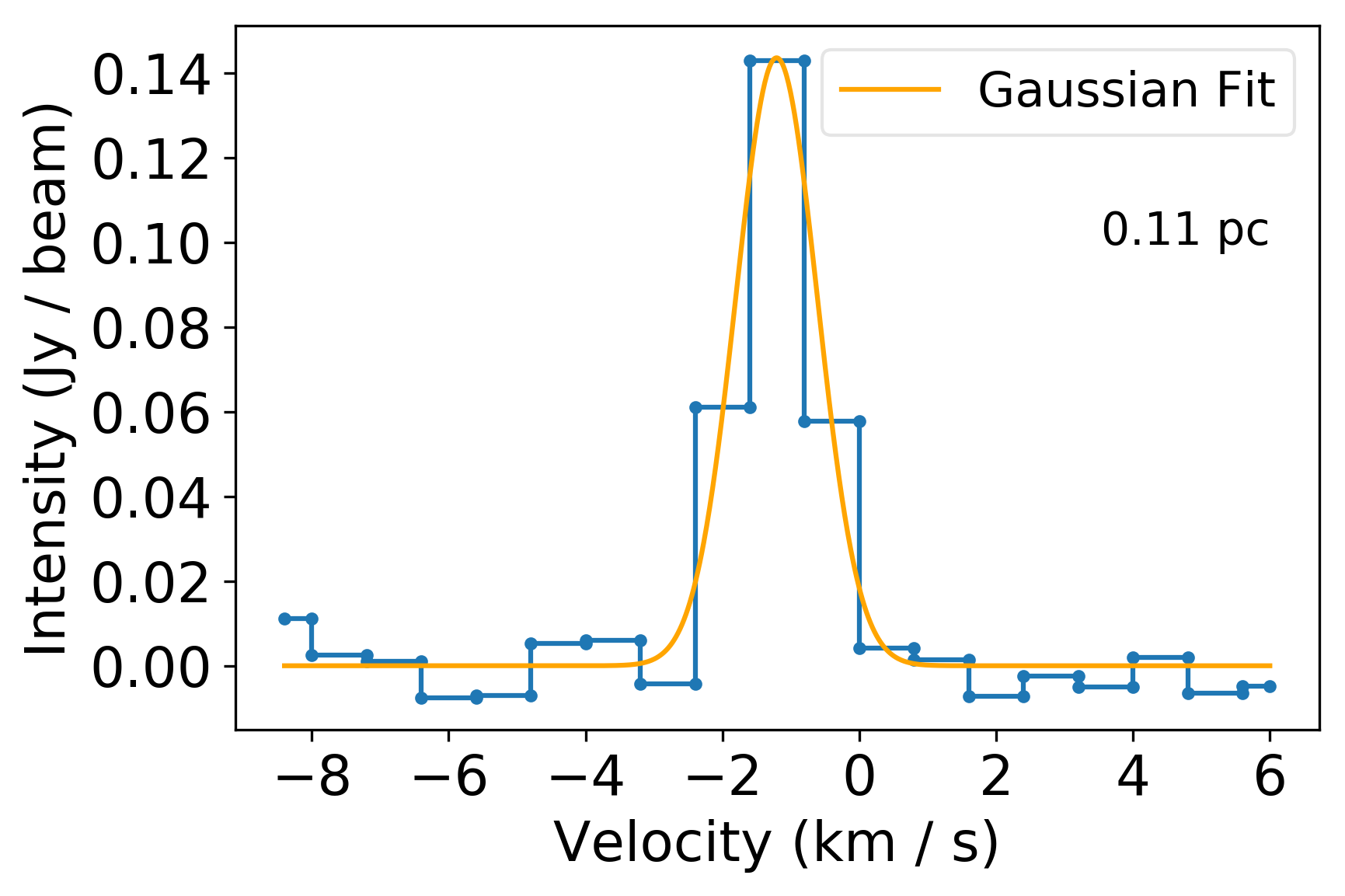}
	\end{subfigure}
	\begin{subfigure}{0.19\linewidth}
		\centering
		\includegraphics[width=1\linewidth ,keepaspectratio]
		{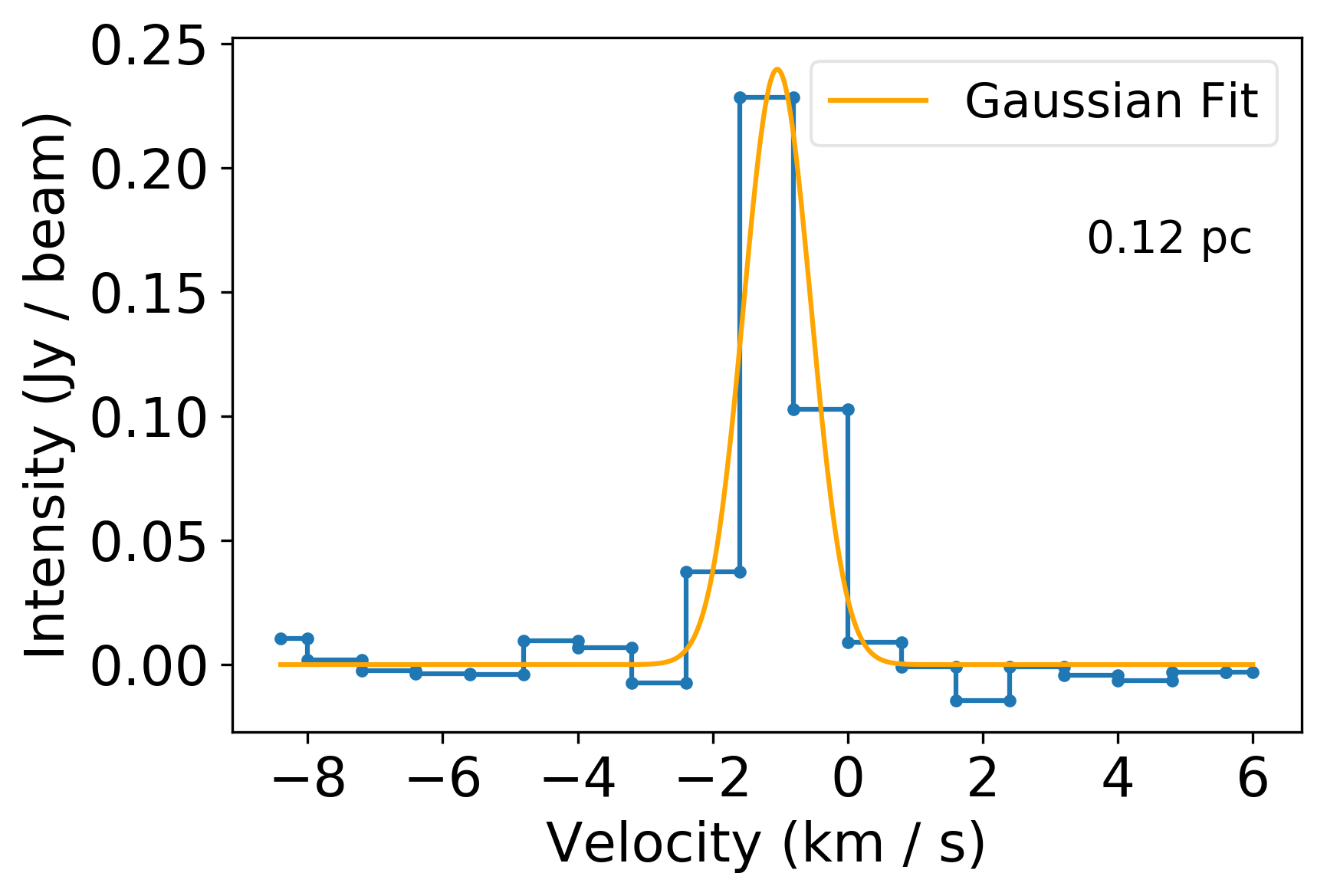}
	\end{subfigure}
\caption[FWHM and peak positions of the Gaussian fits to the spectra within one filament connected to core E in \HCO and the first 10 spectra used for their determination.]{Linewidth and velocity along E1 in \HCO$(1-0)$ (for more details see Fig. \ref{plot_along_chosen_HCO}) and its first ten averaged spectra possibly showing two velocity components. The distance from the core is shown in each spectrum. The green line indicates a double-peak Gaussian fit, while the orange line indicates a single-peak.}
\label{plot_merging_peaks_EHCO}
\end{figure*}

Our method identifies up to four velocity components per spectrum within each filament (Fig. \ref{plot_along_fit_exmpl}). However, while several second peaks are found, a third or fourth peak is only rarely fit. We show the second peaks if there is a reliable amount of data ($\geq$ 10 data points) (e.g., Fig. \ref{plot_along_chosen_HCO}; third peaks are only shown for B$_{\text{rem}}$ in Fig.~B.9 in supplementary material at Zenodo). In some filaments, however, while not enough third peaks are identified to show them in separate plots, there still seem to be three velocity components (e.g., A5 in \HCO, Fig. \ref{plot_along_chosen_HCO}).
In some cases, the identified first and second peaks are clearly separated by their amplitude and velocity into two components, as B2 in HCO$^+$ (Fig.~B.3 in supplementary material at Zenodo). However, in most cases, their separation is difficult, and what might be two (or more) components could also be (at least partly) a large variance in velocity of one component, as A5 in HCO$^+$ (Fig. \ref{plot_along_chosen_HCO}). Similarly, an unidentified second component could resemble a large variance in velocity, as with E1 in HCO$^+$ (e.g., Fig. \ref{plot_merging_peaks_EHCO}). We try to separate components into two plots, but they should always be considered together.
The error bars in the plots show the corresponding parameter errors of the Gaussian fits to the spectra given in the covariance matrix of the curve\_fit function.

For several filaments, we find a projected velocity gradient close to the core in their main velocity component (i.e., the highest amplitude peaks). From 0 to 0.1\,pc projected distance, i.e., along 5 beam sizes from the core, we find changes in the gas velocity inside the filament between 0.4 and 2.4\,km\,s$^{-1}$. While 0.4\,km\,s$^{-1}$ sounds at first sight small compared to the velocity resolution of 0.8\,km\,s$^{-1}$, we point out that these velocity differences are measured from the Gaussian fits to the spectra, and Gaussian peak values can be determined to much higher accuracy than the actual velocity resolution (down to the resolution divided by the signal-to-noise ratio, e.g., \citealt{reid1988}). We find at least one filament with such a gradient toward almost every core, with core B (2nd row in Fig. \ref{plot_along_chosen_HCO} and Fig.~B.3 in supplementary material at Zenodo) as the one exception for HCO$^+$, and core E as the one exception for \HttCO\ (Fig.~B.12 in supplementary material at Zenodo). For core F the S/N close to the core is too low for an analysis. All gradients are presented in Table \ref{tab_vel}. Note that the filaments converge at the core positions, and their first spectra at 0\,pc distance are often the same for several filaments (marked in Table \ref{tab_vel}, e.g., A3 to A5 in \HCO).

While this is difficult to interpret in absolute terms due to unknown inclination angles of the individual filaments, the gas inside such filaments may speed up toward the core. This is discussed in Section \ref{sect_disc}. There is one point to note about these gradients: visual inspection shows that the large velocity gradients in core E of \HCO\ (Fig.~\ref{plot_merging_peaks_EHCO}) and core A of \HttCO\ may originate from two velocity components gradually merging together. The fit might pick up only one of the peaks at first, then fit both peaks as one with a peak position in between the two real positions together with a growing linewidth. We also find such behavior farther from the core. Lastly, only the second peak, which has a higher amplitude than the first farther from the core, might be fit. However, we cannot clearly see in the spectra whether this is the case, or if there is a different reason for the velocity gradients in core E of \HCO\ and core A of \HttCO.

Further from the cores, some filaments show a velocity gradient not
only close to the core, but over the whole length of the filament
(e.g., B2 and D2 in \HCO, Fig.~B.3 in supplementary material at Zenodo
DOI. In contrast to this, we also find filaments with an
approximately constant velocity over most of their length (e.g., B1 in
\HCO, Fig.~B.3 in supplementary material at Zenodo). We refrain
from further interpretation of these features.

\begin{figure*}
	\centering
		\begin{subfigure}{0.33\linewidth}
		\centering
		\includegraphics[width=1\linewidth ,keepaspectratio]
		{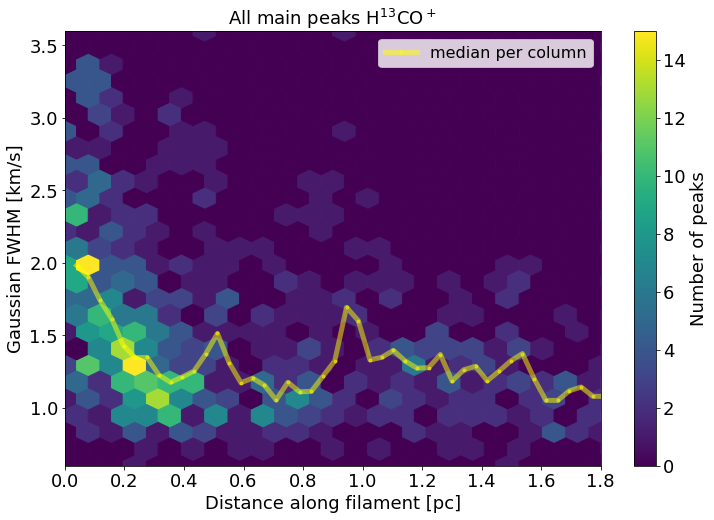}
	\end{subfigure}
	\begin{subfigure}{0.33\linewidth}
		\centering
		\includegraphics[width=1\linewidth ,keepaspectratio]
		{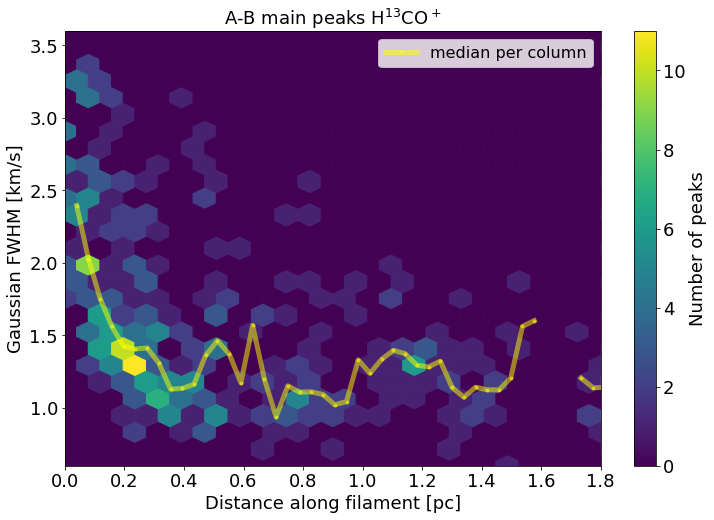}
	\end{subfigure}
	\begin{subfigure}{0.33\linewidth}
		\centering
		\includegraphics[width=1\linewidth ,keepaspectratio]
		{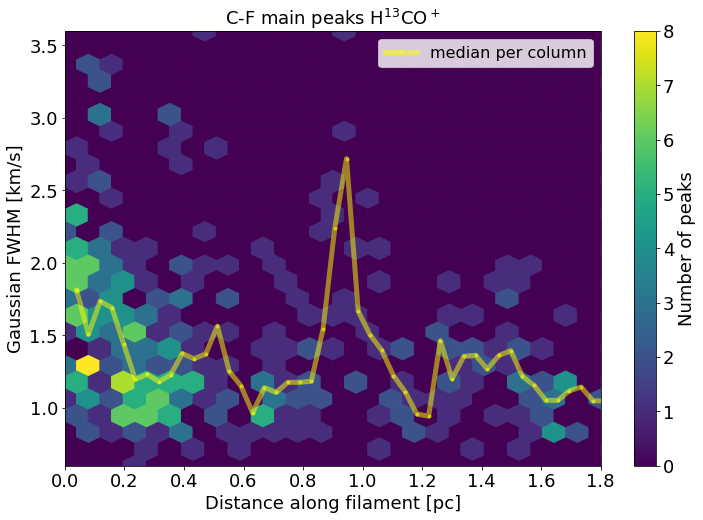}
	\end{subfigure}
\caption[Distribution of spectral linewidth along filaments.]{2D histograms showing the distribution of the linewidths of spectra (i.e., the FWHM of the Gaussian fits to their main peaks) within filaments over their distance from the core they are connected to in \HttCO$(1-0)$. The yellow line and dots show the median value of each column. \textit{Left}: All cores. \textit{Middle}: Evolved cores (A, B). \textit{Right}: Younger cores (C-F). The spike in the mean value for cores C-F at $\sim$\,1pc is likely caused by unresolved multiple components and low statistical sampling.}
\label{plot_hexbin_lw}
\end{figure*}

The observed typical linewidths along the filaments are generally narrower for \HttCO, i.e., between 1.2 and 2.1 \kms\ compared to those for \HCO, which range between 1.6 and 2.5 \kms. We point out that the lower end of the measured linewidths is limited by our spectral resolution (0.8\,km\,s$^{-1}$). Nevertheless, the relative linewidth differences between the two species should be real. The linewidths of the main components at the core are around 2 \kms\ for all filaments in all cores, with the exceptions of two filaments in core B (see Fig.~B.3 in supplementary material at Zenodo) with linewidths around 3 \kms\ (B1) and 4 \kms\ (B3) in \HCO\ and 1 filament in core A (see Fig.~B.10 in supplementary material at Zenodo) around 3 \kms\ in \HttCO. 
We observe the linewidths decreasing over the distance for one or more filaments in each core, sometimes down to $\sim$ 1 \kms\ (e.g., A6 in \HttCO, Fig. \ref{plot_along_chosen_H13CO}). To visualize this tendency, we create 2D histograms of the linewidths of the main peaks over the distance along all filaments, and separated into evolved cores A and B, and younger cores C to F (Fig. \ref{plot_hexbin_lw}). While we do not find a clear trend, there seems to be a tendency of larger linewidths closer to the core, and this tendency is more obvious for the two evolved cores than for the younger ones. 
While increased linewidth toward a core can also be produced by feedback processes, such as molecular outflows or expanding H{\sc ii} regions, a larger linewidth toward the cores fits into the picture of a gravitationally accelerated accretion flow toward the cores (see also discussion section \ref{sect_disc}).

To summarize, we find velocity gradients of 0.4 to 2.4\,\kms\ over 0.1 pc in \HCO, and of 0.7 to 1.7 \kms\ over 0.1 pc in \HttCO. We find such gradients in at least one filament in all regions, with the exceptions of core B in \HCO\ and core E in \HttCO. However, we find gradients in these cores in the respective other data. 

The linewidth is generally narrower for \HttCO\ than for \HCO, with mean values between 1.6 and 2.5 \kms\ in \HCO\ and between 1.2 and 2.1 \kms\ in \HttCO. We find a tendency of larger linewidths closer to the core. We further discuss these results in Section \ref{sect_disc}.

\subsection{Filament properties derived with FilChaP}
\label{subsect_res_Filchap}

\begin{figure*}[h]
	\centering
	\begin{subfigure}{0.33\textwidth}
  		\centering
		\includegraphics[width=0.97\linewidth ,keepaspectratio]
		{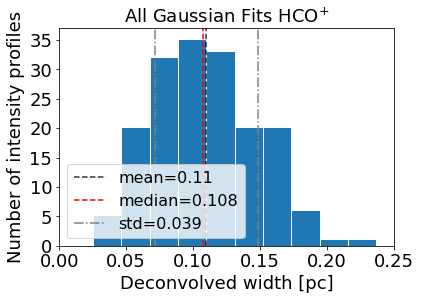}
	\end{subfigure}%
	\begin{subfigure}{0.33\textwidth}
		\centering
		\includegraphics[width=0.97\linewidth ,keepaspectratio]
		{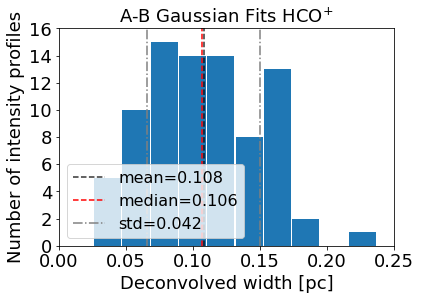}
	\end{subfigure}
	\begin{subfigure}{0.33\textwidth}
		\centering
		\includegraphics[width=0.97\linewidth ,keepaspectratio]
		{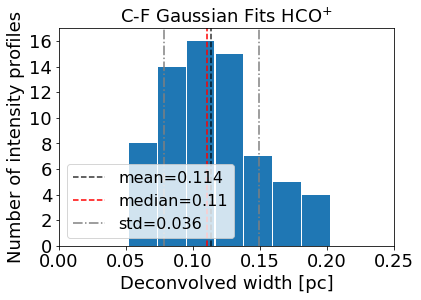}
	\end{subfigure}
	\vskip 1pt
		\begin{subfigure}{0.33\textwidth}
  		\centering
		\includegraphics[width=0.97\linewidth ,keepaspectratio]
		{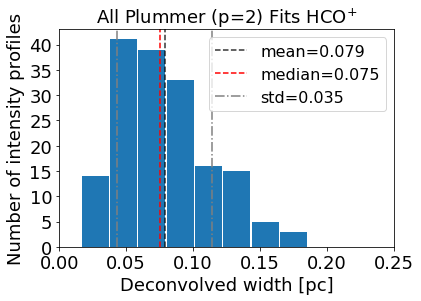}
	\end{subfigure}%
	\begin{subfigure}{0.33\textwidth}
		\centering
		\includegraphics[width=0.97\linewidth ,keepaspectratio]
		{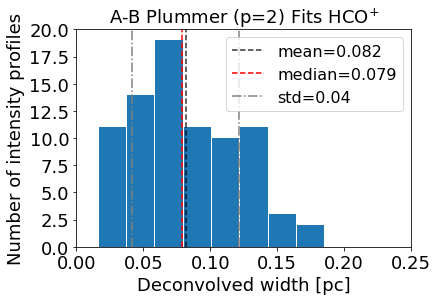}
	\end{subfigure}
	\begin{subfigure}{0.33\textwidth}
		\centering
		\includegraphics[width=0.97\linewidth ,keepaspectratio]
		{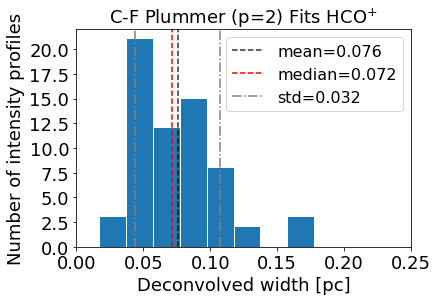}
	\end{subfigure}
		\vskip 1pt
		\begin{subfigure}{0.33\textwidth}
  		\centering
		\includegraphics[width=0.97\linewidth ,keepaspectratio]
		{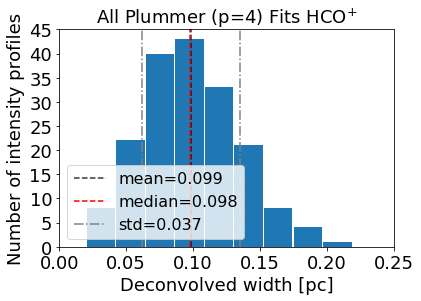}
	\end{subfigure}%
	\begin{subfigure}{0.33\textwidth}
		\centering
		\includegraphics[width=0.97\linewidth ,keepaspectratio]
		{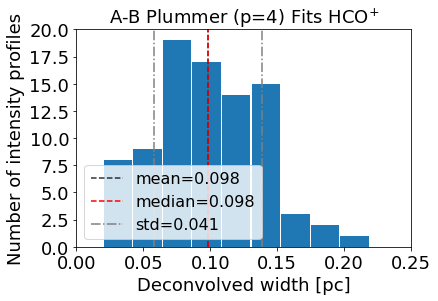}
	\end{subfigure}
	\begin{subfigure}{0.33\textwidth}
		\centering
		\includegraphics[width=0.97\linewidth ,keepaspectratio]
		{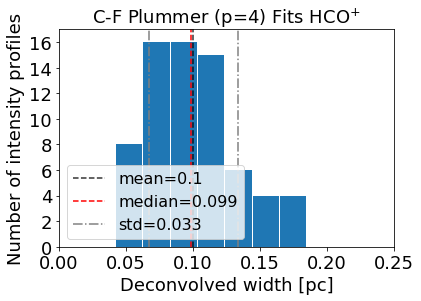}
	\end{subfigure}
	\vskip 1pt
	\begin{subfigure}{0.33\textwidth}
  		\centering
		\includegraphics[width=0.97\linewidth ,keepaspectratio]
		{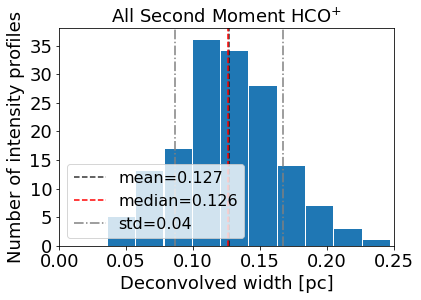}
	\end{subfigure}%
	\begin{subfigure}{0.33\textwidth}
		\centering
		\includegraphics[width=0.97\linewidth ,keepaspectratio]
		{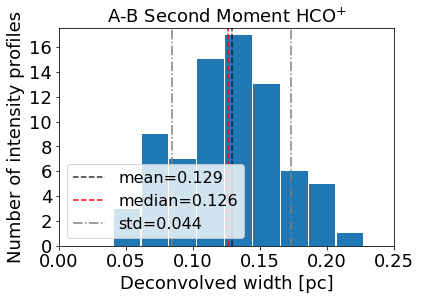}
	\end{subfigure}
	\begin{subfigure}{0.33\textwidth}
		\centering
		\includegraphics[width=0.97\linewidth ,keepaspectratio]
		{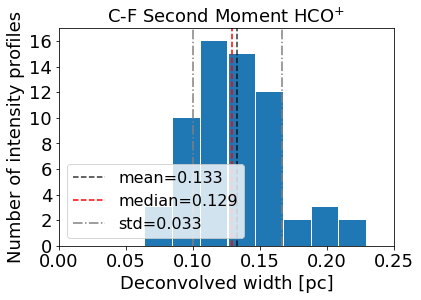}
	\end{subfigure}
\caption[Filament widths in \HCO\ of all analyzed filaments, and separated into younger and evolved cores, for all methods.]{All calculated de-convolved filament widths resulting from the different fits performed with FilChaP on the combined \HCO$(1-0)$ data. \textit{Left column}: All regions including cores A to F and the remote filaments close to A and B. \textit{Middle column}: Only filaments directly connected to the evolved regions A and B. \textit{Right column}: Only filaments directly connected to the younger regions C, D, E, and F. \textit{Rows}: Different fits, i.e., Gaussian, Plummer $p$ = 2 and $p$ = 4, and second moment, from top to bottom.}
\label{plot_filchap_AB_CF}
\end{figure*}

\begin{figure*}
	\centering
	\begin{subfigure}{0.33\textwidth}
  		\includegraphics[width=0.97\linewidth ,keepaspectratio]
		{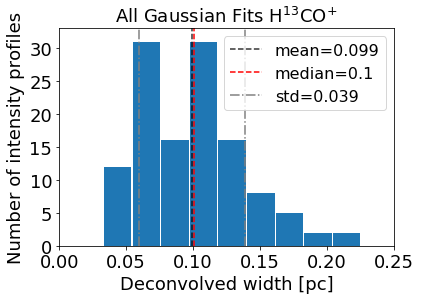}
	\end{subfigure}%
	\begin{subfigure}{0.33\textwidth}
		\includegraphics[width=0.97\linewidth ,keepaspectratio]
		{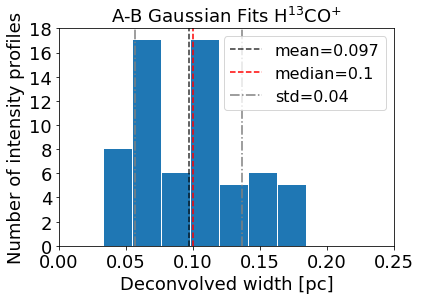}
	\end{subfigure}
	\begin{subfigure}{0.33\textwidth}
		\includegraphics[width=0.97\linewidth ,keepaspectratio]
		{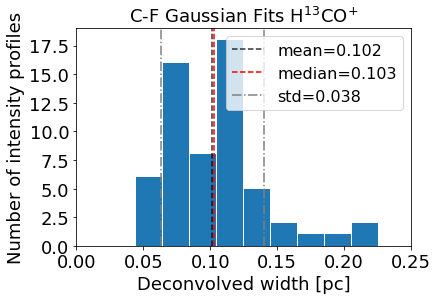}
	\end{subfigure}
	\vskip 1pt
		\begin{subfigure}{0.33\textwidth}
  		\includegraphics[width=0.97\linewidth ,keepaspectratio]
		{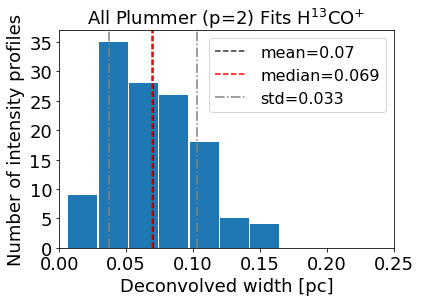}
	\end{subfigure}%
	\begin{subfigure}{0.33\textwidth}
		\includegraphics[width=0.97\linewidth ,keepaspectratio]
		{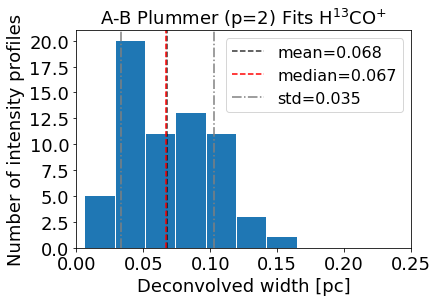}
	\end{subfigure}
	\begin{subfigure}{0.33\textwidth}
		\includegraphics[width=0.97\linewidth ,keepaspectratio]
		{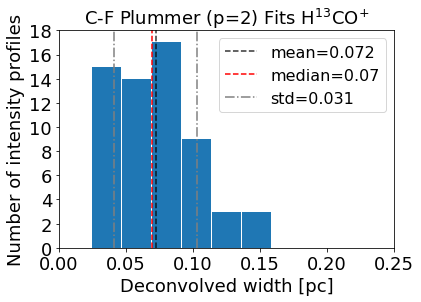}
	\end{subfigure}
		\vskip 1pt
		\begin{subfigure}{0.33\textwidth}
  		\includegraphics[width=0.97\linewidth ,keepaspectratio]
		{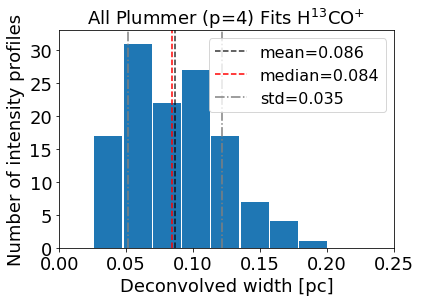}
	\end{subfigure}%
	\begin{subfigure}{0.33\textwidth}
		\includegraphics[width=0.97\linewidth ,keepaspectratio]
		{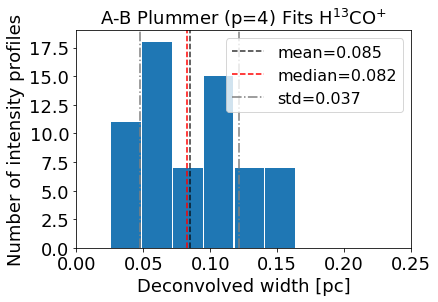}
	\end{subfigure}
	\begin{subfigure}{0.33\textwidth}
		\includegraphics[width=0.97\linewidth ,keepaspectratio]
		{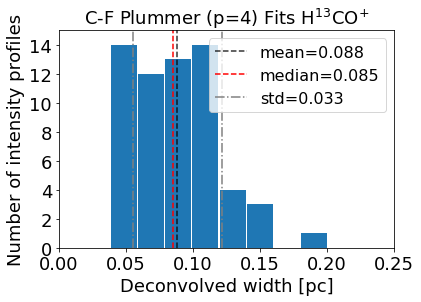}
	\end{subfigure}
	\vskip 1pt
	\begin{subfigure}{0.33\textwidth}
  		\includegraphics[width=0.97\linewidth ,keepaspectratio]
		{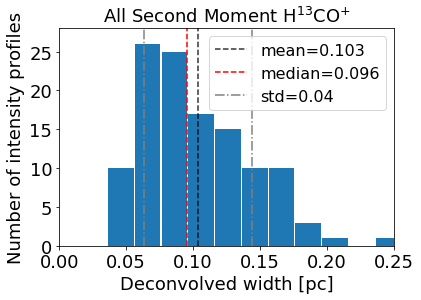}
	\end{subfigure}%
	\begin{subfigure}{0.33\textwidth}
		\includegraphics[width=0.97\linewidth ,keepaspectratio]
		{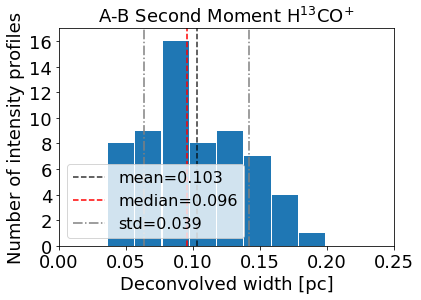}
	\end{subfigure}
	\begin{subfigure}{0.33\textwidth}
		\includegraphics[width=0.97\linewidth ,keepaspectratio]
		{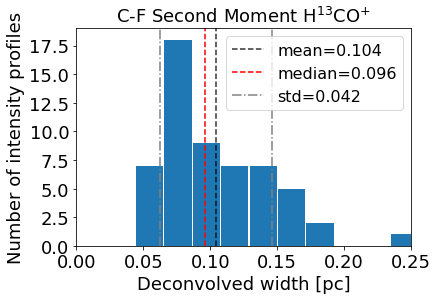}
	\end{subfigure}
\caption[Filament widths in \HttCO\ of all analyzed filaments, separated into younger and evolved cores, for all methods.]{All calculated de-convolved filament widths resulting from the different fits done with FilChaP on the combined \HttCO$(1-0)$ data. \textit{Left column}: All regions A to F. \textit{Middle column}: Only filaments directly connected to the evolved regions A and B. \textit{Right column}: Only filaments directly connected to the younger regions C, D, E, and F. \textit{Rows}: Different fits, i.e., Gaussian, Plummer $p$ = 2 and $p$ = 4, and second moment, from top to bottom.}
\label{plot_filchap_AB_CF_H13CO+}
\end{figure*}

For all selected filaments, we determined the filament width from the combined NOEMA+30\,m data, as described in Section \ref{subsect_meth_Filchap}. We created histograms for the different methods and for all regions, and calculate mean, median, and standard deviations of the filament parameter distributions. First, the histograms of the de-convolved filament width including all filament slices analyzed, i.e., in the case of \HCO\ also including the filaments from the remote areas, are shown in Fig. \ref{plot_filchap_AB_CF} for \HCO\ and in Fig. \ref{plot_filchap_AB_CF_H13CO+} for \HttCO. In the same figures, we show the histograms for the different methods of only the filament slices belonging to the evolved cores A and B (second column in the figures), and of only the filament slices belonging to the younger cores C to F (third column).
All other histograms can be found in the supplementary material at Zenodo (Figs. B.18 to B.29). All mean, median, and standard deviation values for \HCO, \HttCO, and all different methods can be found in Tables \ref{tab_Filchap_all_AF} and \ref{tab_Filchap_all_rem}.

The means of the linewidth distributions for HCO$^+$ filaments in all different methods (first column of Fig. \ref{plot_filchap_AB_CF}), range from 0.08 to 0.13\,pc while the corresponding standard deviations are approximately 0.04\,pc. For the evolved regions A and B in \HCO\ (second column in the figure \ref{plot_filchap_AB_CF}), the mean and standard deviation values of the distribution have the same range as for the one including all filament slices, with the values only differing by $\lesssim$ 0.003 pc. For the younger regions, they differ slightly more, by up to $\sim$ 0.006\,pc compared to the distribution including all filaments, but still ranging from 0.08 to 0.13\,pc, with standard deviations closer to 0.03\,pc. The lower standard deviation could simply be an effect of the sample size for the younger cores being smaller, with, for example, 69 slices for the Gaussian fit of the younger cores, compared to 82 slices for the same fit of the evolved cores.

In \HttCO\ the means of the width distribution of all filament slices for the different methods range from 0.07 to 0.10 pc, with standard deviations from 0.03 to 0.04 pc. This range stays the same for the evolved cores A and B, as well as the younger cores C to F, with the values only differing by $\lesssim$ 0.003 pc. The standard deviations are also similar for evolved and younger cores. 

The lowest mean and median values are always a result of the Plummer (Eq. \ref{eq_plummer}) ($p$ = 2) fit, for both \HCO\ and \HttCO. The highest values are a result of the second moment (Eq. \ref{eq_2ndmoment}) in \HCO\ with significant differences to the other methods. In \HttCO, however, the second moment gives similar results to the Gaussian (Eq. \ref{eq_Gauss}) fits. In the Appendix (Figs. \ref{plot_filchap_AB_CF_chi} and \ref{plot_filchap_AB_CF_chi_H13CO+}) we show histograms of the goodness of the different fits (reduced $\chi^2$). We find that for a Plummer ($p$ = 4) fit the mean of the distributions of all filaments in \HCO\ is closest to 1, and the median has the lowest value. Since many values of reduced $\chi^2$ $<$ 1 are included, this is difficult to interpret (see section \ref{subsect_meth_Filchap}). It could also be argued that the peak of the histograms at reduced $\chi^2$ $\sim$ 1 is most pronounced within the distribution for the Gaussian fits (Eq. \ref{eq_Gauss}).

For \HttCO, the values for all fits are similar, and none of the distributions has a pronounced peak at reduced $\chi^2$ $\sim$ 1. This will be discussed in Section \ref{sect_disc}.
Nevertheless, all fits lead to filament widths around 0.1 pc, with small systematic differences.

In general, we find smaller filament widths for \HttCO\ than for \HCO\ for all methods, with differences between the two usually $\lesssim$ 0.03 pc, but up to 0.06 pc (second moment in core C). The only exception is core E, where we find similar or slightly larger mean and median values of the width distributions in \HCO\ and \HttCO\ for all methods except the second moment. Again, the values range around 0.1\,pc for both tracers. Since H$^{13}$CO$^+$ has a lower optical depth than the main isotopologue, it likely represents the real physical width of the filaments better.

\begin{figure*}
	\centering
	\begin{subfigure}{1\textwidth}
  		\centering
		\includegraphics[width=0.99\linewidth ,keepaspectratio]
		{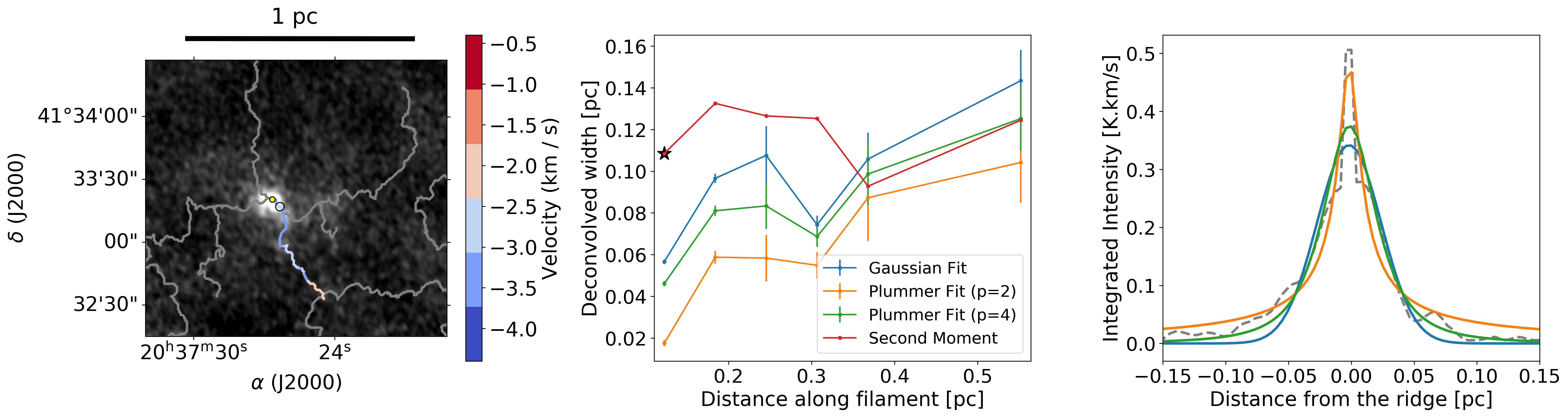}
	\end{subfigure}%
	\vskip 1pt
	\begin{subfigure}{1\textwidth}
  		\centering
		\includegraphics[width=0.99\linewidth ,keepaspectratio]
		{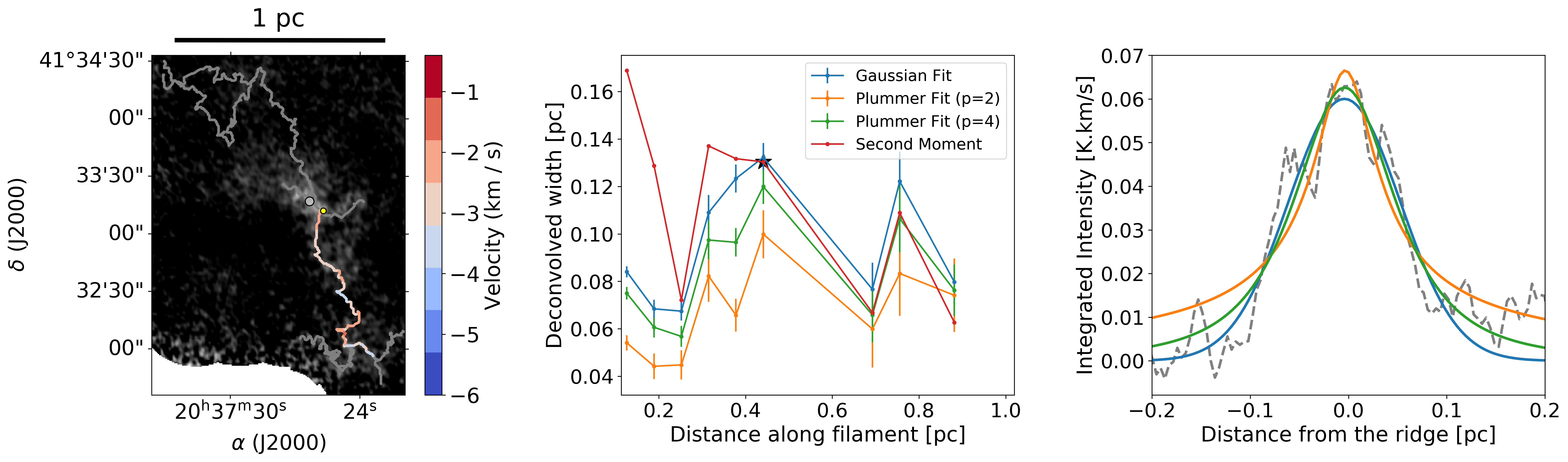}
	\end{subfigure}%
        \caption[Filament width (all methods) of slices of two filaments
          of core D in \HCO\ and \HttCO\ along the
          filament.]{De-convolved widths of filament slices of core D
          in \HCO$(1-0)$ (D2) and in \HttCO$(1-0)$ (D4) determined
          with FilChaP. \textit{Rows}: \HCO$(1-0)$ and
          \HttCO$(1-0)$. Left: Core D. The white dot within the core
          shows the coordinates of the continuum source; the yellow
          dot shows the filament “origin," corresponding to a distance
          of zero. The filament corresponding to the plot is colored
          by the velocity of the channel it was identified in. The
          green color indicates the other analyzed filaments in this region; the ones not shown here can be found in the
          supplementary material available at Zenodo (Figs. B.30 to
          B.45). Middle: De-convolved width of the filament slices
          plotted over the distance from the core (or more
          precisely, the filament “origin," yellow dot in the
          image). A distance of zero thereby corresponds to the slice
          closest to the core. The error bars show the de-convolved FWHM
          errors resulting from the errors of the corresponding
          parameters of the respective fit to the radial profile given
          in the covariance matrix of the curve\_fit function used in
          FilChaP. The black star marks the filament width resulting
          from the spectrum shown in the right plot. Right: Example of
          an intensity profile, as averaged by FilChaP (dashed gray
          line), and its fits at the position of the filament marked
          with a black star in the middle plot.}
\label{plot_Filchap_alongD_chosen}
\end{figure*}

Furthermore, FilChaP calculates the width for filament slices, meaning
we can also plot this width along the length of the respective
filament. We show one filament in core D in \HCO\ and \HttCO\ each as
an example in Fig. \ref{plot_Filchap_alongD_chosen}. The error bars in
these plots show the de-convolved FWHM errors. The errors are
calculated from the corresponding parameters of each fit to the radial
profile as given in the covariance matrix of the \texttt{curve\_fit}
function that was used in FilChaP. As mentioned above, we see the
Plummer (Eq. \ref{eq_plummer}) ($p$ = 2) fit generally giving the
smallest filament width, while the second moment often gives the
largest. We find no general trend of the linewidth along filaments in
any of the cores.  All plots showing the filament width along each
filament can be found in the supplementary material at Zenodo
(Figs. B.30 to B.45).

\section{Discussion}
\label{sect_disc}

\subsection{Do filaments connect the large and small scales?}

In the following we discuss our results as answers to the questions posed in the introduction. Fig. \ref{plot_overlay_cdens} shows an overlay of the filaments identified in the combined NOEMA and 30\,m data with the filaments identified in the Herschel data in a more extended region around DR20. One sees that the large-scale filaments from the Herschel data connect to some of the small-scale filaments in the CASCADE data, indicating a hierarchy of filamentary structures, which connects large to small scales. Furthermore, we can see multiple velocity components in most filaments in \HCO\ and \HttCO, which indicates unresolved spatial substructure, and therefore a hierarchy in filament networks as well. We also find examples of such multiple components right at the core, which then merge together further from the core (Fig.~\ref{plot_merging_peaks_EHCO}). This may also indicate a richer network of filaments closer to cores, but could partly also be an effect of higher S/N closer to the cores.

\subsection{Is gas flowing along the filaments into the cores?}

In Section \ref{subsect_res_vel} we investigate the gas velocities within selected filaments in \HCO\ and \HttCO. For several filaments in most cores, in \HCO\ and \HttCO, we find projected velocity gradients, i.e., the absolute velocity increasing toward the core with gradients of $\sim$ 0.4 to 2.4\,\kms\ over 0.1\,pc.
Since our data are only 2D in space, it is not unambiguous to interpret such gradients. The velocity of many analyzed filaments is not linear or constant further from the core, and similar or even steeper gradients can be seen there as well. 
We also find filaments with rather constant velocities close to the core (e.g., core B \HCO). Furthermore, we find that sometimes the reason for these gradients may be multiple merging velocity components (Fig.~\ref{plot_merging_peaks_EHCO}).
However, we do find velocity gradients close to the core for most filaments. We suggest that this could be a sign of gas flowing along them into the cores. These results are consistent with similar findings reported in the literature (e.g., \citealt{csengeri2011,henshaw2014,peretto2014}).

Considering the linewidths within filaments, we find generally larger linewidths for \HCO\ than for \HttCO, but all filaments in both emission lines show a linewidth of about 2 \kms\ in the spectrum closest to the core. Furthermore, all filaments in \HttCO\ have a linewidth down to $\sim$ 1 \kms\ (close to the spectral resolution of 0.8\,km\,s$^{-1}$) further away from the core. We notice this tendency of smaller linewidths further away from the cores (Sect.~\ref{subsect_res_vel} and Fig.~\ref{plot_hexbin_lw}). The tendency seems to be slightly clearer for the evolved cores A and B. 
Since there are already signs of star formation in the younger cores, this difference could indicate that the linewidth increase is caused by feedback, but it could also be caused by a gravitational accelerated flow toward the core center. While the linewidth in itself cannot discriminate that scenario, the additionally observed velocity gradients toward the core indicate that inflow as well as feedback may contribute to the increased linewidth toward the center.

\subsection{Do filament properties depend on the evolutionary stage of star formation?}

We investigated several filament properties separately for evolved and younger cores (Fig.~\ref{plot_dr20_overview}). Considering the gas velocity within filaments (see Section \ref{subsect_res_vel}), we find no clear difference between them. While we find a velocity gradient close to the core in most filaments, we do not observe it in the ones connected to core B in \HCO; however, for the same core in \HttCO a gradient is identified. Similarly, for core E we do not find a gradient in \HttCO, but we do find it in \HCO. 
We also investigate the filament width for the younger and more evolved cores separately (Section \ref{subsect_res_Filchap}, Fig. \ref{plot_filchap_AB_CF} and \ref{plot_filchap_AB_CF_H13CO+}). However, we find no differences in the mean and median values of their distributions. 
Furthermore, we looked for changes of filament width along the filaments in the different cores. However, we find no clear trends in general, and again no clear differences between younger and more evolved cores.
We also took a look at the goodness-of-fit values, to check if the “best” fit function differs for the younger and more evolved cores, but we cannot find a clear difference.
Therefore, it seems that filament properties do not depend strongly on the evolutionary stage, but the dataset in this work is too small to draw a firm conclusion.

\subsection{Is there a typical filament width?} 

Many studies about filament widths, using different star formation regions, tracers, and calculation methods, find resulting mean values around 0.1\,pc (e.g., \citealt{arzoumanian2019}), while a recent meta-analysis by \citet{hacar_initial_2022} finds a broader filament width distribution. However, there are not yet many studies focusing on high-mass star-forming regions, or various evolutionary stages of star formation. Therefore, we estimated the filament width distribution for each core, as well as for a combination of all younger cores, and all those more evolved (see Section \ref{subsect_res_Filchap}, especially Fig. \ref{plot_filchap_AB_CF} and \ref{plot_filchap_AB_CF_H13CO+}, and Table \ref{tab_Filchap_all_AF}). For all cores and the various combinations, as well as for all used methods, the filament width distribution has a mean close to 0.1 pc with standard deviations around 0.02 and 0.06\,pc (depending on the tracer). We find small variations between methods, where a Plummer ($p$ = 2) fit always gives the narrowest filament widths, and the second moment always the broadest. Hereby the difference between these two methods in the median is up to 0.08 pc (core C in \HCO). Furthermore, in \HttCO\ we find lower values for all cores (except core E), where the difference to \HCO\ is 0.06 pc (second moment core C). 
In summary, although the filament widths have a relatively wide spread, they are in general distributed around the often discussed 0.1 pc.
As described in the previous paragraph, we find no clear difference between the younger and more evolved cores.
As discussed by \cite{andre_filamentary_2014}, there may be a physical reason behind this typical filament width, for example additional material accretion onto filaments while they contract, which would make the central filament diameter on the order of the effective Jeans length, i.e., $\sim$ 0.1 pc.

The systematic differences we find in the filament widths for the different methods, where a Plummer ($p$ = 2) fit always gives the smallest widths, and the second moment always gives the largest are comparable to \cite{suri_carma-nro_2019}. They describe that the second moment has the strongest correlation with the fitting range of the radial profile, and that Plummer ($p$ = 2) only fits the very inner structure of the radial profiles and excludes wings. 
We show histograms of the goodness-of-fit values for the different fits as given by FilChaP in the Appendix (Figs. \ref{plot_filchap_AB_CF_chi} and \ref{plot_filchap_AB_CF_chi_H13CO+}). The mean values of the distribution of all filaments in \HCO\ is closest to 1 for the Plummer ($p$ = 4) fit, and the median has the lowest value for this fit as well. However, many values of reduced $\chi^2$ $<$ 1 are included, which are not ideal and make this difficult to interpret (see section \ref{subsect_meth_Filchap}).  
It could be argued that for our data Plummer ($p$ = 4) (Eq. \ref{eq_plummer}), with its mean of the distribution closest to 1, describes the filament radial profile best, but on the other hand the more pronounced peak of the histograms at reduced $\chi^2$ $\sim$ 1 may indicate better fits with the Gaussian (Eq. \ref{eq_Gauss}). It might be interesting to note that both Gaussian and Plummer ($p$ = 4) often give similar filament widths.
For \HttCO\ the goodness-of-fit values show less variance between the fits, even though the filament widths do show variance.

\section{Conclusions}
\label{sect_concl}

With the aim to study gas flows from large ($>$\,pc) to small ($<0.1$pc) scales in high-mass star formation, we investigated filaments identified with DisPerSE in the high-mass star-forming region DR20 of Cygnus X, using the high-resolution combination of NOEMA interferometry and IRAM 30 m single-dish data (MIOP-CASCADE), as well as larger scale Herschel hydrogen column density data \citep{marsh_multitemperature_2017}. After outlining the parameters of the data, we described the methods used, i.e., DisPerSE for filament identification, Gaussian fits to the spectra within filaments to determine peak gas velocities and linewidths, and FilChaP for filament width estimation, and explained our assumptions and decisions when using them. 
Our first result is the comparison of the filaments identified in CASCADE and the larger-scale hydrogen column density data, which shows overlap in several regions. This may indicate a connection from large to small scales via filaments. 

We selected a number of filaments around each continuum source, i.e., core, in the CASCADE data, and determined the peak velocities and linewidths of the spectral lines within them. This analysis shows projected velocity gradients of 0.4 to 2.4 \kms\ over a 0.1 pc distance from the core in at least one filament in every region, with only one core in each, \HCO\ (core B) and \HttCO\ (core E), as exceptions. We find that this velocity gradient may in some cases arise from merging velocity components but speculate that it generally indicates gas flowing toward the cores. 
Furthermore, we notice a tendency for smaller linewidths farther from the cores in all data, which is more pronounced in \HttCO\ than \HCO\ , especially for the evolved cores A and B. This tendency also fits into the picture of a gravitationally accelerated accretion flow toward the cores.
With FilChaP we find filament width distributions with median values between 0.06 and 0.14 pc, depending on the method, the tracer, and the core. The smallest width for both \HCO\ and \HttCO\ is always given by the Plummer ($p$ = 2) fit and the largest for \HCO\ by the second moment. For \HttCO\ the second moment and Gaussian fit give similarly large widths. The widths are generally smaller for \HttCO\ than for \HCO. However, we find no clear differences between the evolved and the younger cores. While the filament width distribution has a large spread, the mean and median values are generally around the often discussed value of 0.1\,pc.
Furthermore, we presented the estimated filament width along the length of the filaments but found no clear trends.

Given the wealth of data in CASCADE, this analysis is only the first step. There are many more regions covered by CASCADE in Cygnus X, and larger statistics for the two groups of more evolved and younger star-forming regions may be the key to seeing clearer differences between them. Furthermore, there are many more spectral lines available than the \HCO\ and \HttCO\ used in this work. Since we do find small differences in the filament widths between these two, it would be interesting to look at the results using other tracers, such as N$_2$H$^+$, to trace even higher density closer to the cores. In addition, exploring other filament-finding algorithms that may better differentiate velocity structures during the filament identification process will be important. A problem that remains is that of inclination angles, which make the analysis of absolute gas velocities within filaments difficult. A larger sample size can also help in this regard.

\section*{Data availability}
\label{data_availability}

Supplementary material and figures are provided via Zenodo
at \url{https://doi.org/10.5281/zenodo.17672650}.

\begin{acknowledgements}
This work is based on observations carried out under project number L19MA with the
IRAM NOEMA Interferometer and [145-19] with the 30\,m telescope. IRAM is supported by INSU/CNRS (France), MPG (Germany) and IGN (Spain). D.~S. acknowledges support from the European Research Council under the Horizon 2020 Framework Program via the ERC Advanced Grant No. 832428-Origins.
\end{acknowledgements}

\begin{appendix}

\section{Additional tables}

\begin{table*}
\caption{Linewidths and velocity gradients within filaments.}             
\label{tab_vel}      
\centering                          
\begin{tabular}{c c c c c c c c}        
\hline\hline                 
 & & \multicolumn{3}{c}{\HCO} & \multicolumn{3}{c}{\HttCO} \\
 Core & Filament & Length & Linewidth & Velocity gradient & Length & Linewidth & Velocity gradient \\ 
 & & [pc] & [\kms] & [\kms\ over 0.1 pc] & [pc] & [\kms] & [\kms\ over 0.1 pc] \\
\hline                        
   \multirow{6}{0pt}{A} & 1 & 1.2 & 2.0$^{+1.9}_{-1.1}$ & 0.8 & 0.8 & 1.5$^{+1.9}_{-0.7}$ & 1.7$^{(\#)}$ \\   
   & 2 & 0.3 & 1.7$^{+0.6}_{-0.5}$ & - & 0.5 & 1.6$^{+1.6}_{-0.7}$ & -\\ 
   & 3 & 2.0 & 1.7$^{+1.7}_{-0.9}$ & 1.3 & 0.4 & 1.8$^{+1.6}_{-0.7}$ & A1$^{(*)}$\\
   & 4 & 0.6 & 2.2$^{+1.3}_{-1.4}$ & A3$^{(*)}$ & 3.1 & 1.5$^{+2.2}_{-1.3}$ & -\\
   & 5 & 2.0 & 1.8$^{+2.1}_{-1.0}$ & A3$^{(*)}$ & 2.2 & 1.6$^{+1.3}_{-0.9}$ & -\\
   & 6 & & - & - & 0.8 & 1.4$^{+1.3}_{-0.7}$ & A4$^{(-)}$\\
   \vspace{-15pt}\\
 \hline
   \multirow{4}{0pt}{B} & 1 & 0.4 & 2.5$^{+1.3}_{-1.2}$ & - & 1.0 & 1.8$^{+1.5}_{-0.9}$ & 1.4\\  
    & 2 & 1.0 & 1.6$^{+1.7}_{-0.9}$ & - & 1.3 & 1.6$^{+1.6}_{-0.9}$ & B1$^{(*)}$\\ 
    & 3 & 0.9 & 2.1$^{+1.6}_{-1.1}$ & - & - & - & -\\ 
    & 4 & 1.2 & 2.2$^{+1.5}_{-1.4}$ & - & - & - & -\\ 
    \vspace{-15pt}\\
 \hline
   \multirow{4}{0pt}{C} & 1 & 0.1 & 2.2$^{+0.9}_{-0.5}$ & - & 1.9 & 1.3$^{+1.2}_{-0.4}$ & 0.7\\  
    & 2 & 1.0 & 2.1$^{+1.4}_{-1.1}$ & - & 0.5 & 2.1$^{+1.1}_{-0.9}$ & -\\ 
    & 3 & 0.5 & 1.6$^{+1.2}_{-0.8}$ & 0.5 & 1.6 & 1.6$^{+1.6}_{-0.8}$ & C2$^{(-)}$\\ 
    & 4 & 0.7 & 2.4$^{+1.3}_{-1.5}$ & 0.8 & - & - & -\\ 
    \vspace{-15pt}\\
 \hline
   \multirow{4}{0pt}{D} & 1 & 0.7 & 1.9$^{+2.0}_{-1.1}$ & 0.4 & 1.7 & 1.4$^{+0.7}_{-0.6}$ & -\\ 
    & 2 & 0.6 & 1.8$^{+1.5}_{-0.8}$ & D1$^{(*)}$ & 0.3 & 1.6$^{+1.4}_{-0.8}$ & D1$^{(*)}$\\
    & 3 & 0.4 & 2.2$^{+1.4}_{-1.0}$ & - & 0.2 & 1.2$^{+0.4}_{-0.3}$ & -\\ 
    & 4 & 0.7 & 2.2$^{+1.7}_{-1.4}$ & - & 1.1 & 1.6$^{+2.3}_{-1.0}$ & 0.7\\ 
    \vspace{-15pt}\\
 \hline
   \multirow{4}{0pt}{E} & 1 & 0.5 & 1.9$^{+2.1}_{-1.0}$ & 2.4$^{(\#)}$ & 0.8 & 1.6$^{+2.0}_{-0.8}$ & -\\  
    & 2 & 0.7 & 2.2$^{+1.8}_{-1.1}$ & 2.0$^{(\#)}$ & 2.2 & 1.4$^{+1.4}_{-0.7}$ & -\\ 
    & 3 & 0.9 & 2.3$^{+1.6}_{-1.1}$ & E2$^{(*)}$ & - & - & -\\
    & 4 & 0.4 & 2.4$^{+1.6}_{-1.2}$ & E2$^{(*)}$ & - & - & -\\
    \vspace{-15pt}\\
 \hline
   \multirow{2}{0pt}{F} & 1 & 0.7 & 1.7$^{+2.1}_{-0.8}$ & - & 1.7 & 1.3$^{+1.4}_{-0.5}$ & -\\   
    & 2 & 0.9 & 1.9$^{+2.1}_{-1.0}$ & - & - & -& -\\
    \vspace{-15pt}\\
 \hline
   \multirow{4}{0pt}{A$_{\text{rem}}$} & 1 & 0.7 & 1.7$^{+1.3}_{-0.8}$ & - & - & - & -\\   
    & 2 & 1.2 & 1.6$^{+2.2}_{-0.7}$ & - & - & -& -\\
    & 3 & 0.5 & 2.3$^{+1.5}_{-1.4}$ & - & - & -& -\\
    & 4 & 0.6 & 2.3$^{+1.5}_{-1.4}$ & - & - & -& -\\
    \vspace{-15pt}\\
 \hline
   \multirow{1}{0pt}{B$_{\text{rem}}$} & 1 & 2.2 & 1.9$^{+2.0}_{-1.0}$ & - & - & -& -\\   
\vspace{-10pt}\\

\hline                                   
\multicolumn{8}{c}{(*) has the same spectra (and gradient) close to the core as the mentioned filament.}\\
\multicolumn{8}{c}{(-) has the same spectra (without gradient) close to the core as the mentioned filament.}\\
\multicolumn{8}{c}{(\#) may be the result of two merging velocity components.}

\end{tabular}
\tablefoot{The linewidths are given as mean values and ranges (indicating minimum and maximum values), and the velocity gradients are determined close to each core (within a 0.1 pc distance). All values correspond to the main spectral Gaussian peaks.}
\end{table*}

\begin{table*}
\caption[Filament width mean, median, and standard deviation for all methods of cores A-F in \HCO\ and \HttCO.]{De-convolved filament width statistics.}             
\label{tab_Filchap_all_AF}      
\centering                          
\begin{tabular}{c c c c c c c c c c}        
\hline\hline                 
 & & \multicolumn{4}{c}{\HCO} & \multicolumn{4}{c}{\HttCO} \\
 Core & Method & N & Mean & Median & std & N & Mean & Median & std \\ 
 & & & [pc] & [pc] & [pc] & & [pc] & [pc] & [pc] \\
\hline                        
   \multirow{4}{0pt}{A} & Gaussian & 56 & 0.10 & 0.10 & 0.04 & 46 & 0.10 & 0.10 & 0.04  \\ 
   & Plummer ($p$ = 2)  & 53 & 0.08 & 0.08 & 0.04 & 45 & 0.07 & 0.06 & 0.03  \\ 
   & Plummer ($p$ = 4)  & 59 & 0.09 & 0.09 & 0.04 & 46 & 0.08 & 0.08 & 0.04  \\
   & Second Moment  & 52 & 0.12 & 0.13 & 0.05 & 44 & 0.10 & 0.09 & 0.04  \\
   \vspace{-15pt}\\
\hline                        
   \multirow{4}{0pt}{B} & Gaussian & 26 & 0.12 & 0.12 & 0.04 & 18 & 0.10 & 0.10 & 0.03 \\ 
   & Plummer ($p$ = 2)  & 28 & 0.08 & 0.08 & 0.04 & 19 & 0.07 & 0.07 & 0.03  \\ 
   & Plummer ($p$ = 4)  & 29 & 0.11 & 0.11 & 0.04 & 19 & 0.09 & 0.08 & 0.03  \\
   & Second Moment  & 24 & 0.14 & 0.14 & 0.04 & 18 & 0.11 & 0.10 & 0.04 \\
   \vspace{-15pt}\\
\hline                        
   \multirow{4}{0pt}{C} & Gaussian & 17 & 0.12 & 0.12 & 0.04 & 18 & 0.11 & 0.09 & 0.05  \\ 
   & Plummer ($p$ = 2)  & 15 & 0.07 & 0.06 & 0.03 & 18 & 0.07 & 0.07 & 0.03  \\ 
   & Plummer ($p$ = 4)  & 17 & 0.10 & 0.10 & 0.04 & 17 & 0.09 & 0.08 & 0.04  \\
   & Second Moment  & 14 & 0.14 & 0.14 & 0.04 & 17 & 0.11 & 0.08 & 0.05  \\
   \vspace{-15pt}\\
\hline                        
   \multirow{4}{0pt}{D} & Gaussian & 15 & 0.11 & 0.11 & 0.03 & 21 & 0.09 & 0.08 & 0.03  \\ 
   & Plummer ($p$ = 2)  & 15 & 0.07 & 0.06 & 0.03 & 19 & 0.06 & 0.06 & 0.02  \\ 
   & Plummer ($p$ = 4)  & 15 & 0.09 & 0.08 & 0.03 & 22 & 0.08 & 0.08 & 0.02  \\
   & Second Moment  & 15 & 0.13 & 0.12 & 0.02 & 19 & 0.10 & 0.08 & 0.03  \\
   \vspace{-15pt}\\
\hline                        
   \multirow{4}{0pt}{E} & Gaussian & 32 & 0.11 & 0.11 & 0.04 & 13 & 0.11 & 0.12 & 0.03  \\ 
   & Plummer ($p$ = 2)  & 30 & 0.08 & 0.08 & 0.03 & 16 & 0.09 & 0.08 & 0.04  \\ 
   & Plummer ($p$ = 4)  & 32 & 0.10 & 0.10 & 0.03 & 15 & 0.10 & 0.11 & 0.03  \\
   & Second Moment  & 30 & 0.14 & 0.14 & 0.04 & 13 & 0.11 & 0.12 & 0.04  \\
   \vspace{-15pt}\\
\hline                        
   \multirow{4}{0pt}{F} & Gaussian & 5 & 0.12 & 0.11 & 0.02 & 5 & 0.10 & 0.09 & 0.06  \\ 
   & Plummer ($p$ = 2)  & 4 & 0.08 & 0.06 & 0.03 & 5 & 0.05 & 0.06 & 0.03  \\ 
   & Plummer ($p$ = 4)  & 5 & 0.11 & 0.09 & 0.03 & 5 & 0.08 & 0.07 & 0.04  \\
   & Second Moment  & 4 & 0.11 & 0.11 & 0.02 & 5 & 0.10 & 0.10 & 0.04  \\
   \vspace{-15pt}\\
\hline                        
   \multirow{4}{18pt}{A-B} & Gaussian & 82 & 0.11 & 0.11 & 0.04 & 64 & 0.10 & 0.10 & 0.04  \\ 
   & Plummer ($p$ = 2)  & 81 & 0.08 & 0.08 & 0.04 & 64 & 0.07 & 0.07 & 0.04  \\ 
   & Plummer ($p$ = 4)  & 88 & 0.10 & 0.10 & 0.04 & 65 & 0.09 & 0.08 & 0.04  \\
   & Second Moment  & 76 & 0.13 & 0.13 & 0.04 & 62 & 0.10 & 0.10 & 0.04  \\
   \vspace{-15pt}\\
\hline                        
   \multirow{4}{18pt}{C-F} & Gaussian & 69 & 0.11 & 0.11 & 0.04 & 57 & 0.10 & 0.10 & 0.04  \\ 
   & Plummer ($p$ = 2)  & 64 & 0.08 & 0.07 & 0.03 & 58 & 0.07 & 0.07 & 0.03  \\ 
   & Plummer ($p$ = 4)  & 67 & 0.10 & 0.10 & 0.03 & 59 & 0.09 & 0.09 & 0.03  \\
   & Second Moment  & 63 & 0.13 & 0.13 & 0.03 & 54 & 0.10 & 0.10 & 0.04  \\
   \vspace{-15pt}\\
\hline                                   

\end{tabular}
\tablefoot{"std" refers to the Gaussian standard deviation. The statistics include the chosen filaments connected to each core A to F separately, and in the last two rows together for the evolved cores A and B and the younger cores C to F.}
\end{table*}

\begin{table*}
\caption[Filament width mean, median and std for all methods of the remote filaments in \HCO.]{Deconvolved filament width statistics.}             
\label{tab_Filchap_all_rem}      
\centering                          
\begin{tabular}{c c c c c c c c c c}        
\hline\hline                 
 & & \multicolumn{4}{c}{\HCO} & \multicolumn{4}{c}{\HttCO} \\
 Core & Method & N & Mean & Median & std & N & Mean & Median & std \\ 
 & & & [pc] & [pc] & [pc] & & [pc] & [pc] & [pc] \\
\hline                        
   \multirow{4}{18pt}{A$_{\text{rem}}$} & Gaussian & 15 & 0.10 & 0.09 & 0.03 & & & & \\
   & Plummer ($p$ = 2)  & 14 & 0.08 & 0.08 & 0.03  & & & & \\ 
   & Plummer ($p$ = 4)  & 15 & 0.09 & 0.09 & 0.03  & & & & \\
   & Second Moment  & 13 & 0.10 & 0.10 & 0.04  & & & & \\
   \vspace{-15pt}\\
\hline                        
   \multirow{4}{18pt}{B$_{\text{rem}}$} & Gaussian & 7 & 0.11 & 0.11 & 0.04 & & & & \\ 
   & Plummer ($p$ = 2)  & 7 & 0.07 & 0.08 & 0.02  & & & & \\ 
   & Plummer ($p$ = 4)  & 8 & 0.10 & 0.11 & 0.03  & & & & \\
   & Second Moment  & 6 & 0.10 & 0.10 & 0.03  & & & & \\
   \vspace{-15pt}\\
\hline
   \multirow{4}{13pt}{All} & Gaussian & 173 & 0.11 & 0.11 & 0.04 & 121 & 0.10 & 0.10 & 0.04  \\ 
   & Plummer ($p$ = 2)  & 166 & 0.08 & 0.08 & 0.04 & 122 & 0.07 & 0.07 & 0.03  \\ 
   & Plummer ($p$ = 4)  & 178 & 0.10 & 0.10 & 0.04 & 124 & 0.09 & 0.08 & 0.04  \\
   & Second Moment  & 158 & 0.13 & 0.13 & 0.04 & 116 & 0.10 & 0.10 & 0.04  \\
   \vspace{-15pt}\\

\end{tabular}
\tablefoot{The term "std" refers to the Gaussian standard deviation. The statistics include the remote filament(s) close to core A and B in \HCO$(1-0)$ separately, and all analyzed filaments together in the last row.}
\end{table*}

\section{FilChaP features}
\label{sect_app_filchap}

\paragraph{Fitting range:}

Since filaments are not isolated, but encompassed by background emission and sometimes even crossed by other filaments, the fitting range of the radial profiles influences the resulting filament width \citep{smith_nature_2014,suri_carma-nro_2019}. FilChaP tries to solve this by determining an individual fitting range for each averaged profile. After a baseline subtraction of the profile, the first minima to both sides closest to the main peak with a significance of 3$\sigma$ are detected, and their coordinates are used as beginning and end of the fitting range. However, even though the effect is less severe, the filament width is still not entirely independent of the fitting range \citep[s. Fig. 8 in][]{suri_carma-nro_2019}. 

\paragraph{Velocity range:}
To minimize the background emission to a certain degree, the intensity profiles calculated by FilChaP are not integrated over the entire velocity range of the data. Rather, the velocity inside the current slice of filament is determined, and the integrated intensity of the radial profile is calculated using the intensity of the two neighbouring velocity channels, i.e., one channel after and one before, of the filament's peak velocity. This is done to exclude emission that is not connected to the filament slice currently studied.

\paragraph{Average length:}
In previous studies the radial profiles have often been averaged over the filament length, to increase the signal-to-noise ratio (S/N). This not only leads to spatial information along the filament being lost, it also results in an overestimation of the filament width \citep{suri_carma-nro_2019}. Therefore, \cite{suri_carma-nro_2019} choose to average only over a length of three beam sizes, to still obtain a smooth radial profile, but minimize the overestimation. Additionally, FilChaP therefore makes it possible to study changes in width along filaments. This average length can be given as input setting. A comparison of different average lengths can be seen in Fig. 4 of \cite{suri_carma-nro_2019}. 

\section{Additional figures} 
\label{add_figs}

\begin{figure*}[h]
\includegraphics[width=0.99\linewidth ,keepaspectratio]{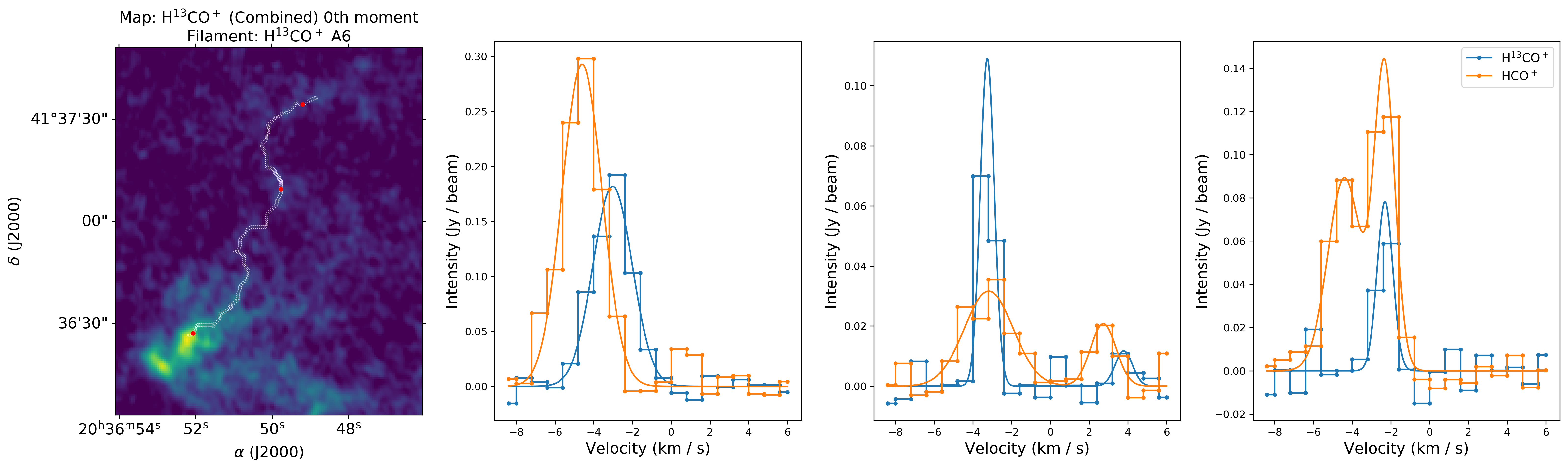}
\caption{Example map and spectra in H$^{13}$CO$^+$(1--0) (blue) and HCO$^+$(1--0) (orange) for three positions along filament A6, as marked by the red dots in the map.}
\label{comparison}
\end{figure*}

\begin{figure*}[h]
	\centering
	\begin{subfigure}{.44\textwidth}
  		\centering
		\includegraphics[width=0.97\linewidth ,keepaspectratio]
		{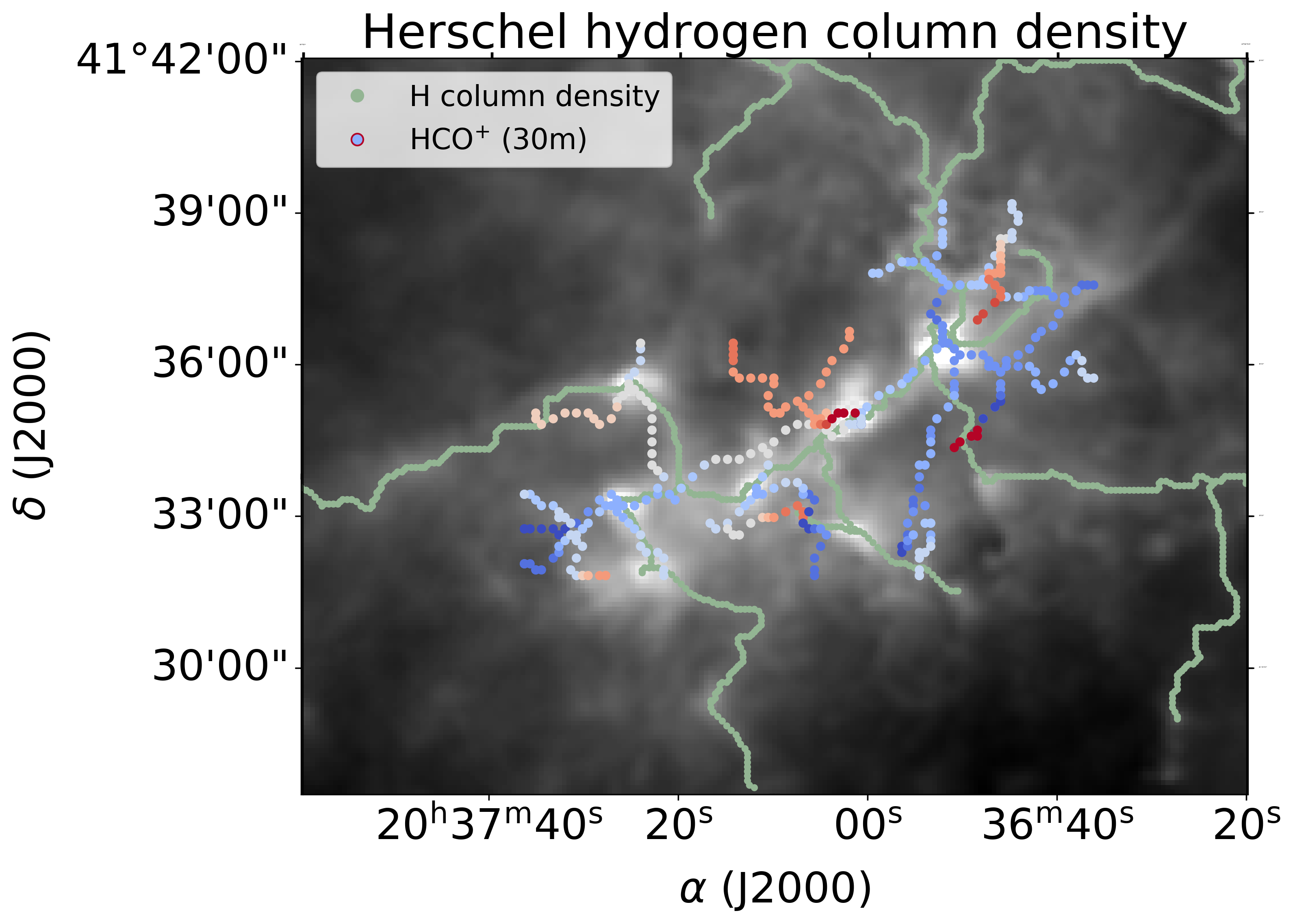}
	\end{subfigure}%
	\begin{subfigure}{.56\textwidth}
		\centering
		\includegraphics[width=0.97\linewidth ,keepaspectratio]
		{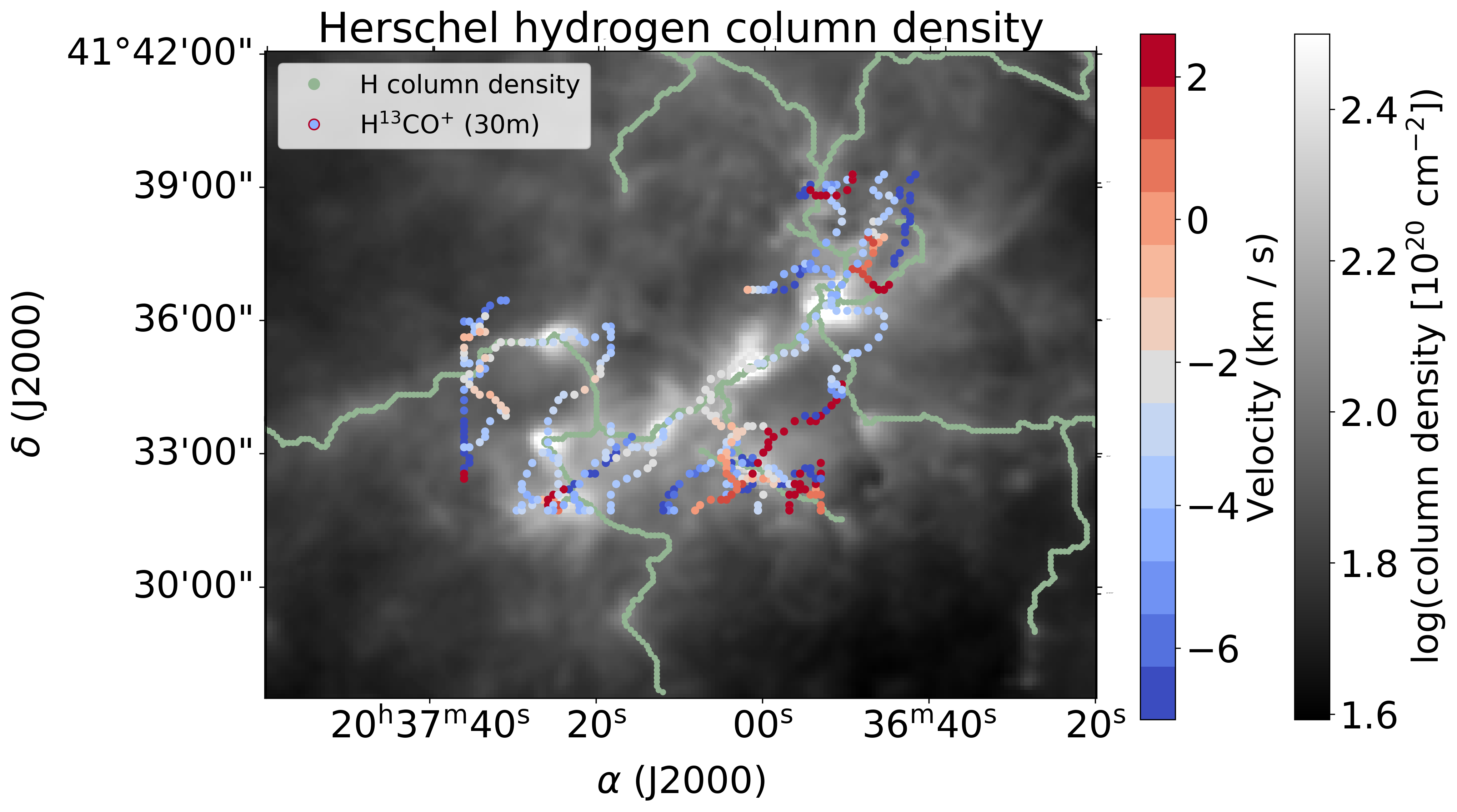}
	\end{subfigure}
\caption[Overlay of MIOP data (30 m \HCO  and \HttCO) filaments and Herschel hydrogen column density filaments and map.]{Filaments identified in the 30 m CASCADE data overlaid on the Herschel hydrogen column density image and its filaments. \textit{Left}: \HCO$(1-0)$. \textit{Right}: \HttCO$(1-0)$.}
\label{plot_overlay_cdens_30m}
\end{figure*}

\begin{figure*}[h]
	\centering
	\begin{subfigure}{.33\textwidth}
  		\centering
		\includegraphics[width=0.97\linewidth ,keepaspectratio]
		{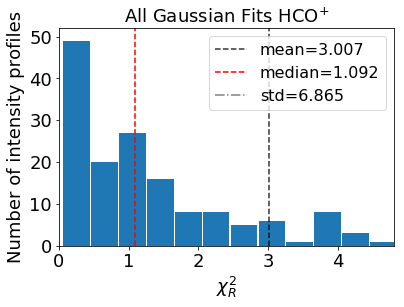}
	\end{subfigure}%
	\begin{subfigure}{.33\textwidth}
		\centering
		\includegraphics[width=0.97\linewidth ,keepaspectratio]
		{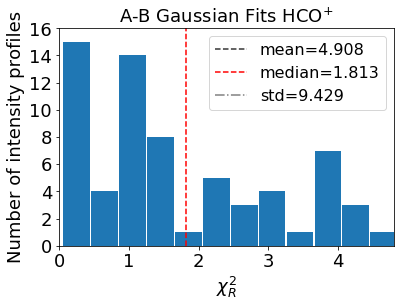}
	\end{subfigure}
	\begin{subfigure}{.33\textwidth}
		\centering
		\includegraphics[width=0.97\linewidth ,keepaspectratio]
		{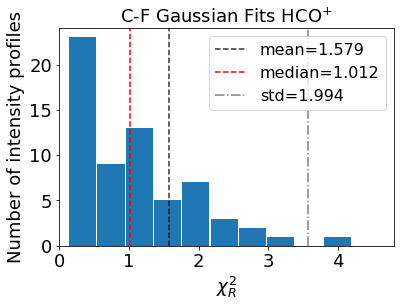}
	\end{subfigure}
		\begin{subfigure}{.33\textwidth}
  		\centering
		\includegraphics[width=0.97\linewidth ,keepaspectratio]
		{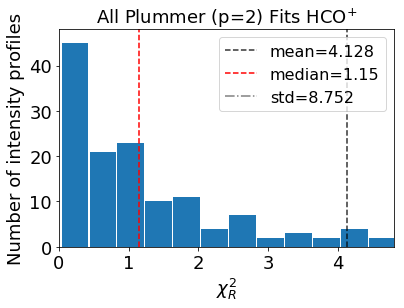}
	\end{subfigure}%
	\begin{subfigure}{.33\textwidth}
		\centering
		\includegraphics[width=0.97\linewidth ,keepaspectratio]
		{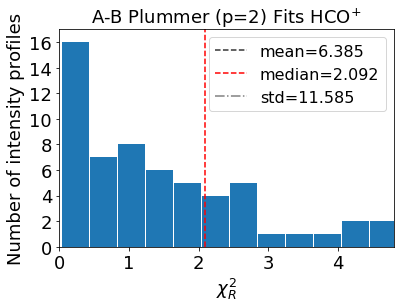}
	\end{subfigure}
	\begin{subfigure}{.33\textwidth}
		\centering
		\includegraphics[width=0.97\linewidth ,keepaspectratio]
		{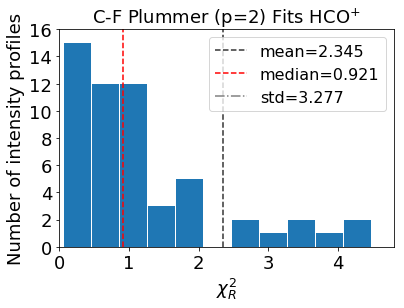}
	\end{subfigure}
		\begin{subfigure}{.33\textwidth}
  		\centering
		\includegraphics[width=0.97\linewidth ,keepaspectratio]
		{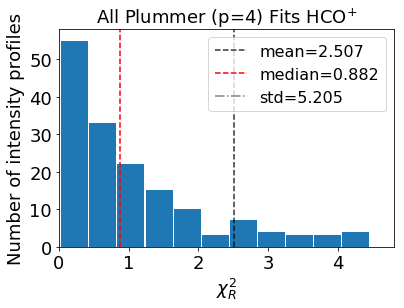}
	\end{subfigure}%
	\begin{subfigure}{.33\textwidth}
		\centering
		\includegraphics[width=0.97\linewidth ,keepaspectratio]
		{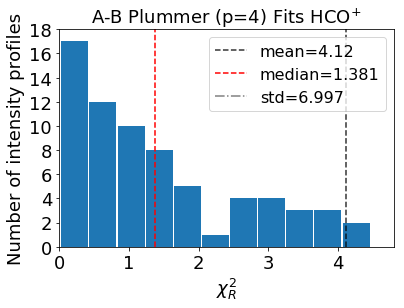}
	\end{subfigure}
	\begin{subfigure}{.33\textwidth}
		\centering
		\includegraphics[width=0.97\linewidth ,keepaspectratio]
		{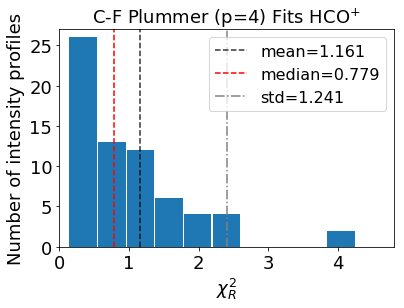}
	\end{subfigure}
\caption[Goodness-of-fit for different FilChaP fits in \HCO.]{Calculated goodness-of-fit values ($\chi^2_R$ $\leq$ 4.8) of the different fits done with FilChaP on the combined \HCO$(1-0)$ data. \textit{Left column}: All regions including cores A to F and the remote filaments close to A and B. \textit{Middle column}: Only filaments directly connected to the evolved regions A and B. \textit{Right column}: Only filaments directly connected to the younger regions C, D, E and F. \textit{Rows}: Different fits, i.e., Gaussian, Plummer $p$ = 2 and $p$ = 4, from top to bottom.}
\label{plot_filchap_AB_CF_chi}
\end{figure*}

\begin{figure*}[h]
	\centering
	\begin{subfigure}{.33\textwidth}
  		\centering
		\includegraphics[width=0.97\linewidth ,keepaspectratio]
		{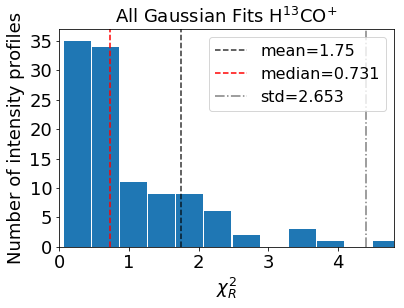}
	\end{subfigure}%
	\begin{subfigure}{.33\textwidth}
		\centering
		\includegraphics[width=0.97\linewidth ,keepaspectratio]
		{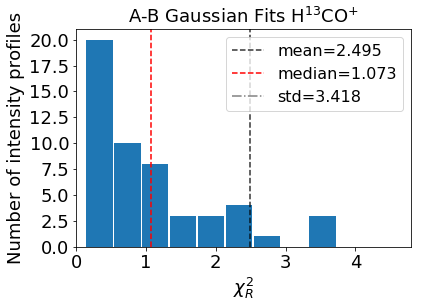}
	\end{subfigure}
	\begin{subfigure}{.33\textwidth}
		\centering
		\includegraphics[width=0.97\linewidth ,keepaspectratio]
		{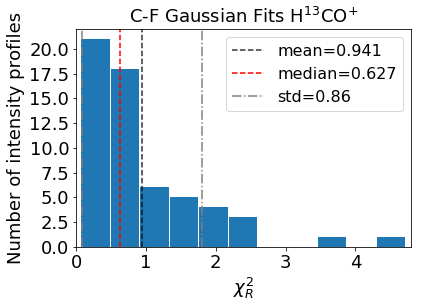}
	\end{subfigure}
		\begin{subfigure}{.33\textwidth}
  		\centering
		\includegraphics[width=0.97\linewidth ,keepaspectratio]
		{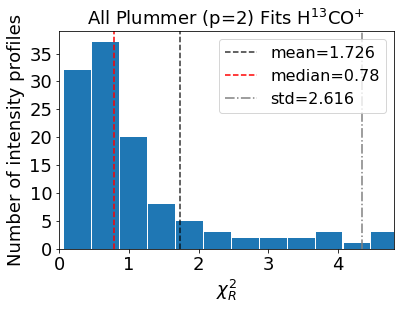}
	\end{subfigure}%
	\begin{subfigure}{.33\textwidth}
		\centering
		\includegraphics[width=0.97\linewidth ,keepaspectratio]
		{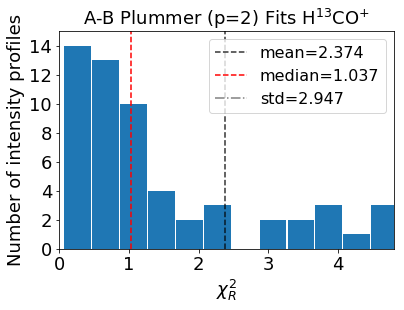}
	\end{subfigure}
	\begin{subfigure}{.33\textwidth}
		\centering
		\includegraphics[width=0.97\linewidth ,keepaspectratio]
		{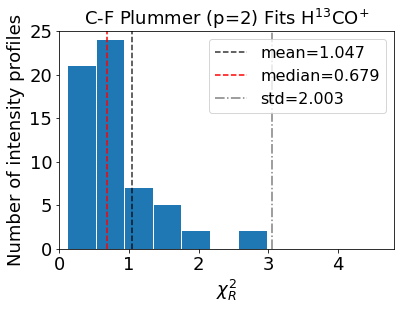}
	\end{subfigure}
		\begin{subfigure}{.33\textwidth}
  		\centering
		\includegraphics[width=0.97\linewidth ,keepaspectratio]
		{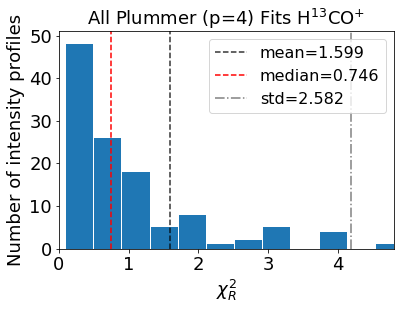}
	\end{subfigure}%
	\begin{subfigure}{.33\textwidth}
		\centering
		\includegraphics[width=0.97\linewidth ,keepaspectratio]
		{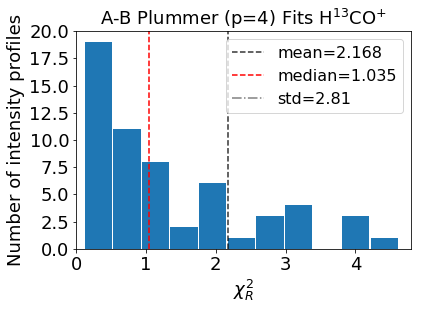}
	\end{subfigure}
	\begin{subfigure}{.33\textwidth}
		\centering
		\includegraphics[width=0.97\linewidth ,keepaspectratio]
		{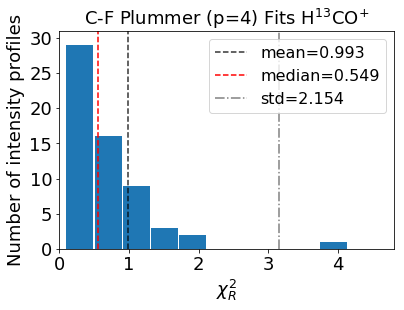}
	\end{subfigure}
\caption[Goodness-of-fit for different FilChaP fits in \HttCO.]{Calculated goodness-of-fit values ($\chi^2_R$ $\leq$ 4.8) of the different fits done with FilChaP on the combined \HttCO$(1-0)$ data. \textit{Left column}: All regions A to F. \textit{Middle column}: Only filaments directly connected to the evolved regions A and B. \textit{Right column}: Only filaments directly connected to the younger regions C, D, E and F. \textit{Rows}: Different fits, i.e., Gaussian, Plummer $p$ = 2 and $p$ = 4, from top to bottom.}
\label{plot_filchap_AB_CF_chi_H13CO+}
\end{figure*}

Additional figures shown are: 

\begin{itemize}

\item Gig.~\ref{comparison} compares a few example spectra for HCO$^+$ and H$^{13}$CO$^+$ to outline real multiple components or optical depth effects.

\item Fig. \ref{plot_overlay_cdens_30m}: Overlay of CASCADE data (30 m \HCO\  and \HttCO) filaments and Herschel hydrogen column density filaments and map (s. beginning of Section 4 in Sawczuck et al. 2026).

\item Fig. \ref{plot_filchap_AB_CF_chi} and Fig. \ref{plot_filchap_AB_CF_chi_H13CO+}: Goodness-of-fit values for the different fits used in FilChaP (s. Section 4.2 in Sawczuck et al. 2026).

\end{itemize}

\end{appendix}

\end{document}